\documentclass[useAMS, usenatbib, usegraphicx]{mn2e} 
\usepackage{amssymb}

\newcommand{\si}{~\mbox{s}^{-1}}
\newcommand{\Myr}{~\mbox{Myr}}
\newcommand{\Gyr}{~\mbox{Gyr}}
\newcommand{\eV}{~\mbox{eV}}

\newcommand{\cMpch}{~h^{-1}~\mbox{comoving Mpc}}

\newcommand{\cmsq}{~\mbox{cm}^{2}}

\newcommand{\cmsqi}{~\mbox{cm}^{-2}}
\newcommand{\cmci}{~\mbox{cm}^{-3}}
\newcommand{\cmc}{~\mbox{cm}^{ 3}}
\newcommand{\kpc}{~\mbox{kpc}}
\newcommand{\K}{~\mbox{K}}

\newcommand{\traphic}{{\sc traphic}}
\newcommand{\testtraphic}{{\sc tt1d}}
\newcommand{\TRAPHIC}{{\sevensize\bf TRAPHIC}}

\newcommand{\gadget}{{\sc gadget-2}}

\newcommand{\ctworay}{{\sc c}$^2${\sc -ray}}
\newcommand{\crash}{{\sc crash}}

\newcommand{\ftte}{{\sc ftte}}

\newcommand{\yr}{~\mbox{yr}}

\newcommand{\invs}{~\mbox{s}^{-1}}

\newcommand{\cms}{~\mbox{cm}^2}

\newcommand{\cloudy}{{\sc cloudy}}

\title[TRAPHIC - multi-frequency and thermal coupling]{Multi-frequency, thermally coupled radiative transfer with TRAPHIC: Method and tests}

\author[A. H. Pawlik and J. Schaye] 
       {Andreas H. Pawlik$^{1}$$^{,2}$\thanks{E-mail: pawlik@astro.as.utexas.edu}
	 and
	 Joop Schaye$^{2}$\thanks{E-mail: schaye@strw.leidenuniv.nl}\\
	 $^{1}$Department of Astronomy and Texas Cosmology Center, The University of Texas at Austin, TX 78712\\
	 $^{2}$Leiden Observatory, Leiden University, P.O. Box 9513, 2300RA Leiden, The Netherlands}

\begin{document}

\date{Accepted; Received; in original form}

\pagerange{\pageref{firstpage}--\pageref{lastpage}} \pubyear{}

\maketitle

\label{firstpage}

\begin{abstract}
We present an extension of \traphic, the method for radiative transfer
of ionising radiation in smoothed particle hydrodynamics simulations
that we introduced in \citet{Pawlik:2008}. The new version keeps all
advantages of the original implementation: photons are transported at
the speed of light, in a photon-conserving manner, directly on the
spatially adaptive, unstructured grid traced out by the particles, in
a computation time that is independent of the number of radiation
sources, and in parallel on distributed memory machines.  We extend
the method to include multiple frequencies, both hydrogen and helium,
and to model the coupled evolution of the temperature and ionisation
balance. We test our methods by performing a set of simulations of
increasing complexity and including a small cosmological reionisation
run. The results are in excellent agreement with exact solutions,
where available, and also with results obtained with other codes if we
make similar assumptions and account for differences in the atomic
rates used. We use the new implementation to illustrate the
differences between simulations that compute photoheating in the grey
approximation and those that use multiple frequency bins. We show that
close to ionising sources the grey approximation asymptotes to the
multi-frequency result if photoheating rates are computed in the
optically thin limit, but that the grey approximation breaks down
everywhere if, as is often done, the optically thick limit is assumed.
\end{abstract}


\section{Introduction}
New telescopes such as Planck\footnote{sci.esa.int/planck/},
LOFAR\footnote{http://www.lofar.org},
MWA\footnote{http://www.haystack.mit.edu/ast/arrays/mwa/},
ALMA\footnote{http://www.almaobservatory.org/} and
JWST\footnote{http://www.jwst.nasa.gov/} will soon open up new
windows onto the epoch of reionisation (e.g., \citealp{Barkana:2001};
\citealp{Ciardi:2005}; \citealp{Fan:2006}; \citealp{Furlanetto:2006}
for reviews of this epoch).  Data collected by these telescopes is
expected to shed light on many unresolved issues in our current
understanding of how galaxies form and evolve and interact with their
surroundings.  Detailed theoretical studies, however, will be needed
to interpret it. Amongst the most promising techniques to perform such
studies are cosmological simulations of reionisation.
\par
Modern simulations of reionisation aim to combine the first-principle
modelling of the gravitational growth of density fluctuations and of
the hydrodynamical evolution of the cosmic gas in the expanding
Universe with recipes for star formation and associated feedback and
to follow also the propagation of ionising radiation emitted by the
first ionising sources.  The computationally efficient, but accurate implementation
of the radiative transfer (RT) is currently one of the biggest
challenges for simulating reionisation.
\par
Computing the ionising intensity throughout the simulation box
requires solving the seven-dimensional (three space coordinates, two
directional coordinates, frequency and time) RT equation. This is a
formidable task, not only because of the high dimensionality of the
problem, but also because of the large number of ionising sources
contained in typical cosmological volumes. To accomplish it, existing
approaches (e.g., \citealp{Abel:1999}; \citealp{Gnedin:2001};
\citealp{Ciardi:2001}; \citealp{Nakamoto:2001};
\citealp{Maselli:2003}; \citealp{Razoumov:2005};
\citealp{Mellema:2006}; \citealp{Susa:2006};
\citealp{Ritzerveld:2006}; \citealp{McQuinn:2007};
\citealp{Semelin:2007}; \citealp{Trac:2007}; \citealp{Pawlik:2008};
\citealp{Aubert:2008}; \citealp{Altay:2008}; \citealp{Petkova:2009};
\citealp{Finlator:2009}; \citealp{Gritschneder:2009}; \citealp{Paardekooper:2010}; \citealp{Hasegawa:2010}; \citealp{Canta:2010}; 
\citealp{Partl:2010}) to transport ionising photons must often resort to a number of approximations.
\par
The accuracy of several ionising (cosmological) RT codes has been
assessed in test simulations that were performed as part of a series
of comparison projects (\citealp{Iliev:2006a};
\citealp{Iliev:2009}). The results of the comparisons are encouraging
and indicate that the participating codes have reached a certain level
of maturity (\citealp{Iliev:2009}). The design of most of the test
simulations was kept simple in order to facilitate comparisons between
different RT codes. More recently, the performance of different RT
codes has been compared in cosmological simulations of reionisation
with an equally promising degree of agreement
(\citealp{Zahn:2010}). However, the inclusion of RT in state-of-the-art
simulations of structure formation remains a tough
computational challenge, as we now explain.
\par
RT codes that are both spatially adaptive and parallel on distributed
memory are still rare (see, e.g., Table~1 in \citealp{Iliev:2006a} and
Table~1 in \citealp{Iliev:2009}). Nearly all reionisation simulations
are therefore performed on uniform grids. Combined with the fact that large simulation boxes are
needed to model representative volumes of the Universe, this means
that the spatial resolution of state-of-the-art RT simulations of
reionisation is typically far below that of the underlying spatially
adaptive hydrodynamical simulations. In fact, many RT simulations of
reionisation ignore hydrodynamical effects altogether and assume the
gas traces the dark matter. Small-scale structure in the cosmic gas is
therefore often ignored or included only in a statistically sense.
\par
Cosmological simulations of reionisation typically contain millions of
star particles (e.g., \citealp{Iliev:2006b}). Large numbers of
ionising sources pose a challenge to simulations of reionisation
because for most of the existing RT methods the computation times
increases linearly with the source number. The usual practice of
reducing the number of ionising sources by combining sources that fall
into the same cell of a superimposed mesh renders reionisation
simulations feasible, but also reduces the spatial resolution at which
the RT is performed.  Note that the inclusion of diffuse ionising
radiation emitted by recombining ions further increases the number of
ionising sources. To reduce the computational effort, this
recombination radiation is therefore usually treated using the
on-the-spot approximation (e.g., \citealp{Osterbrock:1989}), which
assumes it to be re-absorbed in the immediate vicinity of the
recombining ion. However, the validity of this approximation remains
to be assessed (e.g., \citealp{Ritzerveld:2005};
\citealp{Williams:2009}, \citealp{Hasegawa:2010}).
\par
RT simulations of reionisation are still often performed by
post-processing pre-computed static density fields. This static
approximation is appropriate for simulating the initial phase of rapid
growth of ionised regions or the propagation of ionisation fronts on
cosmological scales (see, e.g., the discussion in
\citealp{Iliev:2006b}). Once the speed of ionisation fronts becomes
comparable to the sound speed of the ionised gas, the static
approximation, however, becomes inapplicable and a full
radiation-hydrodynamical treatment is required. In any case, the
static approximation breaks down after about a sound-crossing time, as
the Jeans filtering of the gas can then no longer be ignored (e.g.,
\citealp{Gnedin:2000}). Although radiation-hydrodynamical feedback
from reionisation is known to play a key role, most of the large-scale
reionisation simulations performed to date ignore it.
\par
In Pawlik \& Schaye (2008, hereafter Paper~I) we presented the RT
method \traphic\ (TRAnsport of PHotons In Cones) for use in Smoothed
Particle Hydrodynamics (SPH; \citealp{Gingold:1977};
\citealp{Lucy:1977}) simulations.  \traphic\ can be used to solve both
the time-independent and the time-dependent RT equation in an
explicitly photon-conserving manner. It employs the full
spatial resolution of the underlying SPH simulation because it works
directly on the unstructured grid formed by the discrete set of SPH
particles. It achieves directed transport of radiation on the
irregular distribution of SPH particles by tracing photon packets
inside cones. The solid angle of these cones thereby sets the 
angular resolution at which the RT is performed. 
\traphic\ is by construction parallel on distributed memory machines
if the SPH simulation itself is parallel on distributed memory
machines.  
\par
The computational cost for simulations with \traphic\ is independent 
of the number of ionising sources. It merely scales with the product of the number of spatial
and angular resolution elements, i.e. with the number of SPH particles and
the number of cones needed to tessellate the sky. For comparison, the 
computational cost of conventional ray and
photon tracing methods scales with the product of the number of
spatial resolution elements (gas particles or gas cells) and the
number of sources. Since the number of sources is typically 
proportional to the number of spatial resolution elements,
conventional ray and photon tracing methods face an expensive scaling
with the square of the number of spatial resolution elements.  In
contrast, a relatively small number of angular resolution elements is
typically sufficient to obtain converged results (Paper~I; see also, e.g., 
\citealp{Trac:2007}; \citealp{Paardekooper:2010}), and this, in combination with
the independence of the computational cost of the source number, makes
\traphic\ ideal for simulations containing large numbers of sources
(as is the case for, e.g.,  reionisation simulations) as well as for an
explicit treatment of the diffuse radiation component.
\par
In Paper~I we presented an implementation of \traphic\ for use on
(sets of) static density fields in the SPH code {\sc gadget}
(\citealp{Springel:2005}). We applied this implementation to the
transport of monochromatic ionising radiation in hydrogen-only gas at
a fixed temperature. We demonstrated its excellent performance in
several (static density field) test problems that were designed to
allow a detailed comparison to results obtained with other RT codes.
Here we describe, test and discuss an extension of this
implementation.  The new implementation of \traphic\ allows for the
transport of multi-frequency radiation in primordial gas, i.e., in gas
consisting of both hydrogen and helium. In addition to the computation
of the ionisation state, it also allows for the self-consistent
computation of the temperature of photoionised gas. The new
implementation still solves the RT equation only on static density
fields.  The radiation-hydrodynamical coupling of \traphic\ will be
described in a future work.
\par
Because the computational cost for RT simulations is typically
proportional to the number of frequencies at which the RT equation is
solved, many RT simulations of ionising radiation discretize the RT
problem using only a single frequency bin. In the grey approximation 
(e.g., \citealp{Mihalas:1984}), ionising radiation within this bin is then 
assigned an effective absorption cross-section and an
effective photoenergy that is injected in the gas upon absorption (for examples see, e.g., \citealp{Iliev:2006a}). The 
corresponding photoionisation heating rates can be computed assuming the optically
thin or the optically thick limit. We use our new implementation
to show that RT simulations which employ the grey approximation yield gas temperatures that agree with the exact multi-frequency solution 
only if grey photoheating rates are computed in the optically thin limit and then only
close to the ionising sources. Grey RT 
simulations that compute photoheating rates in the optically thick
limit significantly overestimate the typical temperatures of the photo-heated
gas. A related treatment of this subject and a discussion of its 
astrophysical implications can be found in \cite{AbelHaehnelt:1999}.
\par
The structure of this paper is as follows. In Sec.~\ref{Sec:traphic}
we present a brief review of the main concepts behind \traphic. In
Sec.~\ref{Sec:Theory} we then discuss the equations that govern the
evolution of the ionisation state and temperature of gas exposed to
ionising radiation. With these preparations in hand we are ready to
present our new implementation of \traphic\ in
Sec.~\ref{Sec:Implementation} (and in the appendix). We discuss an
extensive set of tests of this implementation (on static density
fields) in Sec.~\ref{Sec:Tests}. There we also investigate the applicability of the grey 
approximation by comparing simulations using the grey approximation with simulations using multiple 
frequency bins. We conclude with a brief summary in Sec.~\ref{Sec:Summary}.

\section{TRAnsport of PHotons In Cones}
\label{Sec:traphic}
\begin{figure*} 
 \begin{center}
  \includegraphics[trim = 0mm 10mm 0mm 0mm, width=0.49\textwidth, clip=true]{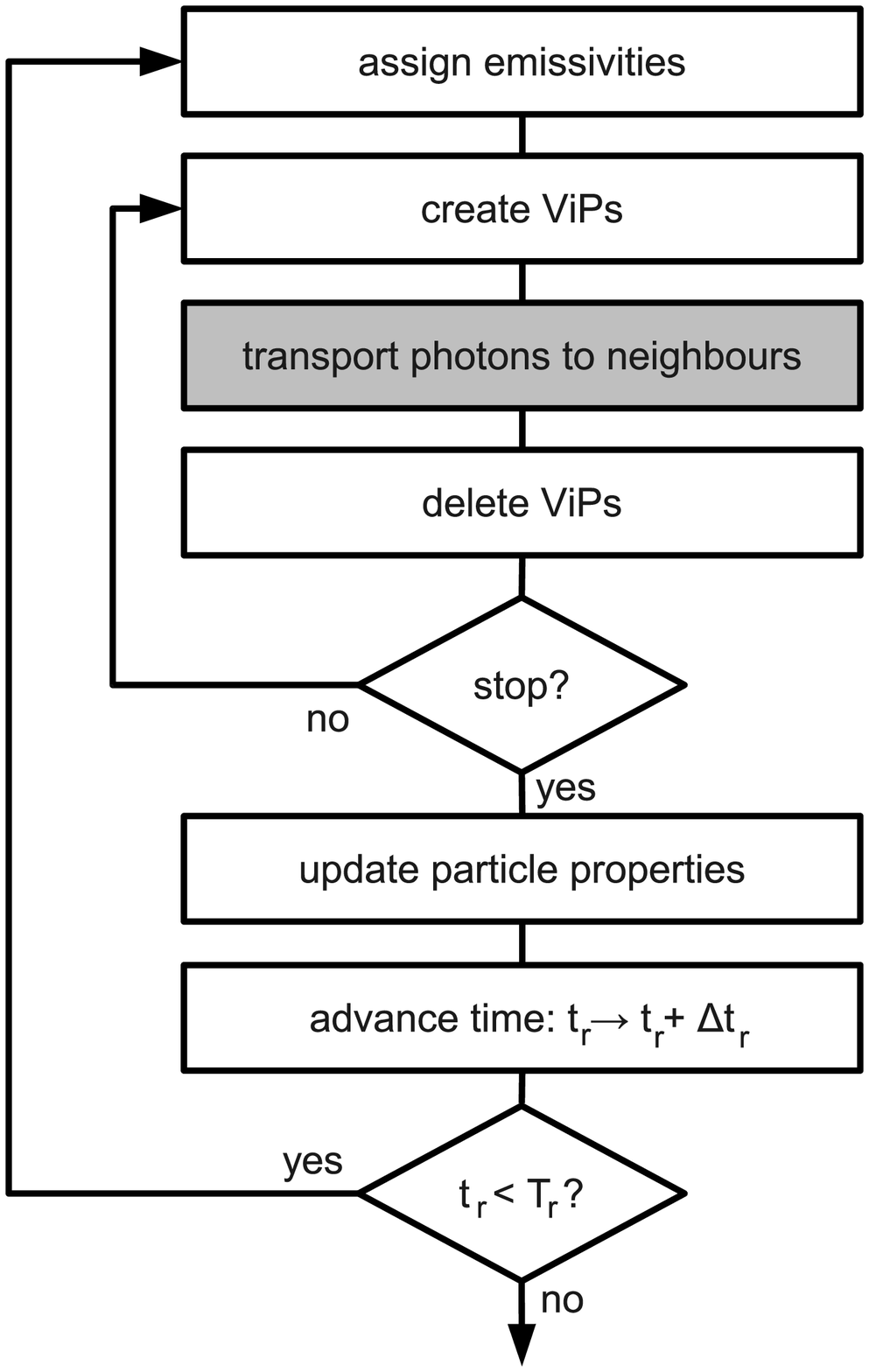} 
  \includegraphics[trim = 0mm 0mm 0mm 0mm, width=0.49\textwidth, clip=true]{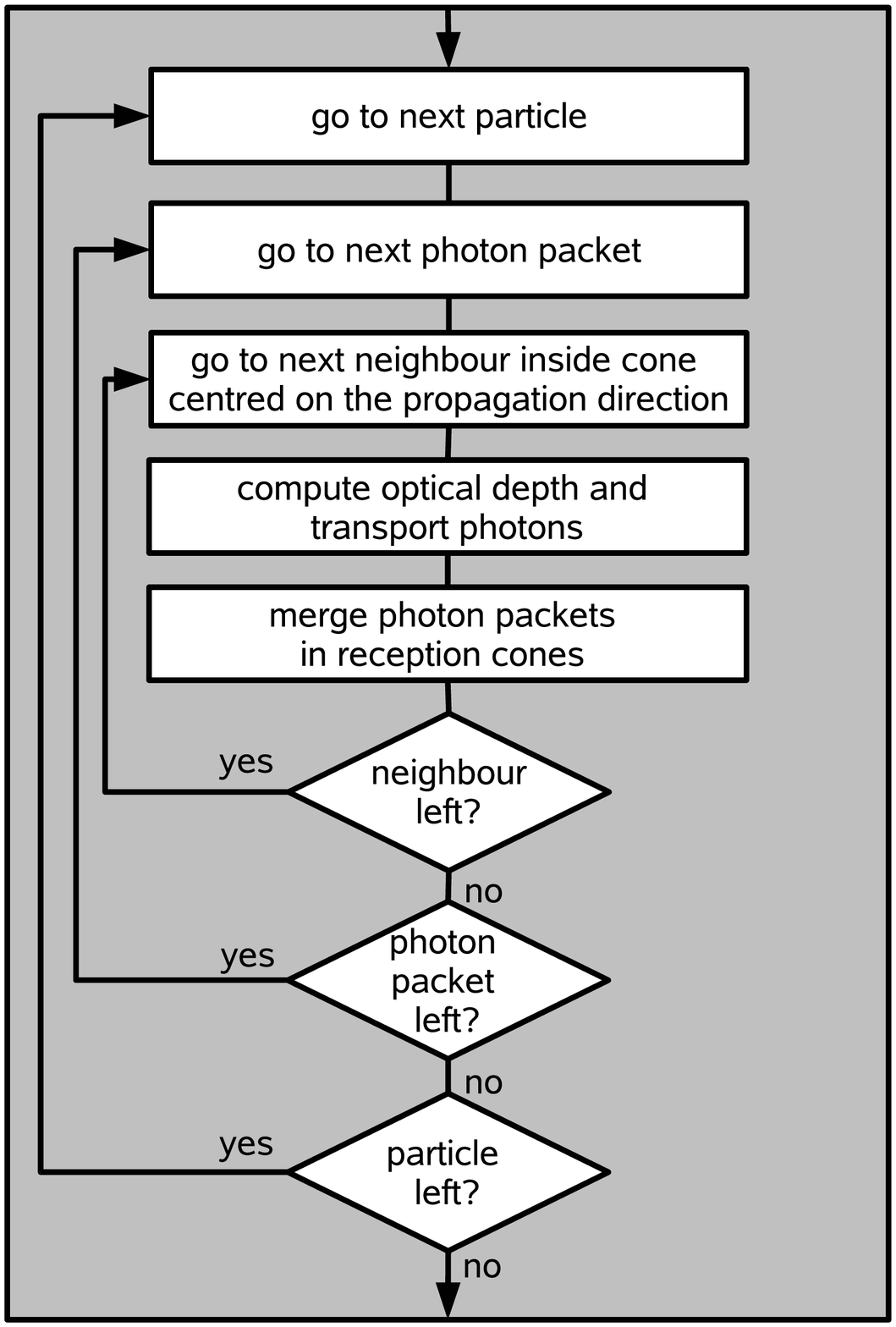} 
 
  \caption{Flow charts. {\it Left chart:} overview of \traphic. The RT simulation starts
    with the assignment of emissivities to source
    particles. Thereafter, photon packets are transported from all
    particles to their neighbours (see right chart for
    details). ViPs, if needed, are created in advance of the transport
    and are deleted immediately afterwards. The transport cycle
    continues until the user-defined stopping criterion is
    satisfied. Then, the properties of the particles are updated
    according to the photon-gas interactions that occurred. Finally,
    the RT time step is advanced. The RT simulation ends at time
    $T_{\rm r}$. {\it Right chart:} details of the transport of
    photons to neighbours (grey box in the left chart). Photons
    in photon packets are distribute amongst the $\tilde{N}_{\rm ngb}$
    neighbouring particles inside $N_{\rm c}$ (tessellating) emission
    or (regular) transmission cones with solid angle $4\pi/ N_{\rm c}$
    centred on the corresponding propagation directions. The optical
    depth to each neighbour is computed and the fraction of
    transmitted and absorbed/scattered photons is determined. Multiple
    photon packets received by individual neighbours are merged in
    $N_{\rm c}$ (tessellating) reception cones separately for each
    frequency $\nu$, which limits the number of photon packets stored
    at each particle to at most $N_{\rm c} \times N_\nu$.}
  \label{Fig:FlowChart}
\end{center}
\end{figure*} 
In this section we briefly summarise the basic concepts underlying the
RT method \traphic\ and introduce some of the notation that will be
frequently employed in the following sections. The reader is referred
to the original description in Sec.~4 of Paper~I for more
details as that description remains valid. The extensions presented in
this work concern only the application of \traphic\ to the transport
of ionising photons, which will be described in
Sec.~\ref{Sec:Implementation}.
\par
To introduce essential notation we briefly recall that SPH is a
Lagrangian numerical method to solve the Euler equations of fluid
dynamics through the representation of continuum fluids by discrete
sets of particles (for reviews see, e.g., \citealp{Monaghan:2005};
\citealp{Springel:2010}).  Any property, say $A_i$, of any given
particle $i$ is determined by performing a weighted average, or
smoothing, $A_i = \sum_j m_j/\rho_j A_j W_{ij} (h)$ of the
corresponding property $A_j$ of all other particles $j$, where $m_j$
and $\rho_j$ are the mass and density of particle $j$, and $W_{ij}(h)$
is the SPH kernel that depends on the SPH smoothing length $h$.
\par
In the following we make the common assumption that the kernel
$W_{ij}$ is compact so that there is a finite number $N_{\rm ngb}$ of
neighbouring particles $j$ within a sphere of radius $h$ around
particle $i$.  If the smoothing lengths, which represent the spatial
resolution elements of the SPH simulation, are allowed to vary in
space such that the number of neighbours $N_{\rm ngb}$ remains fixed,
then SPH enables simulations whose spatial resolution adapts to the
fluid geometry. It is this feature that makes simulations with a large
dynamic range possible and that is perhaps the main reason for the
numerous and successful applications of SPH to solve multi-scale astrophysical
problems such as galaxy formation and reionisation.
\par
\traphic\ transports photons directly on the SPH particles
(i.e.\ without interpolation to a superimposed numerical grid) and
hence the full dynamic range of the SPH simulation is employed.  The
photon transport can be decomposed into the emission of photon packets
by source particles followed by their directed propagation on the
irregular set of SPH particles. We now briefly describe both these
parts in turn.
\par
Photon packets, each of which carries photons of a characteristic
frequency $\nu$, are emitted from source particles to their
$\tilde{N}_{\rm ngb}$ neighbouring SPH particles (residing in a sphere
of radius $\tilde{h}$ centred on the source) using a tessellating set
of $N_{\rm c}$ emission cones. The number of neighbours
$\tilde{N}_{\rm ngb}$ is a parameter that determines the spatial
resolution and is usually matched to the number of neighbours $N_{\rm
ngb}$ (residing in the sphere of radius $h$) used in the computation
of the SPH particle properties, i.e., $\tilde{N}_{\rm ngb} \lesssim
N_{\rm ngb}$.  The number of cones $N_{\rm c}$ is a parameter that
determines the angular resolution of the RT. Emission cones are
necessary to achieve an isotropic emission despite the generally
highly irregular distribution of SPH particles (see App.~A in
Paper~I). The last parameter that controls the emission of photon
packets by source particles is the number $N_\nu$ of frequency bins
that are used to discretize the associated radiation spectrum.
\par
Each of the emitted photon packets (of frequency $\nu$) has an
associated propagation direction. This propagation direction is chosen
to be parallel to the central axis of the corresponding emission
cone. After emission, the photon packets are traced downstream along
their propagation directions. The packets thereby remain confined to
the solid angle of the emission cone into which they were originally
emitted thanks to the use of transmission cones with solid angle
$4\pi/N_{\rm c}$. The transmission cones prevent the unconfined
diffusion of photon packets on the unstructured grid of SPH particles
that would otherwise occur and they ensure that the transport is
directed. Because the transmission cones are defined locally (i.e., at
the positions of the transmitting particles) and because photon
packets are only transmitted to the subset of the $\tilde{N}_{\rm
ngb}$ nearest neighbours that are inside the transmission cones, the
angular resolution at which photon packets are transferred is
independent of the distance from the source (even though the surface
areas implied by the solid angles of the original emission cones increase 
with the square of the distances from the corresponding sources). 
As a result, the sharpness of the shadows cast by opaque
objects such as dense neutral clumps and filaments is independent of
their distances from the sources.
\par
Virtual particles (ViPs) are introduced to accomplish the photon
transport along directions for which no neighbouring SPH particle in
the associated emission or transmission cones could be found. The
properties of the ViPs (like, e.g., their densities) are determined
through SPH interpolation from the $\tilde{N}_{\rm ngb}$ neighbouring
SPH particles. Their name refers to the fact that ViPs are temporary
constructs that are only invoked to accomplish the transport of photon
packets in empty cones. They do not hold permanent information, they
do not affect the SPH simulation and they are deleted as soon as they
have fulfilled this task.
\par
The photon transport is supplemented with a photon packet merging
procedure that respects the chosen angular resolution and renders the
RT computation time independent of the source number.  The merging is
done by binning photon packets in angle (according to their
propagation directions) using $N_{\rm c}$ (tessellating) reception
cones. Binned photon packets define a single new photon packet per
reception cone whose propagation direction is given by the weighted
sum\footnote{We note that the expression in Paper~I (Sec.~4.2.3) for
the propagation direction $\mathbf{n}_{{\rm m}, k}$ of this new photon
packet contains a typo.  The
expression used in that publication, $\mathbf{n}_{{\rm m}, k} = \sum
w_{p} \mathbf{n}_p / \sum w_{p} $, where $\mathbf{n}_p$ are the unit
vectors that represent the propagation directions of photon packets
that are to be merged and $w_p$ are associated weights, does generally
not result in a unit vector, which is inconsistent with the employed
notation. However, a unit vector representing the propagation direction of the
merged photon packet can be obtained by an additional
explicit normalisation, i.e.\ $\mathbf{n}_{{\rm m}, k} \to
\mathbf{n}_{{\rm m}, k} / |\mathbf{n}_{{\rm m}, k}|$.}  of the
propagation directions of photon packets received in that cone. The
merging is done separately for photon packets of different
frequencies. Thus, it does not change the mean free path of the photon
packets, which is important for simulations that contain radiation
sources with a broad range of spectral properties (e.g., quasars and
stellar sources). Thanks to the merging, at each particle at most
$N_{\rm c} \times N_\nu$ photons packets need to be stored\footnote{In
principle it is possible to choose the solid angle of the transmission
cones independently of the angular resolution $4\pi / N_{\rm c}$
implied by the emission/reception cone tessellation. The solid angle
of the transmission cones is the main parameter that determines the
angular resolution of the photon transport, while the number $N_{\rm
c}$ of emission/reception cones controls the number of photon packets
that need to be stored at each particle. Choosing the solid angle of
transmission cones smaller than $4\pi / N_{\rm c}$ would thus result
in a higher angular resolution while keeping the memory needed to
store photon packets unchanged.  We have successfully tested this
option by repeating Test~4 in Paper~I with $N_{\rm c} = 8$ and a
transmission cone solid angle of $4\pi / 128$. We found the results of
this simulation to be indistinguishable from the simulation that
employed $N_{\rm c} = 128$ and transmission cone solid angles of $4\pi
/ 128$. In this work, however, we will not make use of this cone
decoupling option.}.
\par
The photon transport is performed using RT time steps $\Delta t_{\rm
r}$ (see Sec.~5.2.3 in Paper~I for a detailed discussion of the size
of time steps in reionisation simulations).  During each such time
step, photons are propagated and their interactions with the gas are
computed until a certain stopping criterion is satisfied. The
criterion depends on whether one aims to solve the time-independent or
the time-dependent RT equation. In the first case, photons are
propagated until they are absorbed or have left the computational
domain. In the second case, photon clocks associated with each photon
packet are used to synchronise the packet's travel time with the
simulation time such that photon packets travel at the speed of light.
\par
After each time step, the state of the SPH particles is updated
according to the interactions (absorptions, scatterings) with photon
packets they experienced. In Sec.~\ref{Sec:Implementation} we will
explain, for the example of absorption of ionising radiation, how to
combine the photon transport discussed here with the evaluation of the
interactions according to the optical depth encountered by photon
packets. Finally, the RT time is advanced, which concludes the
algorithm. A schematic summary of the RT method is depicted in the
flow chart in Fig.~\ref{Fig:FlowChart}.  

\section{Ionising photons in primordial gas: theory}
\label{Sec:Theory}
In this section we outline the physical processes that determine the
evolution of the ionisation state (Sec.~\ref{Sec:Ionisation}) and
temperature (Sec.~\ref{Sec:Heating}) of primordial gas exposed to
ionising radiation. We discuss the underlying equations and present
the references to atomic data required to evaluate them.  The
description of the numerical implementation used to solve these
equations is deferred to Sec.~\ref{Sec:Implementation}. 
\par
Readers familiar with the physics of ionisation, recombination,
heating and cooling may wish to skip Secs.~\ref{Sec:Ionisation} and
\ref{Sec:Heating} and refer to them only when needed.  For those
readers we have summarised the physical processes that we include in
the computations of the ionisation and thermal state of gas in the RT
simulations presented later in this work, together with the references
to the (fits to) atomic data sets employed for their numerical
evaluation, in Table~\ref{Tab:References}. A detailed discussion of
our choice for certain (fits to) atomic data sets and a comparison
with other works can be found in \cite{Pawlik:Thesis}.
\par
We start with some definitions that we will employ throughout the rest
of this work. We consider an atomic gas of total number density $n =
n_{\rm e} + \sum n_\alpha$, where $n_\alpha$ is the number density of
species $\alpha$ and $n_{\rm e}$ is the number density of free
electrons.  The number density $n_\alpha$ is related to the total mass
density $\rho$ through $n_\alpha = X_\alpha \rho / (\mu_\alpha m_{\rm
H})$, where $X_\alpha$ is the mass fraction of species $\alpha$ and
$\mu_\alpha = m_\alpha / m_{\rm H}$ is its mass $m_\alpha$ in units of
the hydrogen mass $m_{\rm H}$. We assume that the gas is of primordial
composition, i.e.\ $\alpha \in \{\rm HI$, $\rm HII$, $\rm HeI$, $\rm
HeII$, $\rm HeIII\}$ and $X_{\rm H} + X_{\rm He}= 1$. We will set
$X_{\rm H} = 0.25$. We will make frequent use of the species number
density fractions with respect to the hydrogen number density,
$\eta_\alpha \equiv n_\alpha / n_{\rm H}$ and the electron fraction
$\eta_{\rm e} = n_{\rm e} / n_{\rm H}$.

\subsection{Ionisation and recombination}
\label{Sec:Ionisation}
The evolution of the ionisation state of primordial gas in the
presence of a photoionising radiation background of mean intensity 
$J_{\nu}(\nu)$ is determined by the set of rate equations
\begin{eqnarray}
\frac{d\eta_{\rm HI}}{dt} &=& \alpha_{\rm HII} n_{\rm e} \eta_{\rm
HII} - \eta_{\rm HI} (\Gamma_{ \gamma \rm HI} + \Gamma_{\rm e HI}
n_{\rm e}) \label{Eq:NeutralFractions1}\\ \frac{d\eta_{\rm HeI}}{dt}
&=& \alpha_{\rm HeII} n_{\rm e} \eta_{\rm HeII} - \eta_{\rm HeI}
(\Gamma_{ \gamma \rm HeI} + \Gamma_{\rm e HeI} n_{\rm e})
\label{Eq:NeutralFractions2}\\ \frac{d\eta_{\rm HeIII}}{dt} &=&
\eta_{\rm HeII} (\Gamma_{ \gamma \rm HeII} + \Gamma_{\rm e HeII}
n_{\rm e}) - \alpha_{\rm HeIII} n_{\rm e} \eta_{\rm HeIII},
\label{Eq:NeutralFractions3}
\end{eqnarray}
supplemented with the closure relations
\begin{eqnarray}
\eta_{\rm HI} + \eta_{\rm HII} &=& 1 \label{Eq:NeutralFractions4}\\
\eta_{\rm HeI} + \eta_{\rm HeII} + \eta_{\rm HeIII} &=& \eta_{\rm He}
\label{Eq:NeutralFractions5}\\ \eta_{\rm HII} + \eta_{\rm HeII} + 2
\eta_{\rm HeIII} &=& \eta_{\rm e} \label{Eq:NeutralFractions6},
\end{eqnarray}
where $\Gamma_{ \gamma \alpha}$ is the photoionisation rate implied by
the mean intensity $J_\nu$ of the ionising background and $\Gamma_{
{\rm e} \alpha}$ and $\alpha_\alpha$ are the collisional ionisation
and recombination rate coefficients for species $\alpha$; $\eta_{\rm
He} \equiv n_{\rm He} / n_{\rm H} = X_{\rm He} (m_{\rm H} / m_{\rm
He}) / (1 - X_{\rm He})$ denotes the helium abundance (by number);
$m_{\rm H}$ and $m_{\rm He}$ are the masses of the hydrogen and helium
atoms, respectively.
\par
The ionisation and recombination rates are discussed in more detail below.
\subsubsection{Ionisation}
The photoionisation rate $\Gamma_{\gamma \alpha}$ determines the
number of photoionisations of species $\alpha$ per unit time and unit
volume $\eta_\alpha n_{\rm H} \Gamma_{\gamma \alpha}$. It is defined
by (e.g., \citealp{Osterbrock:1989})
\begin{equation}
  \Gamma_{\gamma \alpha} = \int_{\nu_\alpha}^\infty~d\nu \frac{4 \pi J_{\nu}(\nu)}{h_{\rm p} \nu} \sigma_{\gamma \alpha} (\nu),
  \label{Eq:PhotoionisationRate}
\end{equation}
where $\alpha \in \{\rm HI$, $\rm HeI$, $\rm HeII\}$, $h_{\rm p}$ is
Planck's constant, $\sigma_{\gamma \alpha} (\nu)$ is the
photoionisation cross-section and $h_{\rm p} \nu_\alpha$ is the
ionisation potential. We use the fits to the photoionisation
cross-section from \cite{Verner:1996}.  Note that $h_{\rm p} \nu_{\rm
HI} = 13.6 \eV$, $h_{\rm p} \nu_{\rm HeI} = 24.6 \eV$ and $h_{\rm p}
\nu_{\rm HeII} = 54.4 \eV$.
\par
The photoionisation rates can be written as
\begin{equation}
\Gamma_{\gamma \alpha} =  \langle \sigma_{\gamma \alpha} \rangle \int_{\nu_\alpha}^\infty~d\nu \frac{4 \pi J_{\nu}(\nu)}{h_{\rm p}\nu},
\label{Eq:AveragePhotoionisationCrossSection}
\end{equation}
where $ \langle \sigma_{\gamma \alpha} \rangle$ is the average, or
 grey, photoionisation cross-section,
\begin{equation}
 \langle \sigma_{\gamma \alpha} \rangle \equiv  \int_{\nu_\alpha}^\infty~d\nu \frac{4 \pi J_{\nu}(\nu)}{h_{\rm p}
   \nu} \sigma_{\gamma \alpha} (\nu) \times \left [\int_{\nu_\alpha}^\infty~d\nu \frac{4 \pi J_{\nu}(\nu)}{h_{\rm p}\nu} \right ]^{-1}.
\label{Eq:Crosssection}
\end{equation} 
We will employ the grey photoionisation cross-section of hydrogen in
some of our RT simulations in Sec.~\ref{Sec:Tests}.  For reference we
note that its value for a blackbody spectrum of temperature $T_{\rm
  bb} = 10^5\K$ is $\langle\sigma_{\gamma \rm HI} \rangle = 1.63
\times 10^{-18} \cmsq$. 
\par
In addition to photoionisations we include collisional ionisation of
HI, HeI and HeII by electron impact. To compute the corresponding
ionisation rates, we employ the fits to the collisional ionisation
coefficients provided by \cite{Theuns:1998}.
\par
\subsubsection{Recombination}
\label{Sec:Recombination}
We write the number of radiative recombinations per unit time and unit
volume of species $\alpha$ (with $\alpha \in \{\rm HII,$ $\rm HeII$,
$\rm HeIII\}$) to energy level $l$ of the recombined species as
$n_{\rm e} n_\alpha\alpha_{\alpha l}$.
\par
Two radiative recombination coefficients are of special interest and
are referred to as case A and case B. The case A recombination
coefficient $\alpha_{{\rm A}\alpha} \equiv \sum_{l = 1} \alpha_{\alpha
l}$ is the sum of all the recombination coefficients $\alpha_{\alpha
l}$. On the other hand, the case B recombination coefficient is
defined as $\alpha_{{\rm B}\alpha} \equiv \sum_{l = 2} \alpha_{\alpha
l}$ and thus does not include the contribution from recombinations to
the ground state. 
\par
The introduction of the case B recombination coefficient is motivated
by the observation that for pure hydrogen gas that is optically thick
to ionising radiation, recombinations to the ground state are
cancelled by the immediate re-absorption of the recombination photon
by a neutral atom in the vicinity of the recombining ion. RT
simulations of ionising radiation in an optically thick hydrogen-only
gas may therefore work around the (generally computationally
expensive) explicit transfer of recombination photons by simply
employing the case B (instead of the full, i.e.\ case A) recombination
coefficient. Note that this {\it on-the-spot} approximation (e.g,
\citealp{Osterbrock:1989}) is only strictly valid when considering the
transport of ionising radiation in optically thick gas, whereas the gas in 
RT simulations typically shows a range of optical depths, from optically thick 
to optically thin.
\par
To keep the description of our method and its test simple and to allow
for a detailed comparison with published reference results, we will
also assume the on-the-spot approximation and use case B recombination
rates. The explicit transport of recombination radiation will be the
subject of future work.  We use the following coefficients to describe
radiative recombinations (Table~\ref{Tab:References}).  For HII and
HeIII radiative recombination, we employ the fits from
\cite{Hui:1997}.  For the HeII radiative recombination coefficient, we
employ the tabulated coefficients of \cite{Hummer:1998} using linear
interpolation in log-log.
\par
We have not yet discussed the dielectronic contribution to the HeII
recombination coefficient. Dielectronic HeII recombination
(e.g.~\citealp{Savin:2000}; \citealp{Badnell:2001} for a review) is
the dominant recombination process for temperatures $T \gtrsim 10^5
\K$.  We therefore add the dielectronic contribution to the HeII
recombination rates, making use of the fit presented in
\cite{Aldrovandi:1973}.
\subsection{Heating and cooling}
\label{Sec:Heating}
Our main goal in this work is to present and test an implementation of
\traphic\ to compute, in addition to the ionisation state, the
evolution of the temperature of gas exposed to ionising radiation. For
the discussion it is helpful to review the relevant thermodynamical
relations, which is the subject of this section.
\par
The internal energy per unit mass for gas composed of monoatomic species
that are at the same temperature $T$ is
\begin{equation}
u = \frac{3}{2} \frac{n k_{\rm B} T}{\rho} =  \frac{3}{2} \frac{k_{\rm B}
  T}{\mu m_{\rm H}}, 
\label{Eq:InternalEnergy}
\end{equation}
where $k_{\rm B}$ is the Boltzmann constant and $\mu = \rho/(n m_{\rm
H})$ is the mean particle mass in units of the hydrogen mass.
\par
From the first law of thermodynamics
\begin{equation}
d (u \rho V) = -P dV + n_{\rm H}^2 (\mathcal{H} - \mathcal{C}) V,
\end{equation}
where $P$ is the pressure, $V$ the volume and $\mathcal{H}$ and $
\mathcal{C}$ are the radiative heating and radiative cooling rates,
normalised such that the rates of energy gain and loss per unit volume
are described by $n_{\rm H}^2 \mathcal{H}$ and $n_{\rm H}^2
\mathcal{C}$, respectively.  Assuming that $d(\rho V) = 0$, as is the
case for an SPH particle, it follows that
\begin{equation}
\frac{d u}{dt} = -\frac{P}{\rho V} \frac{dV}{dt} + \frac{n_{\rm H}^2}{\rho} (\mathcal{H} -  \mathcal{C}).
\label{Eq:dudt}
\end{equation}
\subsubsection{Cooling}
The normalised cooling rate $\mathcal{C}$ is the sum of the normalised
rates of the individual radiative cooling processes. We include all
standard cooling processes: collisional ionisation by electron impact,
radiative and dielectronic recombination, collisional excitation by
electron impact, bremsstrahlung and Compton scattering. The
expressions for the corresponding cooling rates are taken from the
references listed in Table~\ref{Tab:References}.
\subsubsection{Heating}
The normalised heating rate $\mathcal{H}$ is the sum of the rates of
the individual radiative heating processes. In the following we only
consider the contribution from photoionisation heating, which will be
the main contributor to the heating rate for the high-redshift RT
simulations of interest. We, however, note that Compton heating by
X-rays may not be negligible (\citealp{Madau:1999}).
\par
We write the heating rate due to photoionisation as
\begin{equation}
n_{\rm H}^2 h_\gamma =  (\eta_{\rm HI} \mathcal{E}_{\gamma \rm HI} +  \eta_{\rm HeI}
\mathcal{E}_{\gamma \rm HeI} + \eta_{\rm HeII} \mathcal{E}_{\gamma \rm HeII}) n_{\rm H},  
\end{equation}
where
\begin{equation}
  \mathcal{E}_{\gamma \alpha} = \int_{\nu_\alpha}^\infty~d\nu \frac{4 \pi
  J_{\nu}(\nu)}{h_{\rm p} \nu} \sigma_{\gamma \alpha} (\nu) (h_{\rm p}\nu - h_{\rm
  p}\nu_\alpha)
\label{Eq:ExcessEnergy}
\end{equation}
is the heating rate per particle of species $\alpha$. Using
Eq.~\ref{Eq:PhotoionisationRate}, we can write
\begin{equation}
\mathcal{E}_{\gamma \alpha} = \Gamma_{\gamma \alpha} \langle \epsilon_\alpha \rangle,
\end{equation}
where
\begin{eqnarray}
\langle \epsilon_\alpha \rangle &=& \left [\int_{\nu_\alpha}^\infty~d\nu \frac{4 \pi
    J_{\nu}(\nu)}{h_{\rm p} \nu} \sigma_{\gamma \alpha} (\nu) (h_{\rm p}\nu -
  h_{\rm p}\nu_\alpha) \right] \nonumber \\ &\times& \left[ \int_{\nu_\alpha}^\infty~d\nu \frac{4 \pi J_{\nu}(\nu)}{h_{\rm p} \nu} \sigma_{\gamma \alpha} (\nu) \right]^{-1}
\label{Eq:AverageExcessEnergy}
\end{eqnarray}
is the average excess energy of absorbed ionising photons. For
reference, the average excess energy for photoionisation of hydrogen,
assuming a blackbody spectrum of temperature $T_{\rm bb} = 10^5\K$, is
$\langle \epsilon_{\rm HI}\rangle = 6.32 \eV$.
\par
Sometimes, e.g.\ when considering the energy balance of entire
HII regions, one is interested in the total photoheating rate
integrated over a finite volume, assuming all photons entering this
volume are absorbed within it. The average excess energy injected at
each photoionisation in this optically thick limit is also obtained
from Eq.~\ref{Eq:AverageExcessEnergy}, but after setting
$\sigma_{\gamma \alpha} (\nu) = 1$, since all photons are absorbed
(e.g., \citealp[p.135]{Spitzer:1978}),
\begin{eqnarray}
\langle \epsilon_\alpha^{\rm thick} \rangle &=& \left[\int_{\nu_\alpha}^\infty~d\nu \frac{4 \pi J_{\nu}(\nu)}{h_{\rm p} \nu} (h_{\rm p}\nu - h_{\rm p}\nu_\alpha) \right]  \nonumber \\
&\times&\left[ \int_{\nu_\alpha}^\infty~d\nu \frac{4 \pi J_{\nu}(\nu)}{h_{\rm p} \nu} \right]^{-1}.
\label{Eq:AverageExcessEnergyThick}
\end{eqnarray}
For reference, the value of the average excess energy for
photoionisation of hydrogen in the optically thick limit, assuming a
blackbody spectrum of temperature $T_{\rm bb} = 10^5\K$, is $\langle
\epsilon_{\rm HI}^{\rm thick}\rangle = 16.01 \eV$.
\par
In writing Eqs.~\ref{Eq:AverageExcessEnergy} and
\ref{Eq:AverageExcessEnergyThick} we assumed that all of the photon
excess energy is used to heat the gas, corresponding to a complete
thermalization of the electron kinetic energy. In reality, (very
energetic) photo-electrons may lose some of their energy due to the
generation of secondary electrons (e.g., \citealp{Shull:1985};
\citealp{Furlanetto:2010}).
\begin{table*}
\begin{center}
\begin{small} 
  \caption{References to (fits to) the atomic data used to calculate
  photoionisation rates, collisional ionisation rates, recombination
  rates and cooling rates in the simulations presented in this
  work. See \protect\cite{Pawlik:Thesis} for detailed discussions and
  comparisons of our choices in favour of certain
  references. \label{Tab:References}}
\begin{tabular}{lllllll}
\hline
\hline
Photoionisation & HI, HeI, HeII photoionisation cross-sections  & \cite{Verner:1996} & \\
Collisional ionisation & HI, HeI, HeII collisional ionisation rate
coefficients & \cite{Theuns:1998} & \\
\hline
Recombination & HII, HeIII recombination rate coefficients  & \cite{Hui:1997} & \\
& HeII recombination rate coefficient  & \cite{Hummer:1998} & \\
& HeII dielectronic recombination rate coefficient  & \cite{Aldrovandi:1973} & \\
\hline
Collisional ionisation cooling & HI, HeI, HeIII collisional ionisation cooling rate
  & \cite{Shapiro:1987} & \\
Collisional excitation cooling &  HI, HeI, HeIII collisional excitation cooling rate
 & \cite{Cen:1992} & \\
Recombination cooling & HII, HeIII recombination cooling rate (case A and B)  &  \cite{Hui:1997} & \\
& HeII recombination cooling rate (case A and B) &  \cite{Hummer:1998} & \\
& HeII dielectronic recombination cooling rate   &  \cite{Black:1981}& \\
Cooling by bremsstrahlung & Bremsstrahlung cooling rate & \cite{Theuns:1998} & \\
Compton cooling & Compton cooling rate & \cite{Theuns:1998} & \\
\hline
\end{tabular}
\end{small}
\end{center}

\end{table*}

\section{Ionising photons in primordial gas: implementation}
\label{Sec:Implementation}
Here we extend the description of \traphic\ given in
Sec.~\ref{Sec:traphic} to the transport of ionising photons by
describing our implementation of the absorption of ionising photons
and the subsequent computation of the species fractions and gas
temperatures.  
\subsection{Absorption of ionising photons}
\label{Sec:Implementation:Absorption}
The number of ionising photons that are absorbed during the
propagation of a photon packet over distance $d_{ij}$ between
neighbouring particles $i$ and $j$ is given by $\delta N_{\rm abs,
\nu} = \delta N_{\rm in, \nu}[1 - \exp(-\tau(\nu))]$, where $\delta
N_{\rm in, \nu}$ is the number of photons inside the photon packet
before propagation and the optical depth $\tau (\nu)$ is the sum
$\tau(\nu) = \sum_\alpha \tau_{\alpha}(\nu)$ of the optical depths of
each absorbing species $\alpha \in \{\rm HI, HeI, HeII\}$ and
\begin{equation}
\tau_{\alpha} (\nu)\equiv \int_{\mathbf{r}_i}^{\mathbf{r}_j} dr\ \sigma_\alpha (\nu) n_{\alpha}(\mathbf{r}) \approx \sigma_{\alpha} (\nu) n_{\alpha} (\mathbf{r}_j) d_{ij}.
\end{equation} 
The last approximation is reasonable because SPH neighbours will have
similar densities. We consider these photons to
be absorbed by particle $j$.
\par
The number of photons $\delta N_{\rm abs, \alpha} (\nu)$ absorbed by species $\alpha$
is determined from the number of absorbed photons $\delta N_{\rm abs, \nu}$ using
\begin{equation}
\delta N_{\rm abs, \alpha} (\nu) = \frac{w_{\alpha}(\nu)}{\sum w_{\alpha}(\nu)}\delta N_{\rm abs, \nu}, 
\label{Eq:absorptionweights}
\end{equation}
where we choose (un-normalized) weights $w_{\alpha}(\nu) = \tau_{\alpha}(\nu)$ (\citealp{Osterbrock:1989}; see also, e.g., 
\citealp{Trac:2007}). The total number of photons $\Delta
N_{\rm abs, \alpha} (\nu)$ absorbed by species $\alpha$ 
is the sum of the photons absorbed due to the propagation of
photon packets from all neighbouring particles during the RT time
step $\Delta t_{\rm r}$, i.e., $\Delta N_{\rm abs, \alpha} (\nu) =
\sum \delta N_{\rm abs, \alpha} (\nu)$.
\par
We pause to note that other choices for the weights $w_\alpha$ (as employed in, e.g.,
\citealp{Bolton:2004}; \citealp{Maselli:2003}; \citealp{Whalen:2008}) 
will in general imply an unphysical distribution of the
number $\delta N_{\rm abs, \nu}$ of absorbed photons amongst the
individual species. To see this, consider, for example, a gas parcel
with optical depth $\tau$ that contains only hydrogen atoms. Let an
arbitrary fraction $f_A$ of the hydrogen atoms be labelled $A$ and let
the remaining fraction $f_B = 1 - f_A$ of the hydrogen atoms be
labelled $B$. Suppose that the optical depth of the hydrogen atoms
with label $A$ is $\tau_A$. The ratio of the number of photons
absorbed by hydrogen atoms with label $A$ to the number of photons
absorbed by all hydrogen atoms should be
\begin{equation}
\frac{\delta N_{\rm abs, \alpha}(\nu)}{\delta N_{\rm abs, \nu}} =
\frac{\tau_A(\nu)}{\tau(\nu)},
\end{equation}
while Eq.~\ref{Eq:absorptionweights} implies
\begin{equation}
\frac{\delta N_{\rm abs, \alpha}(\nu)}{\delta N_{\rm abs, \nu}} =
\frac{w_A(\nu)}{w_A(\nu) + w_B(\nu)}.
\end{equation}
The two ratios generally agree only if $w_{\alpha} \propto
\tau_\alpha$, where $\alpha$ takes values $A$ and $B$. Other choices
of the weights $w_{\alpha}$ would imply that the
probability for absorption of ionising photons by hydrogen atoms with
label $A$ is different from the probability for absorption of ionising
photons by hydrogen atoms with label $B$. This is unphysical, since there 
is no physical difference between these two types of hydrogen atoms.
\par
As described in Paper~I, photons absorbed by a virtual particle (ViP)
are redistributed amongst the $\tilde{N}_{\rm ngb}$ neighbouring SPH
particles that have been used to compute its species densities. This
is necessary, because ViPs are temporary constructs; physical
properties like the gas species fractions are only defined and stored
for the SPH particles. There, however, is an important change with
respect to the original description. Previously, we assigned to each
of the neighbours a fraction of the absorbed photons that is
proportional to the value of the ViP's SPH kernel $W$ at its
position. In the current version we assign, to each of the neighbours,
a fraction of the photons absorbed by species $\alpha$ that is
proportional to the neighbour's contribution to the SPH estimate of
the density of that species at the location of the considered ViP. The
current version is equivalent to the original version of Paper~I if $X
= 1$ (i.e.\ no helium) and if all neighbours have the same neutral
hydrogen mass.  However, in general this will not be true, in which
case the current version is the only self-consistent one.  We discuss
the differences between the current and the original version in detail
in App.~\ref{Sec:Appendix}.
\par
The number of photons $\Delta N_{\rm abs, \alpha} (\nu)$ that are
absorbed by a given particle during the RT time step $\Delta t_{\rm
r}$ is used to obtain the photoionisation rates $\Gamma_\alpha$ for
that particle directly (i.e., without reference to the mean intensity
$J_\nu$) using
\begin{eqnarray}
\eta_\alpha \mathcal{N}_{\rm H} \Delta t_{\rm r} \Gamma_{\gamma \alpha} &=& \sum_{\nu}\Delta\mathcal{N}_{\rm abs, \alpha} (\nu),
\label{Eq:Photoionizationrate}
\end{eqnarray}
where $\mathcal{N}_{\rm H} \equiv m X_{\rm H} / m_{\rm H}$ is the
number of hydrogen atoms associated with the SPH particle of mass
$m$. The photoionisation rates are then used to advance the species
fractions and the gas temperature as we will describe in the next
section.
\par
\subsection{Integration of the rate equations}
\label{Sec:Implementation:Subcycling}
Here we present our numerical method to solve the equations of the
evolution of the ionisation balance and temperature of gas exposed to
ionising radiation (Eqs.~\ref{Eq:NeutralFractions1} -
\ref{Eq:NeutralFractions6} and Eq.~\ref{Eq:dudt}).  This method is an
extension of the subcycling method described in Paper~I, which we
therefore briefly recall.
\par
In Paper~I we presented a method to follow the ionisation state of a
(hydrogen-only) gas parcel exposed to (hydrogen-) ionising radiation at
fixed temperature. The ionisation rate equations were solved 
at the end of each RT time step $\Delta t_{\rm r}$
by explicit numerical integration (hereafter also referred to as subcycling)
using subcycle steps $\delta t = \min(f \tau_{\rm eq}, \Delta t_{\rm r})$, where
\begin{equation}
\tau_{\rm eq} \equiv \frac{\tau_{\rm ion}\tau_{\rm rec}} {\tau_{\rm
ion}+\tau_{\rm rec}}
\label{Eq:TauEq}
\end{equation}
is the time scale to reach ionisation equilibrium (Eq.~20 in Paper~I),
$\tau_{\rm rec} \equiv 1 / (n_{\rm e} \alpha_{\rm HII})$ is the
recombination time scale, $\tau_{\rm ion} = 1/(\Gamma_{\gamma \rm HI}
+ n_{\rm e} \Gamma_{\rm eHI})$ is the ionisation time scale and $f$ is
a dimensionless factor that controls the integration
accuracy. Subcycling allows the RT time step $\Delta t_{\rm r}$ to be
chosen independently of the values of the ionisation and recombination
time scales on which the species fractions evolve.  A RT time step
$\Delta t_{\rm r}$ limited by the ionisation and recombination time
scales would prevent efficient RT simulations since these time scales
may become very small.
\par
In this work we are interested in the self-consistent computation of
the non-equilibrium ionisation state of gas with an evolving
temperature. As for the case of a non-evolving temperature studied in
Paper~I, we integrate the ionisation rate equations over subcycle
steps $\delta t = \min(f \tau_{\rm eq}, \Delta t_{\rm r})$. The time scale $\tau_{\rm
eq}$ to reach ionisation equilibrium is computed using
Eq.~\ref{Eq:TauEq} with $\tau_{\rm ion} = 1 /
\sum_{\alpha}(\Gamma_{\gamma \alpha} + n_{\rm e} \Gamma_{\rm
e\alpha})$ and $\tau_{\rm rec} = 1 / \sum_{\alpha} n_{\rm e}
\alpha_{\rm \alpha}$.  Recombination and collisional ionisation rates
are determined using the temperature at the beginning of each subcycle
step and the species fractions are advanced in a photon-conserving
manner as detailed\footnote{In Paper~I we only considered ionisation
of gas of pure hydrogen. The corresponding expressions, however,
are straightforward to generalise to include the ionisation of
helium. \label{footnote:subcycling}} in Paper~I.
\par
In addition, the temperature is advanced by evolving the internal
energy according to Eq.~\ref{Eq:dudt} over the same subcycle step
assuming isochoric evolution ($dV = 0$), which is appropriate for a
fixed gas distribution (and thus during a single hydro-step in
radiation-hydrodynamical simulations). We use the mean particle mass
$\mu$ derived from the current species fractions to convert between
temperature (which is required to compute the rate coefficients) and
internal energy using Eq.~\ref{Eq:InternalEnergy}. Note that the
species fractions and the temperature are evolved independently of
each other over a single subcycle step of size $\delta t$. We thus
implicitly assume that during any of these steps the species fractions
and the temperature do not evolve significantly. This assumption is
excellent because the species fractions and the
temperature evolve, by definition, on time
scales large compared with the size of the subcycle steps. The evolutions of the
species fractions and the temperature are coupled at the beginning of
the next subcycle step, where the new species fractions and the new
temperature determine new collisional ionisation, recombination and
cooling rates.
\par
We now describe our numerical implementation of the subcycling. We
limit ourselves to the description of how we advance the internal
energy over a single subcycle step as the implementation of the
subcycling of the species fractions was already described\footnote{See
footnote \ref{footnote:subcycling}} in Paper~I. The internal energy is
advanced by solving a discretized version of the energy equation
(i.e., Eq.~\ref{Eq:dudt} with $dV = 0$). We make use of implicit Euler
integration when the subcycle step is larger than the time scale
$\tau_{\rm u} \equiv u/(du/dt)$ on which the internal energy evolves.
That is, if $\delta t > \tau_{\rm u}$ we advance the internal energy
according to
\begin{equation}
u_{t+\delta t} = u_t + \frac{n_{{\rm H}, t}^2}{\rho_{t}}
(\mathcal{H}_{t+\delta t} - \mathcal{C}_{t + \delta t}) \delta t.
\end{equation}
The last equation is solved iteratively for $u_{t+\delta t}$ by finding the zero of the function 
\begin{equation}
f(u_{t+\delta t}) = u_{t + \delta t} - u_t -  \frac{n_{{\rm H}, t}^2}{\rho_{t}}
(\mathcal{H}_{t+\delta t} - \mathcal{C}_{t + \delta t})\delta t
\label{Chapter:Heating:Sec:Nonequilibrium:Eq:Implicit}
\end{equation}
and setting $\mathcal{H}_{t+\delta t}=\mathcal{H}_{t}$ and assuming $\mathcal{C}_{t + \delta t}=\mathcal{C}_{t}$ during the first iteration.
If, instead, $\delta t < \tau_{\rm u}$, we employ the 
explicit Euler integration scheme,
\begin{equation}
u_{t+\delta t} = u_t + \frac{n_{{\rm H}, t}^2}{\rho_t} (\mathcal{H}_t -
\mathcal{C}_t) \delta t,
\label{Chapter:Heating:Sec:Nonequilibrium:Eq:Explicit}
\end{equation}
Our implementation combines the advantages of the explicit scheme (its
accuracy) with that of the implicit scheme (its stability; see, e.g.,
\citealp{Shampine:1979} and \citealp{NumericalRecipes:1992} for useful
discussions on implicit and explicit integration).
\par
In Paper~I (for the case of a constant temperature), we sped up the
subcycling of the species fractions by keeping the species fractions fixed
once ionisation equilibrium has been reached\footnote{We assume that ionisation 
equilibrium is reached once either the fractional change in all species fractions 
individually becomes smaller than a predefined small value (here we use $10^{-9}$).}. 
We employ a similar recipe here. However, thermal equilibrium is reached on the time scale
$\tau_{\rm u}$, which may be much larger than the time scale
$\tau_{\rm eq}$ to reach ionisation equilibrium. In this case the
temperature continues to evolve after the species fractions have
attained their (quasi-) equilibrium values. The evolution of the temperature
implies an evolution of the recombination and collisional ionisation
rates, and hence an evolution of the equilibrium ionisation
balance. Our recipe for speeding up the subcycling should respect this
evolution.
\par
We therefore proceed as follows. Once ionisation equilibrium has been
reached, we stop the subcycling of the species fractions. Over the
remainder of the time step $\Delta t_{\rm r}$ only the internal energy
is subcycled, which can be done using time steps $\delta_{\rm u} t
\equiv f_{\rm u} \times \tau_{\rm u}$, where $f_{\rm u} < 1$ is a
dimensionless parameter that controls the accuracy of the integration
(we set $f_{\rm u} = f$). This results in a speed-up since typically
$\delta_{\rm u} t \gg \delta t$. After each such subcycle step, we
reset the species fractions to their current equilibrium values. The
equilibrium species fractions are obtained by iteratively solving the
set of equations \ref{Eq:NeutralFractions1} -
\ref{Eq:NeutralFractions6} with $d\eta_{\alpha}/dt = 0$.
\par
In summary, we solve the evolution of the ionisation balance and
temperature using a hybrid numerical method that makes use of both
explicit and implicit Euler integration schemes. The ionisation rate
equation is solved explicitly using the subcycling procedure presented
in Paper~I. This ensures the accurate conservation of photons and
allows one to choose the size of the RT time step independently of the
(often very small) ionisation and recombination time scales, a
pre-requisite for efficient RT simulations.  The temperature is
evolved along with the ionisation balance by following the evolution of
the internal energy of the gas. We use an explicit discretisation
scheme to advance the internal energy if the cooling time is larger
than the size of the subcycle step. For smaller cooling times,
stability considerations lead us to employ an implicit discretisation
scheme to advance the internal energy. Once ionisation equilibrium has
been reached, the subcycling computation is sped up by fixing the
species fractions to their (temperature-dependent) quasi-equilibrium
values. From then on, only the evolution of the internal energy is
subcycled.

\section{Ionising photons in primordial gas: results}
\label{Sec:Tests}
In this section we perform simulations to test our new, thermally coupled implementation of
\traphic. We also use these simulations to discuss the differences in RT simulations performed 
using the grey approximation in the optical thick and thin limits and 
compare simulations using the grey approximation to simulations that 
solve the RT using multiple frequency bins.
\par
We begin in Sec.~\ref{Sec:Tests:Test1} with verifying that the
subcycling method described in
Sec.~\ref{Sec:Implementation:Subcycling} can be successfully employed
to solve for the non-equilibrium ionisation balance and temperature of
gas exposed to ionising radiation. Then, in
Sec.~\ref{Sec:Tests:Reference}, we present a set of reference
solutions for the idealised problem of a single spherically symmetric
expanding HII region that we will employ later to test the performance
of \traphic\ in this same problem. We compare reference solutions
derived in the grey approximation with the exact multi-frequency
results and discuss their differences. Thereafter, in Secs.~\ref{Sec:Tests:Test2} and
\ref{Sec:Tests:Test3}, we investigate \traphic's performance in 
RT simulations of increasing complexity: in
Sec.~\ref{Sec:Tests:Test2} we compute the ionised fractions and
temperatures around a single source in a homogeneous density field and
in Sec.~\ref{Sec:Tests:Test3} we follow the ionising radiation of
multiple sources in a highly inhomogeneous density field. Throughout
we will compare the results obtained with \traphic\ to analytical and
numerical reference results.
\par
The simulations were performed with \traphic\
implemented in a modified version of \gadget\
(\citealp{Springel:2005}).  All simulations were run on static density
fields. We also remind the reader that, to facilitate comparisons with
reference simulations, we do not explicitly follow recombination
radiation but treat it using the on-the-spot-approximation.

\subsection{Test 1: Subcycling}
\label{Sec:Tests:Test1}

\begin{figure*}
\begin{center}
  \includegraphics[trim = 0mm 0mm 0mm 0mm, width=0.49\textwidth]{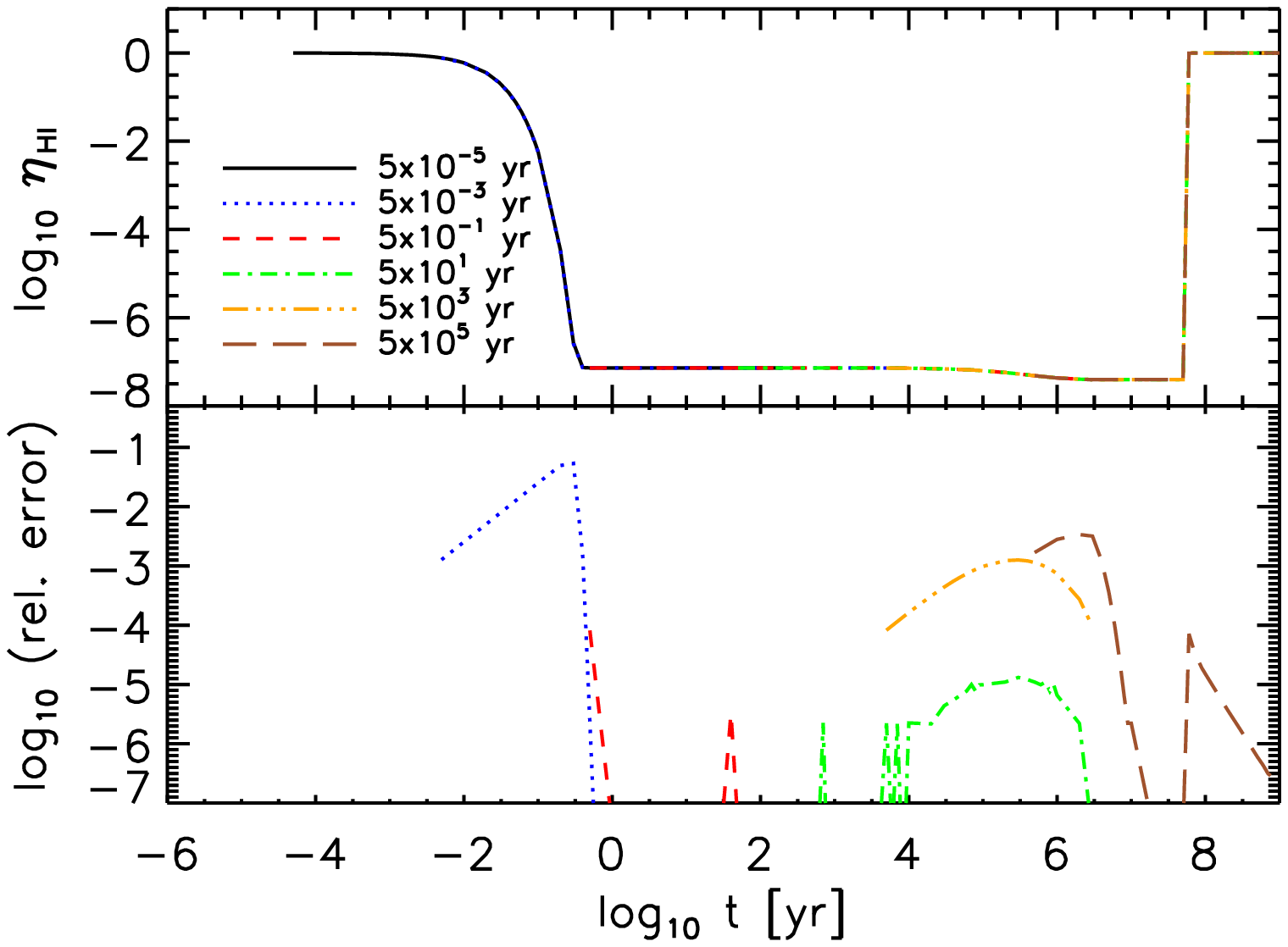}
  \includegraphics[trim = 0mm 0mm 0mm 0mm, width=0.49\textwidth]{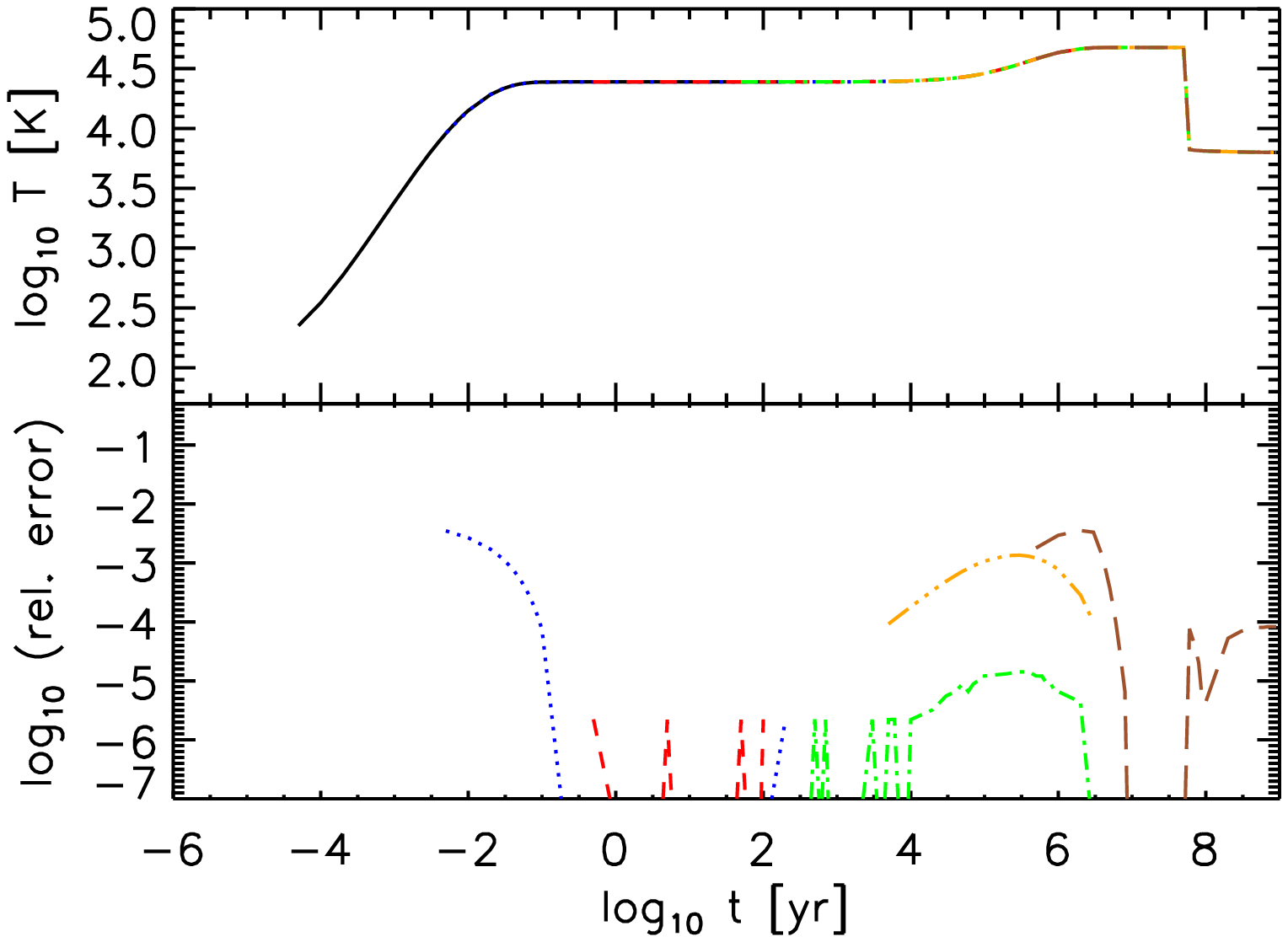}
 \caption{Test 1. Single gas parcel with prescribed photoionisation
 rate. The top panels show the evolution of the neutral hydrogen
 fraction ({\it left}) and temperature ({\it right}) for simulations
 with different time steps $\Delta t$ (in the range $5\times 10^{-5}
 \yr - 5 \times 10^5 \yr$), as indicated in the legend in the top left
 panel.  All simulations start at time $t = 0$ and evolve the gas
 parcel on subcycle steps $\delta t = \min(f \tau_{\rm eq}, \Delta
 t)$, where $\tau_{\rm eq}$ is the time scale on which the gas parcel
 approaches equilibrium and $f = 0.01$ is a dimensionless parameter that controls the accuracy of the subcycling. 
 Note that the first output occurs only at the end of the first time step, i.e., at time
 $t = \Delta t$, which is different for the different simulations.
 The bottom panels show the relative errors of the evolution shown in
 the top panels with respect to the evolution obtained from the
 simulations with the next smaller time step. The relative errors are 
 small, $\lesssim 1\%$, except during a brief phase of rapid
 evolution in the ionised fraction. The relative error can be reduced by 
 choosing a smaller value for the accuracy parameter $f$.
  \label{Fig:Test1:Error}}
\end{center}
\end{figure*}

Here we test the subcycling approach to the computation of the coupled
evolution of the non-equilibrium ionisation balance and temperature of
gas exposed to ionising radiation that we have introduced in
Sec.~\ref{Sec:Implementation:Subcycling}. Our aim is to demonstrate
that, given a flux impinging on a gas parcel (or, equivalently, a
photoionisation rate experienced by this parcel), the subcycling
allows for an accurate computation of the evolution of its ionisation
state and temperature, independent of the size of the RT time step
$\Delta t_{\rm r}$.
\par
The setup of the test is as follows. We simulate the evolution of the
ionisation state of an optically thin gas parcel with hydrogen number
density $n_{\rm H} = 1~\cmci$. For simplicity and clarity of the
presentation, we set the hydrogen mass fraction to $X = 1$ (i.e., no
helium). The simulation starts at time $t = 0$ with a fully neutral
parcel with initial temperature $T = 10^2 \K$. We then apply a
photoionising flux of $F = 10^{12}~\si\cmsqi$ with a blackbody
spectrum of characteristic temperature $T_{\rm bb} = 10^5
\K$. Consequently, the parcel becomes highly ionised and is heated to
a temperature $T \sim 10^4 \K$. After $t = 50 \Myr$ we switch off the
ionising flux and the parcel recombines and cools. The simulation ends
at $t_{\rm end} = 1 \Gyr$.  The test here is identical to Test 0
presented in \cite{Iliev:2006a}, except for the switch-off time
(\citealp{Iliev:2006a} used $t_{\rm end} = 0.5 \Myr$).
\par
We employ a grey photoionisation cross-section
$\langle{\sigma}_{\gamma \rm HI}\rangle = 1.63 \times 10^{-18} \cmsq$
(Sec.~\ref{Sec:Ionisation}), yielding a photoionisation rate
$\Gamma_{\gamma \rm HI} = 1.63 \times 10^{-6}\invs$. We assume that
each photoionisation adds $\langle \epsilon_{\rm HI} \rangle = 6.32
\eV$ to the internal energy of the gas (Sec.~\ref{Sec:Heating}), which
corresponds to the optically thin limit. We solve the equations for
the evolution of the ionisation state and temperature of the gas
parcel by subcycling them on subcycle time steps $\delta t$ over consecutive time intervals $\Delta
t$. Note that in a full RT computation these intervals would
correspond to the RT time steps $\Delta t_{\rm r}$.  Here we
distinguish between $\Delta t$ and $\Delta t_{\rm r}$ only because in this test we
are considering a single gas parcel with prescribed photoionisation
rate and do not perform RT simulations.  The dimensionless parameter
$f$ that controls the size of the subcycling steps $\delta t$ (and hence the accuracy of the subcycling) 
is set to $f=10^{-2}$. When computing Compton cooling rates off the cosmic
microwave background, we assume a redshift $z = 0$.
\par
Fig.~\ref{Fig:Test1:Error} shows the results.  The top left and top
right panels show the evolution of the neutral hydrogen fraction and
the temperature, respectively, for simulations with time steps $\Delta
t = 5 \times (10^{-5}$, $10^{-3}$, $10^{-1}$, $10^{1}$, $10^{3}$,
$10^{5}) \yr$.  Note that in order to limit the computation time, not
all of the simulations have been evolved until the end of the
simulation time. They were stopped once their simulation time
overlapped with that of the simulation with the next larger time step. 
The earliest output of a given simulation is at  $t = \Delta t$ (and hence, in Fig.~\ref{Fig:Test1:Error}, 
different curves start at different times), but all 
simulations started at $t = 0$ and were numerically evolved on subcycle time steps 
$\delta t < \Delta t$ as described in Sec.~\ref{Sec:Implementation:Subcycling}.  
\par
The gas parcel quickly approaches photoionisation equilibrium,
reaching its equilibrium neutral fraction after $\sim 10$
(photo-)ionisation time scales ($\tau_{\rm ion} \equiv \Gamma_{\gamma
\rm HI}^{-1} \approx 0.02 \yr$).  During this period, photoheating
raises its temperature to $T \approx 2 \times 10^4\K$. Around $t \sim
10^5 \yr$, the neutral fraction exhibits a slight decrease.  As noted
in \cite{Iliev:2006a}, this behaviour is caused by the decrease in the
recombination rate due to the rise in temperature that can be observed
at this time. The fact that the temperature still evolves after the
neutral fraction reached its equilibrium value means that thermal
equilibrium is reached on a larger time scale than photoionisation
equilibrium. Note that the thermal equilibrium phase is missed in the test 
simulations of \cite{Iliev:2006a} because these simulations 
were stopped at a much earlier time.
 \par
The observed behaviour can be understood as follows. When thermal
equilibrium is approached from a temperature lower than the
equilibrium temperature, the net cooling rate is approximately given
by the photoheating rate. In photoionisation equilibrium, the
photoheating rate is proportional to the recombination rate.  The
time scale $\tau_{\rm u} \equiv u / (du/dt)$ to reach thermal
equilibrium can therefore be expressed in terms of the recombination
time $\tau_{\rm rec} \equiv 1 / (n_{\rm e} \alpha_{\rm HII})$,
\begin{eqnarray}
\tau_{\rm u} &=& \frac{(3/2) n k_{\rm B} T }{n_{\rm H}^2 h_\gamma} \\
&=& \frac{(3/2) n k_{\rm B} T }{n_{\rm HII} n_{\rm e} \alpha_{\rm HII} \langle
  \epsilon_{\rm HI} \rangle}\\
&=& \frac{(3/2) n k_{\rm B} T }{ \langle \epsilon_{\rm HI} \rangle n_{\rm HII}} \tau_{\rm rec}\\
&\sim& \tau_{\rm rec},
\end{eqnarray}
where in the last step we assumed that the gas is highly ionised,
i.e.\ $n_{\rm HII} \approx n_{\rm H} \approx n / 2$, and that $T
\approx 10^4 \K$.  The recombination time (and hence the cooling time)
is much larger than the time $\tau_{\rm eq} \equiv (\tau_{\rm ion}
\tau_{\rm rec}) / (\tau_{\rm ion} + \tau_{\rm rec})$ to reach
ionisation equilibrium which asymptotes to $\tau_{\rm ion}$ for
$\tau_{\rm ion} \ll \tau_{\rm rec}$ (see also the discussion in
Sec.~5.1 in Paper~I). Here, $\tau_{\rm ion} = \Gamma_{\gamma \rm
HI}^{-1} \approx 0.02 \yr$ and $\tau_{\rm rec} \approx 10^5
\yr$. Accordingly, thermal equilibrium is reached much later than
photoionisation equilibrium.
\par
After thermal equilibrium has been reached, the ionising flux is
switched off and the parcel recombines and cools. Once it has cooled
to a temperature $T \lesssim 10^4 \K$, cooling becomes inefficient.
The temperature of the recombining parcel therefore remains roughly
constant.
\par
In the bottom panels of Fig.~\ref{Fig:Test1:Error} we quantify the
accuracy of our subcycling approach. Ideally, we would like to compare
the numerical results to an exact analytical reference solution. However, 
such a solution exists only for the case of a constant
temperature\footnote{We mention that by repeating the test at fixed
temperature, we have convinced ourselves that the ionisation history
computed using our subcycling recipe follows the analytical solution
very closely (see \citealp{Pawlik:Thesis}).}  (see, e.g., the appendix
in \citealp{Dove:1994}).  Instead, we therefore show the relative
error of the evolutions shown in the top panel with respect to the
evolutions obtained from the simulation with the next smaller time
step.  
\par
For all our choices of the time step $\Delta t$ and for
most of the simulation time the relative errors are small,
$\lesssim 1\%$ or even much smaller. During the initial phase of rapid
evolution the relative error in the ionised fraction briefly becomes
as large as $10\%$. In practice, 
such errors will have little impact on the results of
RT simulations if photons are conserved and if the equilibrium solution is still 
obtained with high accuracy (as is the case). Moreover, relative differences of 
order $10\%$ are already implied by uncertainties in current atomic data 
used to compute the ionisation and recombination rates and the radiative 
heating and cooling rates (as will be discussed in Sec.~\ref{Sec:Tests:Test2:Reference:T}). 
Note that the relative error can be reduced by lowering the numerical factor $f$, which controls the
size of the subcycle steps and hence the integration accuracy. 
We conclude that the results of the subcycling are insensitive to the size of the simulation time step.
\par
In summary, we have demonstrated that our subcycling recipe
accurately computes, independently of the size of the RT time step, 
the combined evolution of the neutral fraction and
temperature of gas exposed to hydrogen-ionising radiation. In the
following sections we will employ the subcycling to compute the
species fractions and temperature of gas particles in RT
simulations.

%%%%%%%%%%%%%%%%%%%%%%%%%%%%%%%%%%%%%%%%%%%%%%%%%%%%%%%%%%%%%%%%%%%%%%%%%%%%%%%%%%%%%%%%%%%%%%%%%%%%%%%%%%%%%
\subsection{HII region expansion. Reference results and comparisons of multi-frequency and grey solutions}
\label{Sec:Tests:Reference}
In the next section (Sec.~\ref{Sec:Tests:Test2}) we will apply our new implementation of \traphic\ to
compute the evolution of the ionisation state and temperature around
an ionising source surrounded by gas of constant density. This is an
idealised test problem designed to facilitate the verification of our
implementation through the direct comparison to results obtained with
an improved version of our one-dimensional (1-d) RT code
(\citealp{Pawlik:2008}; hereafter referred to as \testtraphic, which
stands for TestTraphic1D), which solves the rate equations using the
same techniques (and code) as \traphic, as well as to published
reference results obtained with other RT codes for the same test problem
(\citealp{Iliev:2006a}). In this section we present these reference results. 
We also discuss the applicability of the grey approximation for solving multi-frequency RT problems. 
\par
We start in Sec.~\ref{Sec:Tests:Test2:Reference:fixedT} by presenting
reference solutions obtained with our 1-d RT code \testtraphic\ for
the case of hydrogen-only gas at fixed temperature. This is an
important case because it allows analytical solutions to be derived
against which the numerical results obtained with \testtraphic\ can be
compared. Then, in Sec.~\ref{Sec:Tests:Test2:Reference:T}, we compare
the performance of \testtraphic\ in a simulation of hydrogen-only gas
in which the gas temperature is allowed to evolve due to photoheating
and radiative cooling to published numerical results obtained for the
same problem.  Finally, in Sec.~\ref{Sec:Tests:Test2:Reference:He}, we
discuss results from simulations with \testtraphic\ in which the gas
also contains helium.
\par
\subsubsection{HII region in pure hydrogen gas at fixed temperature}
\label{Sec:Tests:Test2:Reference:fixedT}
In this section we discuss the RT problem of an 
expanding HII region. Despite its simplicity, an analytical solution to this problem 
cannot generally be obtained, even if the gas densities are assumed to 
be non-evolving (as is the case throughout this work). This is because the coupling between the
ionisation and temperature state through the dependence of the
collisional ionisation, recombination and cooling rates on the
temperature and species fractions impedes the evaluation of the
governing differential equations (Eqs.~\ref{Eq:NeutralFractions1} -
\ref{Eq:NeutralFractions6} and \ref{Eq:dudt}).
\par
To provide an approximate point of reference, we recall the
evolution of an HII region at fixed gas temperature, 
for which an analytical solution is known
(under the approximation that the ionised region is fully ionised; we will also ignore
collisional ionisations, although this is not necessary). We have reviewed this solution in Paper~I,
where we showed that the radius of the ionised sphere around a source
of ionising luminosity $\dot{\mathcal{N}}_\gamma$ that is located in a
homogeneous hydrogen-only medium of density $n_{\rm H}$ is given by
\begin{equation}
r_{\rm I}(t) = r_{\rm s} (1-e^{-t/\tau_{\rm s}})^{1/3},
\label{Eq:Ifront}
\end{equation}
where $r_{\rm s} = [3\dot{\mathcal{N}}_\gamma / (\alpha_{\rm B HII}
n_{\rm H}^2)]^{1/3}$ is the Str\"omgren radius and $\tau_{\rm
s}=1/(\alpha_{\rm B HII} n_{\rm H})$ is the Str\"omgren time scale,
which equals the recombination time for fully ionised gas. 
In some of our comparisons we will employ this approximate point of
reference. We will refer to it as an analytical approximation. On the other hand, because
of the lack of an accurate analytical solution, we will 
mostly employ results obtained with our 1-d RT code \testtraphic\ in
our benchmarking below. For this reason, we will first discuss its
performance.
\par  
\begin{figure}
  \includegraphics[trim = 25mm 0mm 30mm 15mm, width=0.49\textwidth]{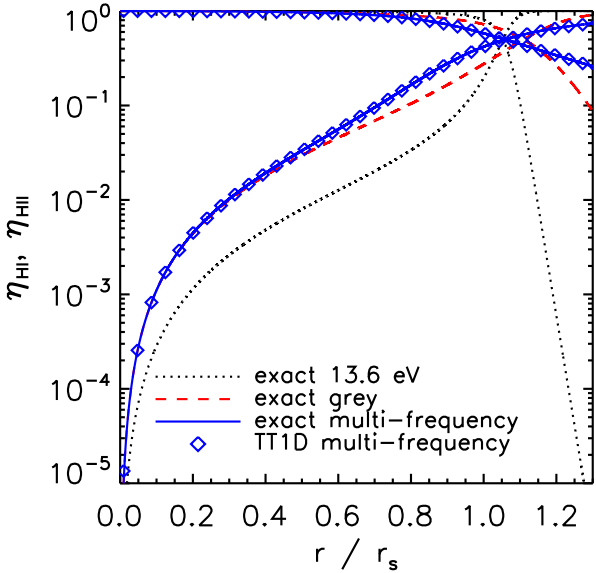}
  \caption{Photoionisation equilibrium profiles of the neutral and
    ionised fraction around a single blackbody source in a homogeneous
    hydrogen-only medium. The numerical result obtained with
    \testtraphic\ (diamonds) shows excellent agreement with the
    analytically computed exact results (Eq.~\ref{Eq:Equilibrium};
    blue solid curve). For comparison, we also show the analytically
    computed exact solutions assuming the grey approximation (red
    dashed curve) and the monochromatic treatment used in Test 1 of
    Paper~I (black dotted curve). The grey approximation agrees with
    the exact solution in the optically thin limit (i.e., in the
    absence of spectral hardening), while the monochromatic treatment
    fails at all distances because it implies photoionisation rates 
    inconsistent with the spectrum of the considered blackbody source. \label{Chapter:Heating:Fig:Multifrequency}}
\end{figure}
We start by verifying our multi-frequency treatment in \testtraphic\
by comparing its performance in a simple HII region test problem to the corresponding equilibrium
solution that can be derived analytically (except for a numerical
evaluation of the integrals involved). The test consists of simulating
the spherically symmetric growth of the ionised region around a single
ionising source in a homogeneous hydrogen-only medium at fixed temperature. The source has a blackbody
spectrum with temperature
$10^5 \K$ and emits radiation with an ionising luminosity
$\dot{\mathcal{N}}_{\gamma} = 5 \times 10^{48} ~\mbox{photons} \invs$.
The gas density is $n_{\rm H} =
10^{-3} \cmci$. The
initial ionised fraction is assumed to be $\eta_{\rm HII} = 0$, and we use a
recombination coefficient $\alpha_{\rm B HII} = 2.59\times 10^{-13}
\cmc\invs$, independent of radius and time.  Collisional ionisation is
not included.  For reference, with the physical parameters mentioned above, the
Str\"omgren time is $\tau_{\rm S} = 122.4 \Myr$ and the Str\"omgren
radius is $r_{\rm S} = 5.4 \kpc$. The spatial resolution, the time step and the number of
frequency bins used in the simulation with \testtraphic\ are chosen
such as to achieve numerical convergence.
\par
In Fig.~\ref{Chapter:Heating:Fig:Multifrequency} we show the neutral
(ionised) fraction profile in photoionisation equilibrium. Diamonds
show the result of the simulation with \testtraphic\ (at $t = 2000
\Myr$). The blue solid curve indicates the exact equilibrium solution
obtained by solving (e.g., \citealp{Osterbrock:1989})
 \begin{equation}
   \frac{\eta_{\rm HI, eq}(r) n_{\rm H}}{4\pi r^2}\int d\nu\
     \dot{\mathcal{N}}_\gamma(\nu) e^{-\tau_{\nu}}\sigma_{\nu} =
     \eta_{\rm HII, eq}^2 (r)n_{\rm H}^2 \alpha_{\rm B HII},
 \label{Eq:Equilibrium}
 \end{equation}
where the frequency-dependent optical depth $\tau_{\nu}(r)$ is given
by
\begin{equation}
   \tau_{\nu}(r) = n_{\rm H}\sigma_{\nu} \int_0^r dr^{\prime}\
   \eta_{\rm HI, eq}(r^{\prime}).
\end{equation}
\par
The simulation result is in excellent agreement with the exact
equilibrium solution, verifying our multi-frequency implementation of
\testtraphic. For comparison, we also show the exact equilibrium
solution assuming that the radiation is monochromatic (dotted black
curve) with a photoionisation cross-section evaluated at the
ionisation threshold, i.e.\ $\sigma_{\gamma \rm HI} = 6.3 \times
10^{-18} \cms$. We also show the exact equilibrium solutions in the
grey treatment, i.e.\ using the average cross-section $\langle
\sigma_{\gamma \rm HI}\rangle = 1.63 \times 10^{-18} \cms$ (dashed red
curve). 
\par
The reason for the differences between the results of the
multi-frequency computation and the results of the grey and
monochromatic computation can be readily understood. The absorption
cross-section for ionising photons is a strongly decreasing function
of the photon energy. The ionising photons with the lowest energy are
therefore preferentially absorbed, which leads to an increase in the
typical photon energy with distance. This effect is referred to as
spectral hardening. Because the photon mean free path is inversely
proportional to the absorption cross-section, spectral hardening
increases the width of the ionisation front with respect to that
obtained in the absence of spectral hardening. Note that spectral
hardening only becomes important for large optical depths, which
explains why the grey approximation reproduces the multi-frequency
solution at small distances where the optical depth is low. The
monochromatic approximation, on the other hand, implies an
inappropriate value for the photoionisation rate and hence fails to
describe the present multi-frequency problem at all distances from the
source.  Note that both the grey and the monochromatic approximation
will provide a better description of the multi-frequency problem for
sources with a softer radiation spectrum.
\par
\subsubsection{HII region in pure hydrogen gas with an evolving temperature}
\label{Sec:Tests:Test2:Reference:T}
Having demonstrated the validity of our multi-frequency treatment with
\testtraphic, we now repeat the test problem from the previous section 
but this time we account for the self-consistent evolution of the gas temperature due to photoheating and radiative cooling.  
The physical parameters for the test are taken from
\cite{Iliev:2006a}.  We consider an ionising source embedded in a
homogeneous hydrogen-only density field with number density $n_{\rm H}
= 10^{-3} \cmci$. The source
emits $\dot{\mathcal{N}}_{\gamma} \equiv \int d\nu\
\dot{\mathcal{N}}_\gamma(\nu) = 5 \times 10^{48} ~\mbox{ionising
photons} \invs$ with a blackbody spectral shape
$\dot{\mathcal{N}}_\gamma(\nu)$ corresponding to a blackbody
temperature $T_{\rm bb} = 10^5 \K$. The test described here is identical to Test 1 in Paper~I, except that
now the gas temperature is allowed to vary due to heating and cooling
processes as described in Sec.~\ref{Sec:Heating} (with Compton cooling
off the redshift $z=0$ cosmic microwave background included) and that
collisional ionisation is included. The gas is assumed to have an
initial ionised fraction $\eta_{\rm HII} = 1.2\times 10^{-3}$
(approximately corresponding to the ionised fraction implied by
collisional ionisation equilibrium at the temperature $T =10^4 \K$). Its
initial temperature is set to $10^2 \K$. As before, 
the spatial resolution, the time step and the number of
frequency bins used in the simulation with \testtraphic\ are chosen
such as to achieve numerical convergence.
\par
\begin{figure}
  \begin{center}
    \includegraphics[trim = 0mm 0mm 50mm 0mm, width=0.49\textwidth]{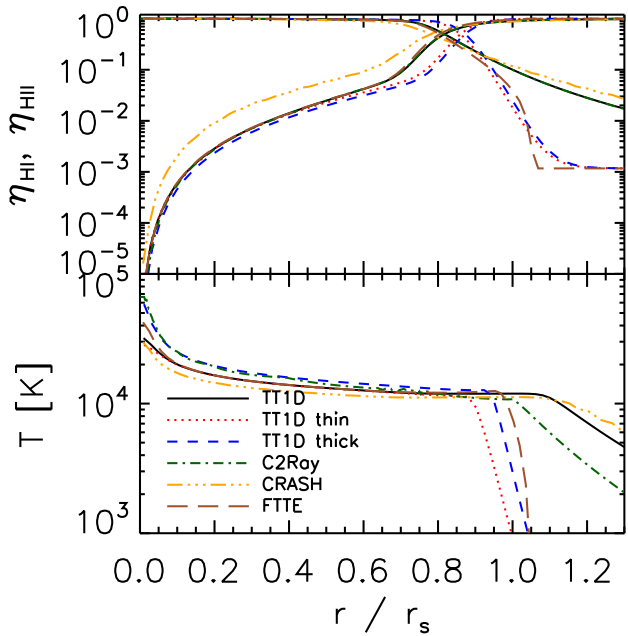}
    \caption{Test~2 (using \testtraphic). Comparison of the grey
      approximations with the full multi-frequency solution. The
      figure shows spherically averaged profiles of neutral (ionised)
      fraction (top) and temperature (bottom) at time $t=100\Myr$. The
      black solid curve shows the multi-frequency solution. The red
      dotted (blue dashed) curve shows its grey approximation assuming
      photoheating rates computed in the optically thin (thick)
      limit. The grey and multi-frequency simulations show clear
      differences close to and beyond the ionisation front, where the
      large optical depth causes spectral hardening of the emitted
      blackbody spectrum. For reference, we also show results obtained
      with other RT codes as published in
      \protect\cite{Iliev:2006a}. The green dash-dotted and black
      solid curves in the top panel are difficult to distinguish
      because they fall nearly on top of each other. The differences
      between these results at large distances mainly reflect the
      differences in the numerical treatment of multi-frequency
      radiation in these codes. Most of the differences close to the
      ionising source have their origin in the use of different
      assumptions for computing photoheating rates, as the comparison
      to the results obtained with \testtraphic\
      reveals. \label{Chapter:Heating:Fig:Comparison}
    }
  \end{center}
\end{figure}

Fig.~\ref{Chapter:Heating:Fig:Comparison}
shows the neutral (ionised) fraction and temperature profiles at time
$t = 100 \Myr$ using \testtraphic\ (black solid curves). We compare 
these multi-frequency results to results obtained using the grey 
approximation, i.e. using an average cross-section $\langle
\sigma_{\gamma \rm HI}\rangle = 1.63 \times 10^{-18} \cms$ and grey photoheating rates. 
We employ grey photoheating rates computed in the
optically thin limit (red dotted curves), according to which each photoionisation adds
$\langle \epsilon_{\rm HI}\rangle=6.32 \eV$ 
to the internal energy of the gas, and in the optically thick limit (blue dashed curves), in which 
case each photoionisation adds 
$\langle\epsilon^{\rm thick}_{\rm HI}\rangle=16.01 \eV$ 
to the internal energy of the gas (Sec.~\ref{Sec:Heating}).
We employ the labels {\it grey thin} and {\it grey thick}  to distinguish the 
two grey simulations from each other. 
We also show the results obtained (in three-dimensional RT simulations) with the RT codes \ctworay\
(\citealp{Mellema:2006}), \crash\ (\citealp{Ciardi:2001};
\citealp{Maselli:2003}) and \ftte\ (\citealp{Razoumov:2005}) for the
same test problem, as published in \cite{Iliev:2006a}.
\par

The differences in the neutral fractions between the grey and the
multi-frequency simulations that we have discussed above for
Fig.~\ref{Chapter:Heating:Fig:Multifrequency} are again clearly
visible (top panel of Fig.~\ref{Chapter:Heating:Fig:Comparison}). The
{\it grey thin} simulation yields results that asymptote to those
obtained in the multi-frequency simulation at small distances from the
ionising source. At large distances, i.e.\ near the ionisation front
and beyond, on the other hand, the multi-frequency simulation implies
significantly larger ionised fractions than those implied by this
grey simulation. This is because the photon mean free path is larger
in the multi-frequency simulation than in the grey simulations due to
spectral hardening, leading to a smoother transition of the neutral
fraction between the highly ionised gas interior to and the neutral
gas far ahead of the ionisation front.
\par
The {\it grey thick} simulation yields neutral fractions that are very
similar to those found in the {\it grey thin} simulation. However, the {\it
grey thick} simulation yields slightly lower neutral
fractions than the {\it grey thin} simulation, since it yields
slightly larger temperatures, and thus smaller recombination rates,
throughout the ionised region (bottom panel of
Fig.~\ref{Chapter:Heating:Fig:Comparison}). In contrast to the {\it
grey thin} simulation, the neutral fractions obtained in the {\it grey
thick} simulation therefore do not asymptote to those obtained in the
multi-frequency simulation at small distances to the ionising
source. Instead, they remain systematically too small.
\par
The differences between the grey and multi-frequency simulations (and
between the {\it grey thin} and {\it grey thick} simulations) become
particularly apparent when inspecting the corresponding temperature
profiles. The multi-frequency simulation yields substantially higher
gas temperatures ahead of the ionisation front. This {\it pre-heating}
is a simple consequence of the increase in the photon mean free path
above that in the grey simulations. As already noted, 
at fixed radii the {\it grey thick} simulation
shows systematically higher gas temperatures than the {\it grey
thin} simulation. The reason is that in the optically thin limit the
contribution of high-energy photons to the photoheating rate is
reduced due to the weighting by the absorption cross-section
$\sigma_{\rm HI}(\nu)$, which is a strongly decreasing function of the
photon energy. Observe that the temperatures (like the neutral
fractions) obtained in the {\it grey thin} simulation asymptote to
those obtained in the multi-frequency simulation at small distances to
the ionising source, while the temperatures in the {\it grey
thick} simulation are too high.
\par
We summarise our discussion of the differences between the grey and
multi-frequency simulations for the present problem by noting that the
use of the grey approximation leads to neutral fractions and
temperatures that generally are very different from those obtained in
detailed multi-frequency simulations. At large optical depths, the
neutral fractions are systematically too high and the temperatures too
low due to the lack of spectral hardening.  The grey treatment yields
neutral fractions and temperatures that asymptote to those obtained in
the corresponding multi-frequency simulation at small distances to the
ionising source when photoheating rates are computed in the optically
thin limit, i.e.\ using Eq.~\ref{Eq:AverageExcessEnergy}. When
computing photoheating rates in the optically thick limit, i.e.\
using Eq.~\ref{Eq:AverageExcessEnergyThick}, the neutral fractions and
temperatures do not asymptote to the correct values at small distances
to the ionising source, i.e.\ the values in the
multi-frequency simulation. Consequently, when one invokes the grey
approximation to compute the thermal structure of ionised regions, one
should compute photoheating rates in the optically thin
limit. Photoheating rates in the optically thick limit should only be
employed when considering the thermal balance of an ionised region as
a whole. Ideally, one would perform detailed multi-frequency
simulations and simply dispense with the grey approximation.
\par

We now discuss the results of our simulations with \testtraphic\ with
respect to those obtained with \ctworay, \crash\ and \ftte\ for the
same test problem (\citealp{Iliev:2006a}). We note that the simulation
with \crash\ employed multiple frequency bins, while the one with
\ftte\ was done using a single frequency bin and computing
photoionisation and optically thick photoheating in the grey
approximation (Alexei Razoumov, private communication). Finally,
\ctworay\ used a hybrid method (Garrelt Mellema, private
communication): the absorption of ionising radiation was computed as a
function of frequency, but each photoionisation injected the same
amount of energy, regardless of the frequency of the absorbed
photon. This method thus accounts fully for the spectral hardening of
the radiation but ignores it when computing photoheating rates.
\par
There are noticeable differences in the results obtained with these
three codes. At large distances from the ionising source, i.e.\ close
to and beyond the ionisation front, most of these differences may be
attributed to differences in the multi-frequency implementation,
leading to differences in the spectral hardening of the emitted
blackbody spectrum. At these distances, the neutral fractions obtained
in our grey simulations agree closely with those obtained with \ftte,
while the neutral fractions obtained in our multi-frequency
simulations closely agree with (and are, in fact, nearly identical to) those obtained with \ctworay, as
expected from our discussion above. We note that the fact that the
neutral fractions obtained with \crash\ are systematically too large
may indicate that the radiation field was too poorly sampled (see
\citealp{Maselli:2003}, in particular their Fig.~2, for a thorough
discussion).
\par
The results, however, show also significant differences in the
neutral fractions and temperatures close to the ionising source, where
the gas is close to optically thin and the emitted blackbody radiation
spectrum is not severely deformed due to spectral hardening. Some of
these differences can be attributed to the fact that the different
codes employ different expressions for cross-sections, recombination
and cooling rates. As demonstrated in \cite{Iliev:2006a} (their
Fig.~4), different recombination and cooling rates may only
account for differences in the neutral fraction and temperature of at
most $\lesssim 10 \%$. We have verified this by employing the rates
used with the different codes (Table~2 in \citealp{Iliev:2006a}) in
simulations with \testtraphic.
\par
Most of the differences close to the ionising source may instead be
traced back to the use of different assumptions underlying the
computation of the photoheating rates. In fact, the temperatures
obtained with \crash\ are in very good
agreement\footnote{\cite{Maselli:2009} have repeated this test with a
more recent version of \crash\ with improved sampling of the Monte
Carlo photon field. They find slightly larger temperatures (their
Fig.~3), which further improves the agreement with the temperatures
found with \testtraphic.} with the temperatures obtained in our
multi-frequency and grey thin simulations, while the temperatures
obtained by \ftte\ and \ctworay\ are in excellent agreement with the
temperatures in our grey thick simulation.
\par
\begin{figure}
\begin{center}
  \includegraphics[width=0.44\textwidth]{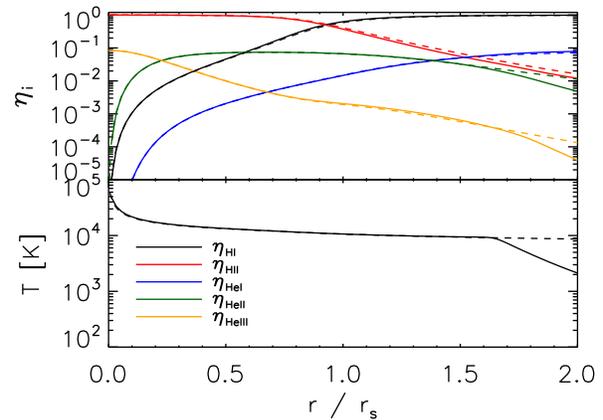}
  \caption{Comparison of the results from a simulation with
    \testtraphic\ (solid curves) with those from a simulation with
    \cloudy\ (dashed curves) for the HII region test problem described
    in Sec.~\ref{Sec:Tests:Test2} (assuming $X = 0.75$, $Y = 1 - X$).
    Note that the results computed with \cloudy\ correspond to
    equilibrium ($t\to\infty$), while the results computed with
    \testtraphic\ correspond to a time $t = 2000 \Myr$ (which is much
    larger than the recombination time on which the gas reaches
    thermal and ionisation equilibrium). The results obtained with
    \testtraphic\ are nearly indistinguishable from those obtained
    with \cloudy\ for all radii at which equilibrium has been reached.}
  \label{Fig:Reference:Helium}
\end{center}
\end{figure}
\subsubsection{HII region in gas containing hydrogen and helium with an evolving temperature}
\label{Sec:Tests:Test2:Reference:He}
Finally, we test the ability of \testtraphic\ to accurately compute
the ionisation and temperature structure in gas containing both
hydrogen and helium by comparing results obtained with
\testtraphic\ with results obtained with the photoionisation code 
\cloudy\ (version 08.00; last described by \citealp{Ferland:1998}) 
for the same test problem. Note that \cloudy\ assumes ionisation equilibrium.
\par
As before we consider an ionising source with blackbody spectrum of
temperature $T=10^5\K$ with an ionising luminosity of $5\times 10^{48}
\invs$ in gas of density $n_{\rm H} = 10^{-3}$, but we now assume $X =
0.75$ and $Y = 1 - X$. \cloudy\ includes considerably more physics
than \testtraphic. To facilitate a direct comparison, we therefore
keep the setup of the simulations as simple as possible: we assume
that there is no radiative coupling between hydrogen and helium, i.e.,
photons emitted due to recombination of helium do not lead to
ionisations of hydrogen, and compute recombinations in the Case A
(Sec.~\ref{Sec:Recombination}) limit.
\par
Fig.~\ref{Fig:Reference:Helium} shows the ionised fractions and
temperatures computed with \testtraphic\ at time $t = 2000 \Myr$. It
also shows the results obtained with \cloudy, which correspond to a
time $t \to \infty$.  The agreement between \testtraphic\ and \cloudy\
is excellent. The results only differ at radii where (in the
simulation with \testtraphic) equilibrium has not yet been
reached. Thus, \testtraphic\ yields equilibrium ionisation and
temperature profiles that are almost indistinguishable from those
obtained with a well-tested and widely employed state-of-the-art
photoionisation code.
\par 

\subsection{Test 2: HII region expansion. \TRAPHIC}
\label{Sec:Tests:Test2}
In this section we apply our new implementation of \traphic\ to
compute the evolution of the ionisation state and temperature around
an ionising source surrounded by gas of constant density.
This idealised test problem captures the main characteristics of a
thermally coupled RT simulation that we wish to verify: conservation
of the number of ionising photons, which ensures that the final
ionised region attains the correct size, and conservation of the
associated energy, which, together with an accurate implementation of
the relevant cooling processes, ensures that the ionised region
settles into the correct thermal structure. The physical parameters for the test are identical to those employed to obtain 
the reference solutions presented in Sec.~\ref{Sec:Tests:Test2:Reference:T} but, for definiteness, we repeat the 
problem description here. 
\par
We consider an ionising source embedded in a
homogeneous hydrogen-only density field with number density $n_{\rm H}
= 10^{-3} \cmci$. The source
emits $\dot{\mathcal{N}}_{\gamma} \equiv \int d\nu\
\dot{\mathcal{N}}_\gamma(\nu) = 5 \times 10^{48} ~\mbox{ionising
photons} \invs$ with a blackbody spectral shape
$\dot{\mathcal{N}}_\gamma(\nu)$ corresponding to a blackbody
temperature $T_{\rm bb} = 10^5 \K$. The test described here is identical to Test 1 in Paper~I, except that
now the gas temperature is allowed to vary due to heating and cooling
processes as described in Sec.~\ref{Sec:Heating} (with Compton cooling
off the redshift $z=0$ cosmic microwave background included) and that
collisional ionisation is included. The gas is assumed to have an
initial hydrogen ionised fraction $\eta_{\rm HII} = 1.2\times 10^{-3}$
(approximately corresponding to the ionised fraction implied by
collisional ionisation equilibrium at the temperature $T =10^4 \K$). Its
initial temperature is set to $10^2 \K$. 
\par
First, in Sec.~\ref{Sec:Tests:Test2:Traphic}, we consider the case in
which radiation is transported using a single frequency bin in the
grey optically thin approximation in pure hydrogen gas.  We employ the grey approximation
to allow for a more direct comparison with the results presented in
Paper~I. In Sec.~\ref{Sec:Tests:Test2:TraphicMultifreq} we will 
also briefly discuss the performance of \traphic\ in
multi-frequency simulations in gas containing both hydrogen and helium.
\par

\begin{figure*}
  \begin{center}
    \begin{minipage}[c]{0.95\linewidth}
      \includegraphics[trim = 45mm 20mm 20mm 0mm, width=0.24\textwidth]{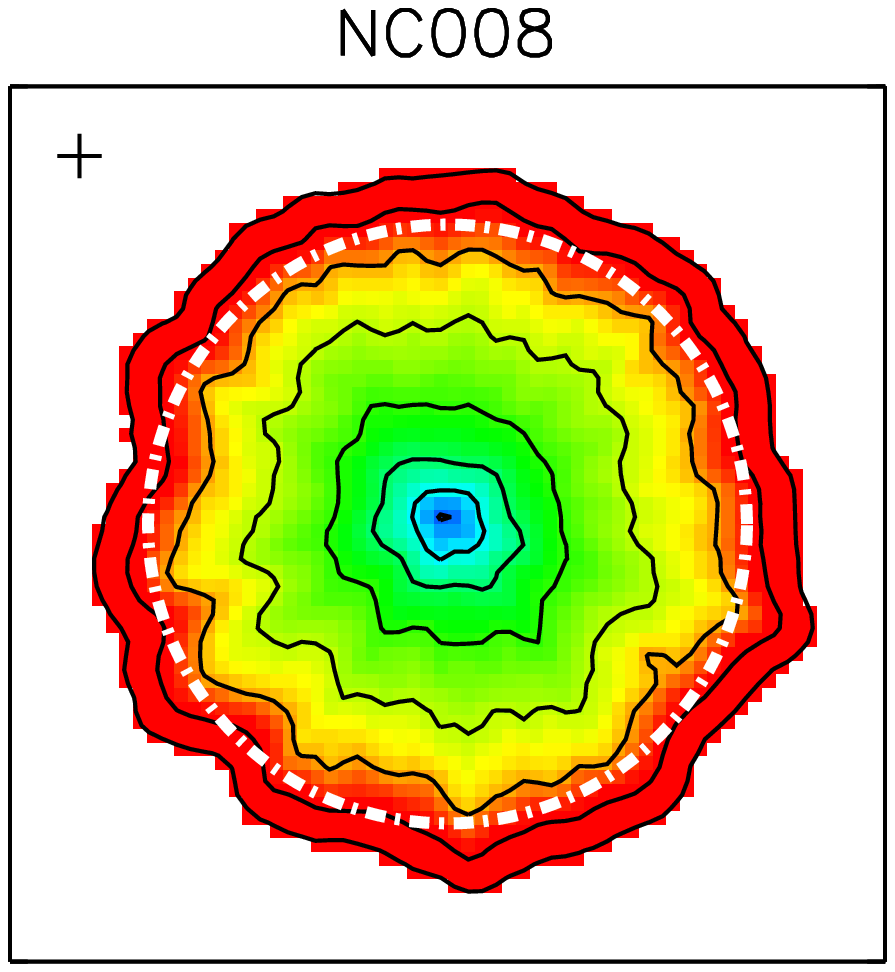}
      \includegraphics[trim = 45mm 20mm 20mm 0mm, width=0.24\textwidth]{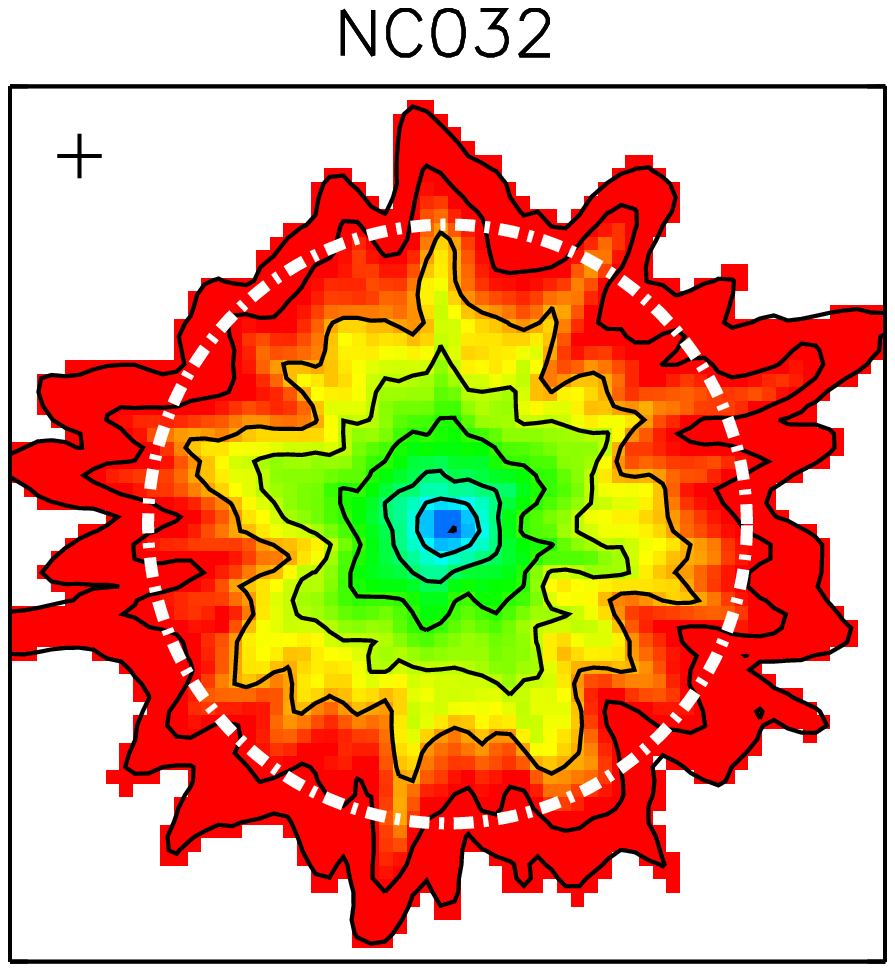}
      \includegraphics[trim = 45mm 20mm 20mm 0mm, width=0.24\textwidth]{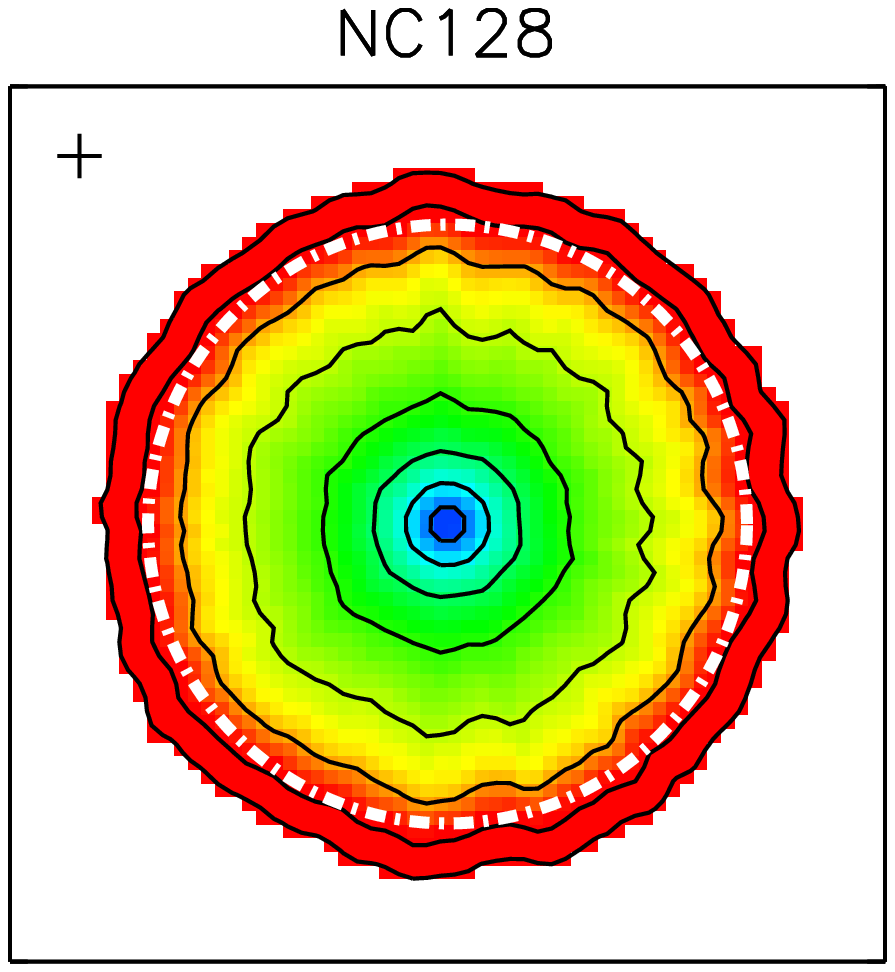} 
      \includegraphics[trim = 45mm 20mm 20mm 0mm, width=0.24\textwidth]{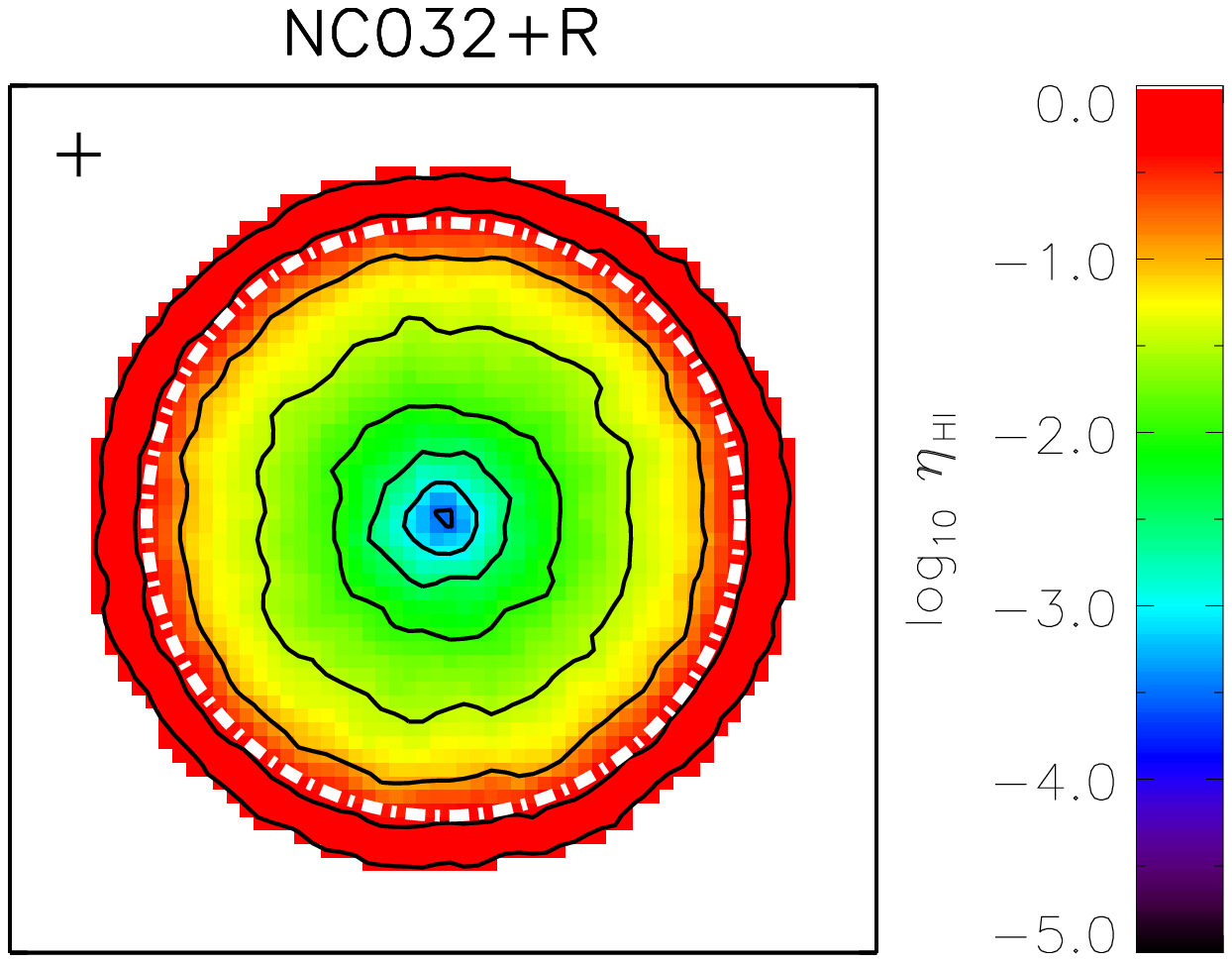}\\
      \includegraphics[trim = 45mm 20mm 20mm 0mm, width=0.24\textwidth]{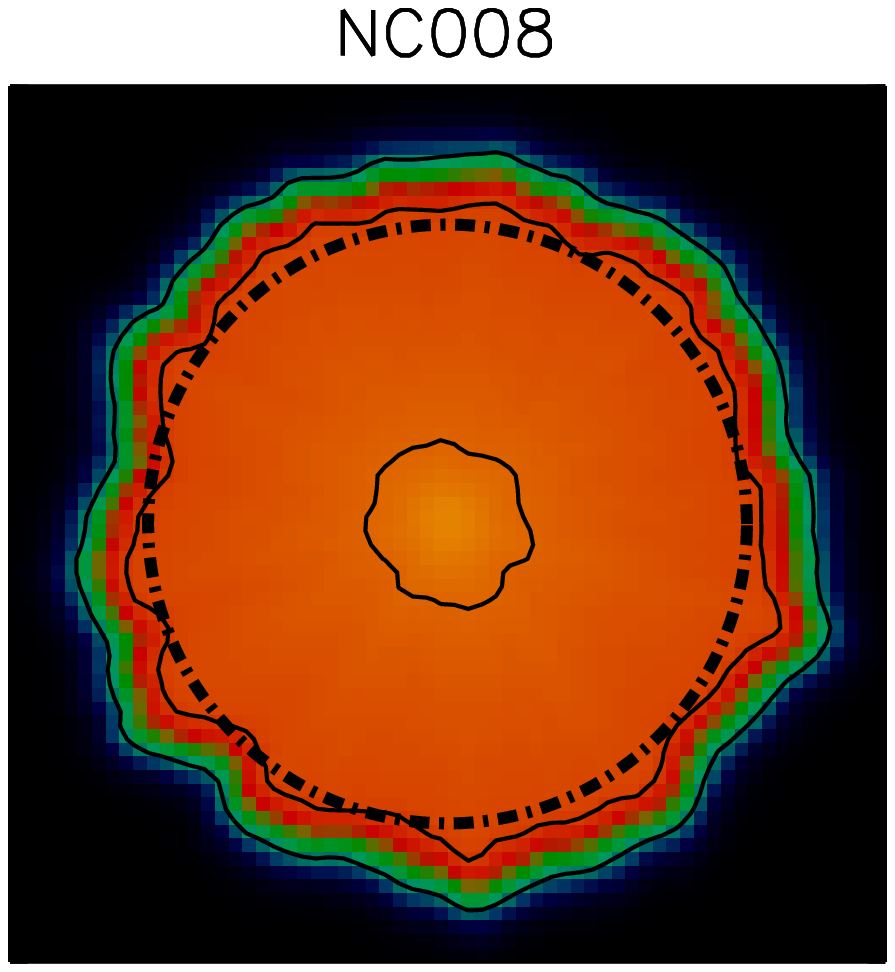}
      \includegraphics[trim = 45mm 20mm 20mm 0mm, width=0.24\textwidth]{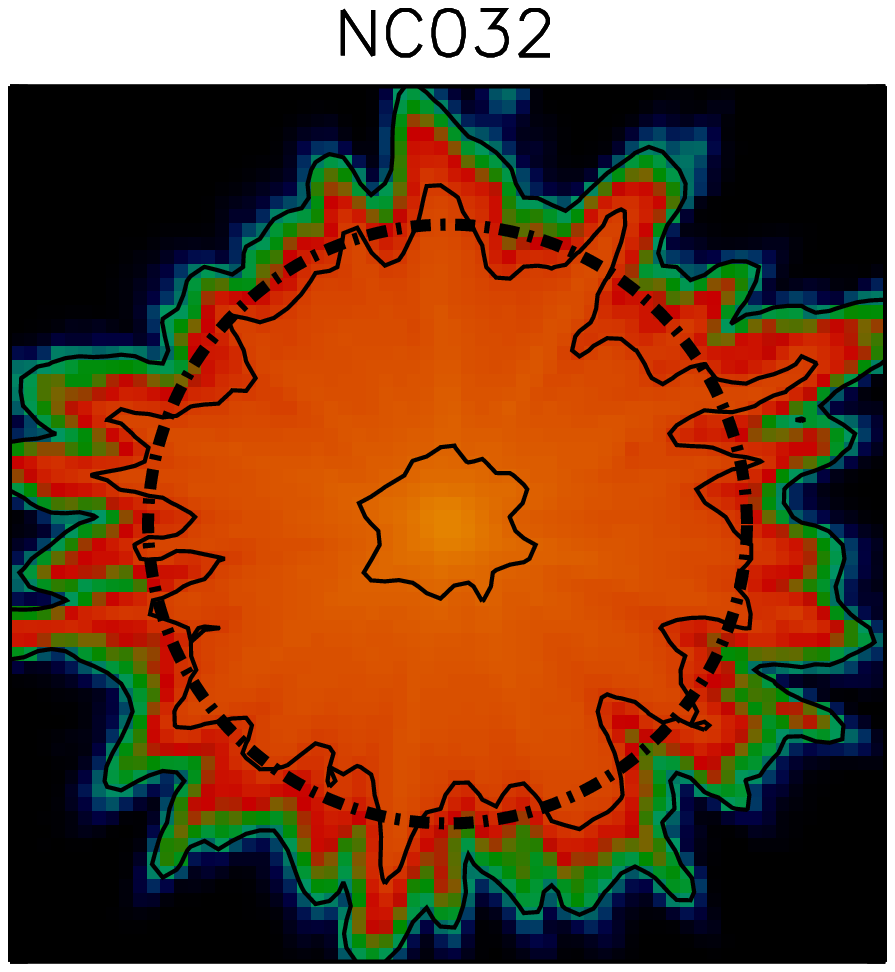}
      \includegraphics[trim = 45mm 20mm 20mm 0mm, width=0.24\textwidth]{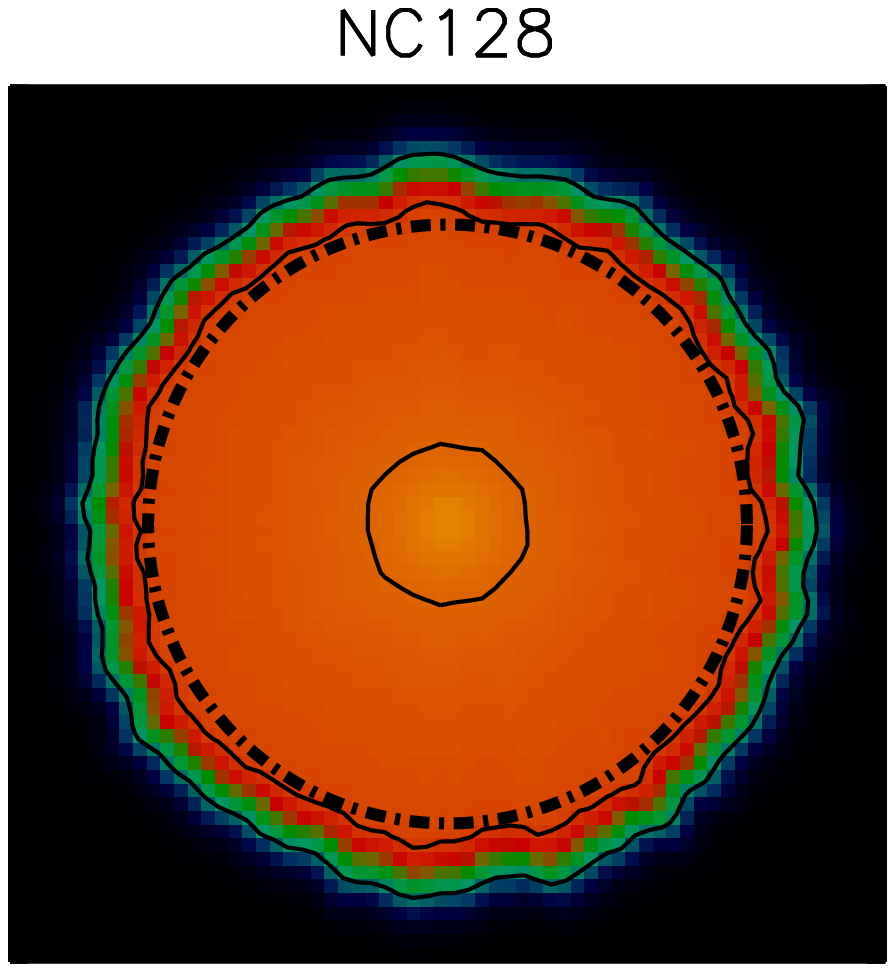}
      \includegraphics[trim = 45mm 20mm 20mm 0mm, width=0.24\textwidth]{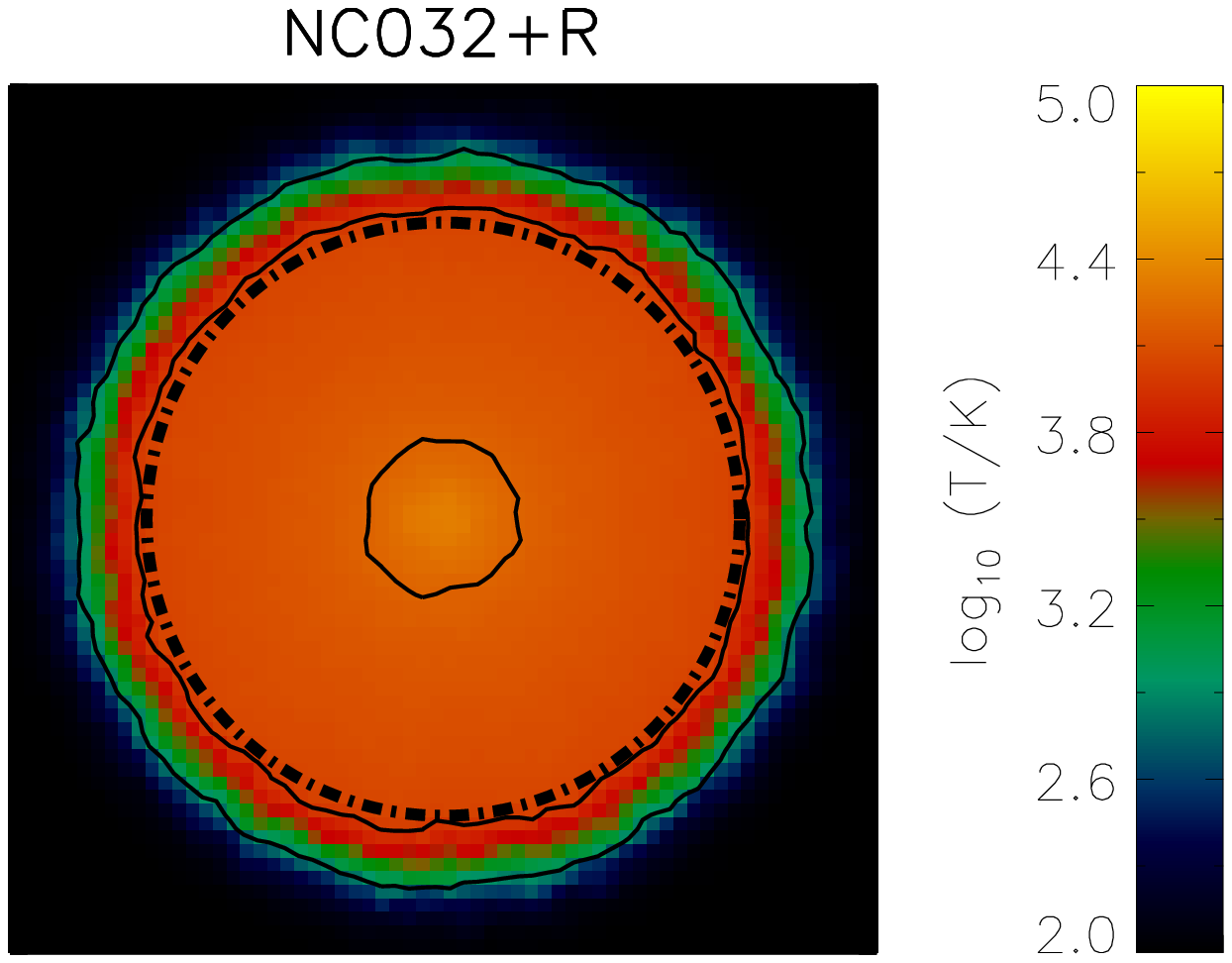}

\end{minipage}\vfill
\begin{minipage}[c]{0.95\linewidth}
  \caption{Test~2 (using \traphic). Neutral fraction ({\it top row})
    and temperature ({\it bottom row}) at time $ t = 100 \Myr$ in a
    slice through the centre of the simulation box. \textit{From left
    to right:} angular resolution $N_{\rm c} = 8, 32$, $128$ (all
    without resampling) and $32$ (with resampling of the particle
    positions after every 10th RT time step, as indicated by the
    letter `R' in the panel titles).  The dot-dashed circles indicate
    the position of the ionisation front, calculated using the
    analytical approximation (Eq.~\ref{Eq:Ifront}). Contours show
    neutral fractions of $\eta_{\rm HI} = 0.9, 0.5$, $\log_{10}
    \eta_{\rm HI} = -1, -1.5, -2, -2.5, -3, -3.5$ and temperatures
    $\log_{10} (T / {\rm K}) = (3, 4, 4.2)$ (from the outside in).
    The crosses in the top row indicate the spatial resolution
    $\langle 2\tilde{h} \rangle$.\label{Fig:Test2:Slices}}
\end{minipage}\vfill
\end{center}
\end{figure*}

\begin{figure*}
\begin{center}
\begin{minipage}[c]{0.9\linewidth}
\begin{center}

  \includegraphics[trim = 0mm 0mm 45mm 0mm,width=0.31\textwidth]{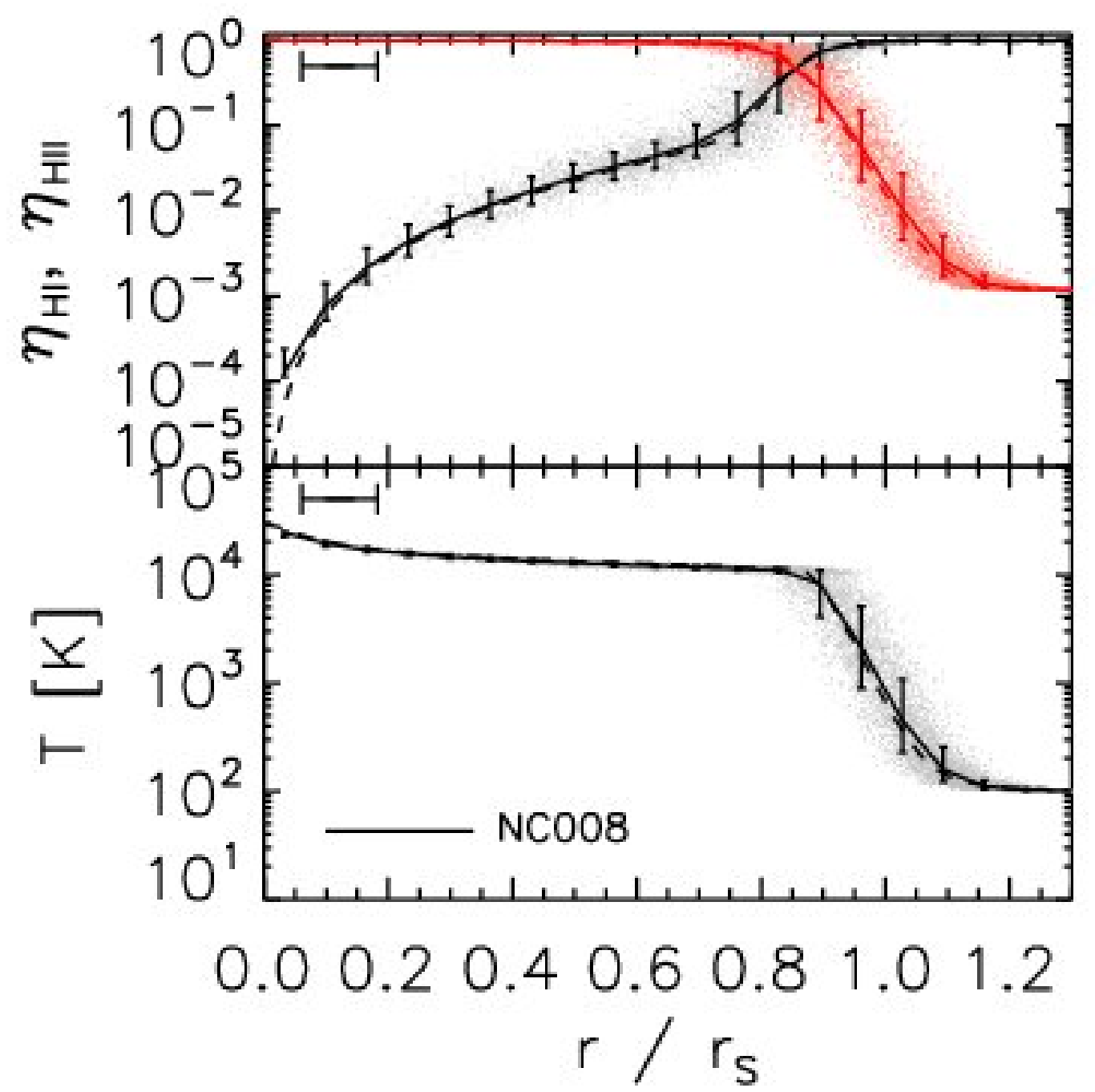}
  \includegraphics[trim = 0mm 0mm 45mm 0mm,width=0.31\textwidth]{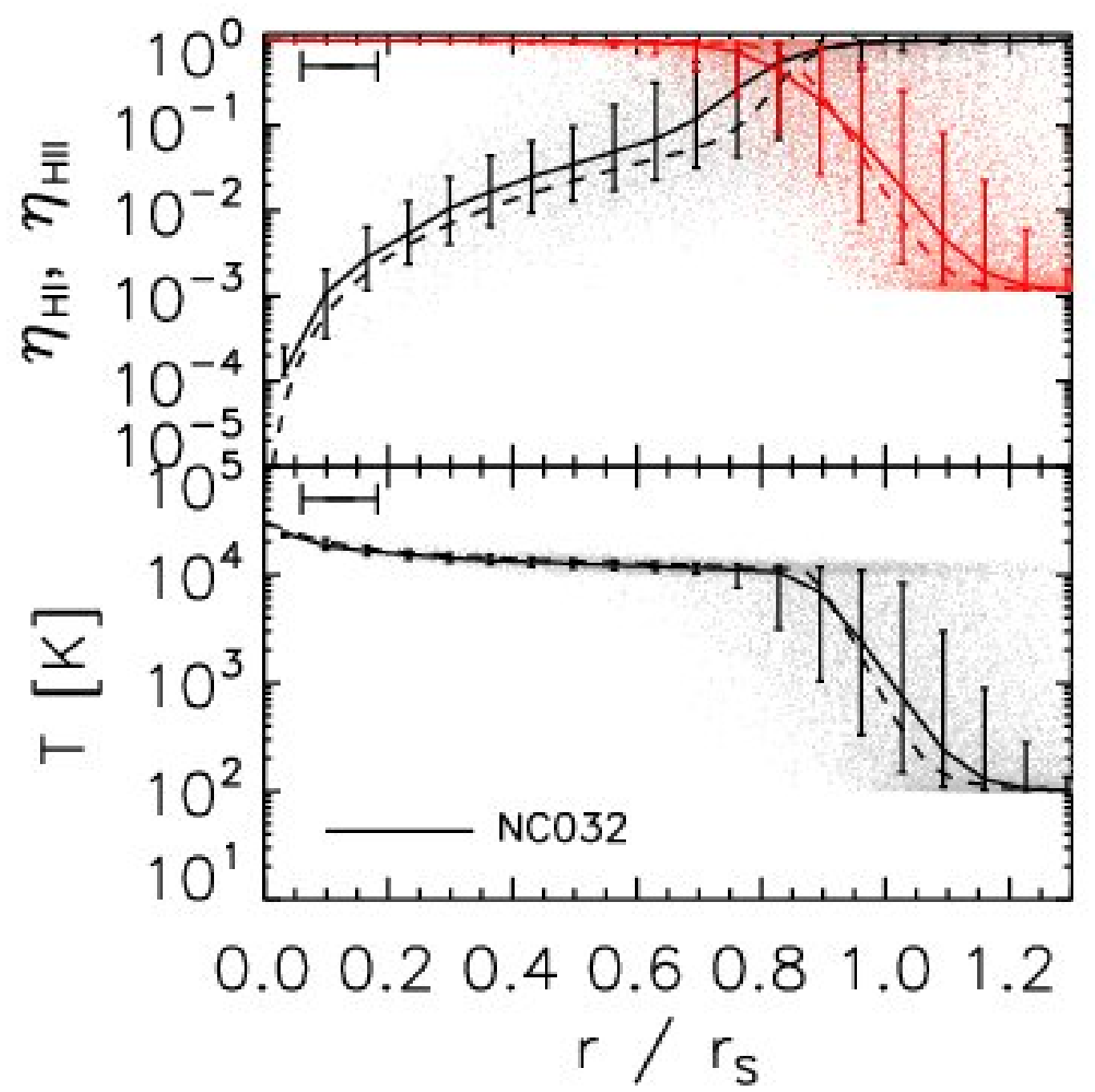}
  \includegraphics[trim = 0mm 0mm 45mm 0mm,width=0.31\textwidth]{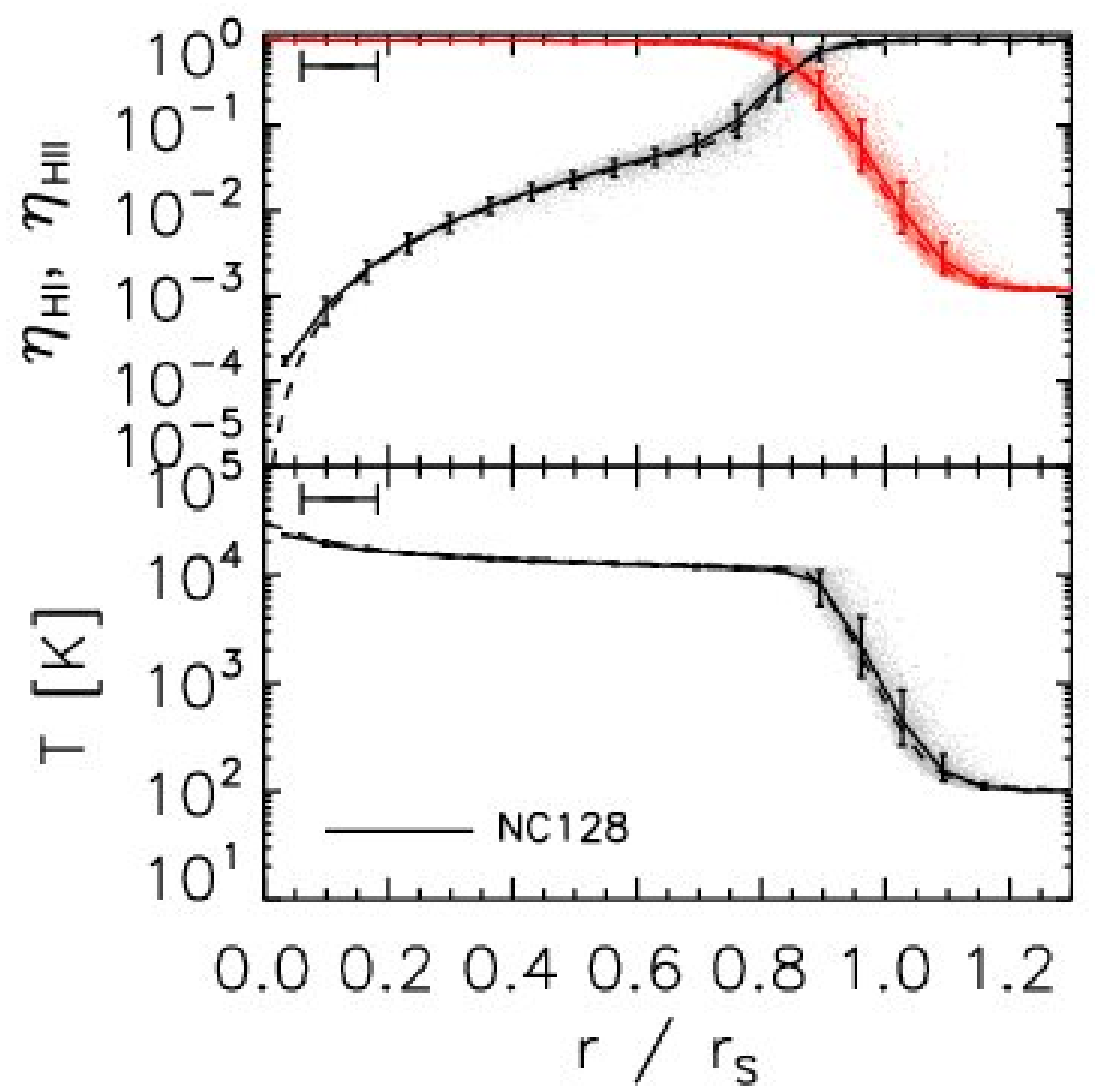}\\
  \includegraphics[trim = 0mm 0mm 45mm 0mm,width=0.31\textwidth]{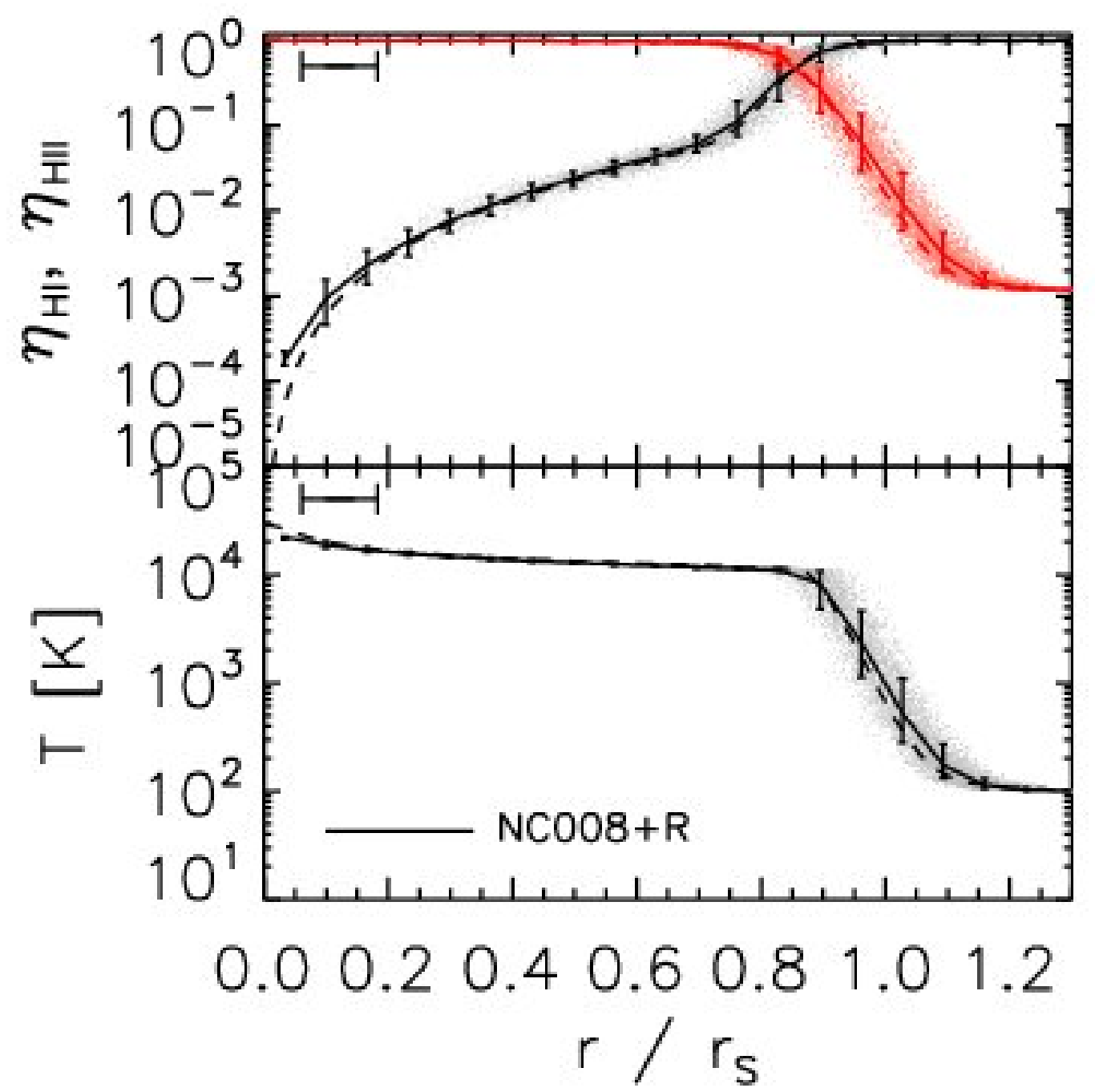}
  \includegraphics[trim = 0mm 0mm 45mm 0mm,width=0.31\textwidth]{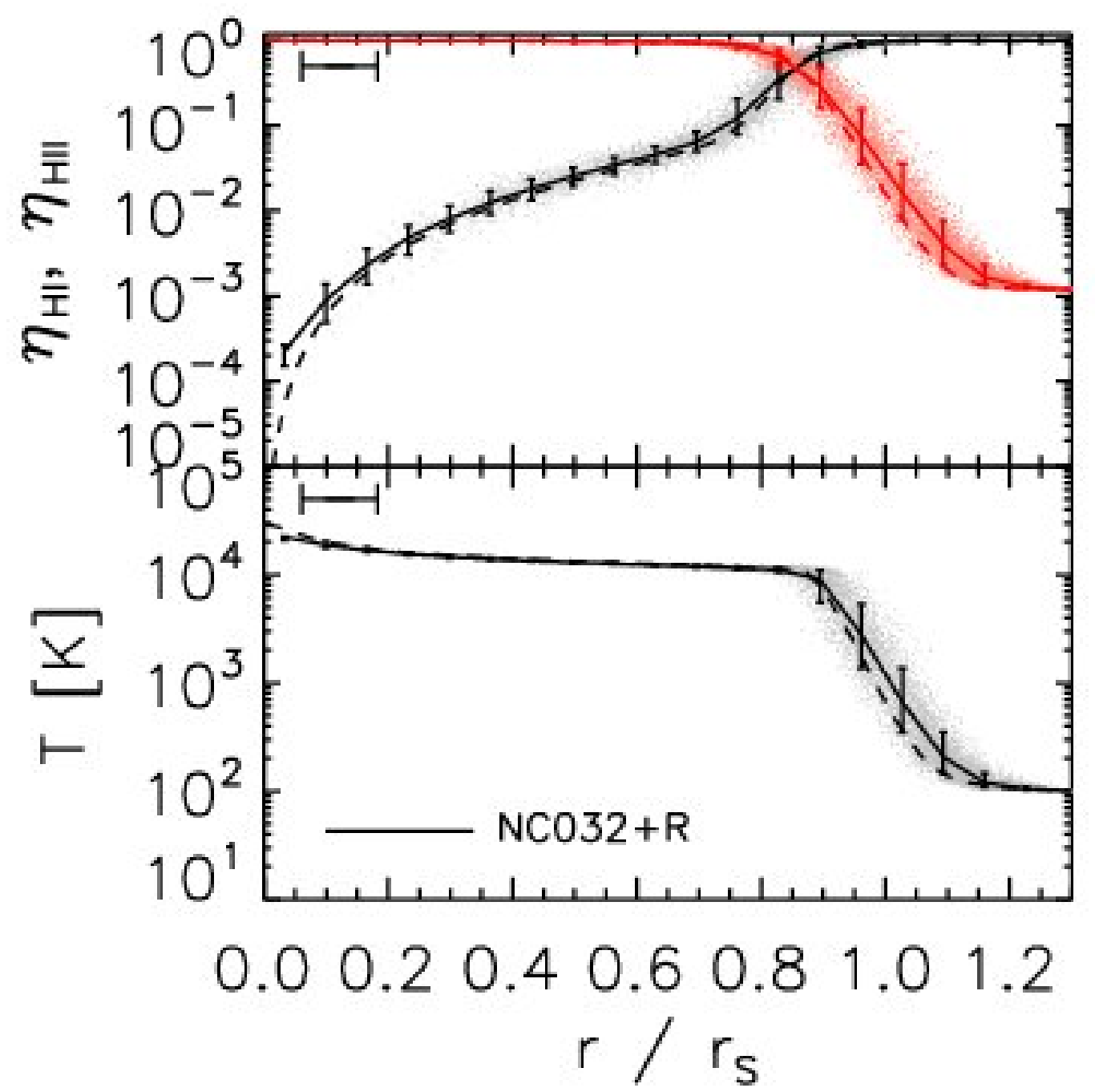}
  \includegraphics[trim = 0mm 0mm 45mm 0mm,width=0.31\textwidth]{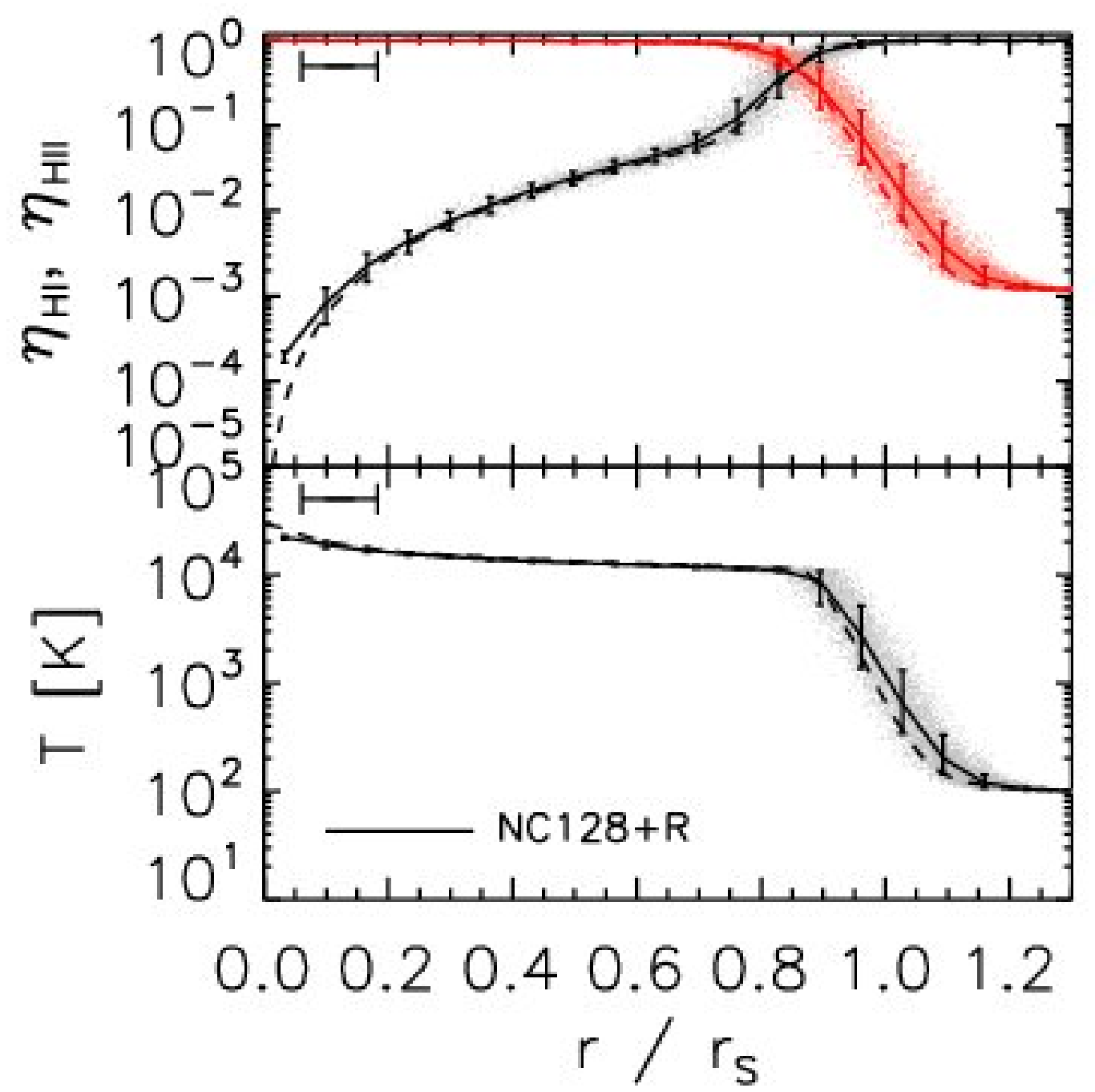}

\end{center}
\end{minipage}\vfill
\begin{minipage}[c]{0.9\linewidth}
  \caption{Test~2 (using \traphic). Scatter plots and profiles of the
    neutral (ionised) fraction and temperature at time $ t = 100 \Myr$
    for simulations with angular resolution $N_{\rm c} = 8$ (left),
    $32$ (middle) and $128$ (right). \textit{Top row:} No
    resampling. \textit{Bottom row:} Resampling of the particle
    positions after every 10th RT time step (indicated by the letter
    `R' in the legends). Each dot represents the neutral fraction
    (ionised fraction, temperature) of a single SPH particle (only 
    a randomly chosen subset of $10\%$ of all particles is shown). Solid
    curves show the median neutral fraction (ionised fraction,
    temperature) in spherical bins around the ionising source. The
    vertical error bars enclose $68.3\%$ of the particles in each
    bin. Dashed curves indicate the reference solution obtained with
    our 1-d RT code \testtraphic. The horizontal error bars in the
    upper left corners indicate the spatial resolution. The results of
    all simulations are in excellent agreement with the reference
    solution. Without the resampling, the results are noisier if
    $N_{\rm c} \approx \tilde{N}_{\rm ngb}$ (top middle
    panel).\label{Fig:Test2:Profiles}
  }
\end{minipage}\vfill
\end{center}
\end{figure*} 
\subsubsection{HII region expansion: grey thin, hydrogen-only}
\label{Sec:Tests:Test2:Traphic} 

The numerical realisation of the initial conditions is similar to that
used for Test 1 in Paper~I. The ionising source is located at the
centre of a simulation box with side length $L_{\rm box} = 13.2
\kpc$. The box boundaries are photon-transmissive, i.e., photons
leaving the box are lost from the computational domain. We assign each
SPH particle a mass $m = n_{\rm H} m_{\rm H} L_{\rm box}^3/ N_{\rm
SPH}$, where $N_{\rm SPH}$ is the total number of SPH particles. The
positions of the SPH particles are chosen to be glass-like (e.g.,
\citealp{White:1996}). Glass-like initial conditions imply a more
regular distribution of particles in space when compared to that
obtained from a Monte Carlo sampling of the density field. The SPH
smoothing kernel is computed and the SPH densities are found using the
SPH formalism implemented in \gadget, with $N_{\rm ngb} = 48$.
\par

Photons are transported using a single frequency bin assuming the grey approximation
in the optically thin limit. We therefore employ 
a photoionisation cross-section $\langle \sigma_{\gamma \rm HI}\rangle = 1.63 \times 10^{-18}
\cms$ (Sec.~\ref{Sec:Ionisation}) and assume that each photoionisation adds
$\langle \epsilon_{\rm HI} \rangle = 6.32 \eV$ to the thermal energy
of the gas (Sec.~\ref{Sec:Heating}). The RT time step is set to $\Delta t_{\rm r} = 10^{-2}
\Myr$ to facilitate a comparison to Test 1 in Paper~I. For the same
reason, we limit ourselves to solving the time-independent RT equation and propagate photons during each time step only
from a given particle to its direct neighbours. All
simulations presented in this section employ $N_{\rm SPH} = 64^3$ SPH
particles, which are evolved for a total of $500 \Myr$.
Some of our simulations employ the resampling technique introduced in
Paper~I to reduce artifacts due to the particular setup of the initial
conditions. Briefly, each SPH particle is, within its spatial
resolution element whose size is determined by the diameter of the SPH
kernel, $2h$, from time to time\footnote{The particle distribution is resampled every 10th RT
time step. Our results are insensitive to the precise frequency 
with which the resampling is applied. We note that the choice for the resampling frequency is problem-dependent and 
hence the resampling frequency must usually be determined experimentally using convergence tests.
In simulations with many sources, in which SPH particles receive photons from many different directions, 
artefacts due to the particular arrangement of SPH particles are typically 
much less prominent, as discussed in Paper~I (Sec.~5.3.3; see also, e.g., the related discussion on cell randomization
in \citealp{Trac:2007}). Hence, realistic simulations will typically not require resampling.} offset randomly from its initial position. For comparison,
we repeat all simulations without employing this technique. We perform
simulations with different angular resolutions.
Figs.~\ref{Fig:Test2:Slices} and \ref{Fig:Test2:Profiles} show our
results.
\par

In Fig.~\ref{Fig:Test2:Slices} we present slices through the centre of
the simulation box showing the neutral fraction (top row) and
temperature (bottom row) at time\footnote{The reason why we do not
show the slices at the end of the simulations, i.e.\ at time $t = 500
\Myr$, as we did in the corresponding Test 1 in Paper~I, is that the
simulation box is slightly too small to contain the whole ionised
sphere at this time (because of the smaller photoionisation
cross-section that is employed here).} $t = 100 \Myr$. In each row,
the three left-most panels show results from simulations with angular
resolution $N_{\rm c} = 8,32$ and $128$ and no resampling of the
particle positions applied and the right-most panel shows results from
a simulation with angular resolution $N_{\rm c} = 32$ and resampling
of the particle positions every 10th RT time step. In each panel we
indicate, as a point of reference, the analytical approximation for
the position of the ionisation front (Eq.~\ref{Eq:Ifront}) by a
dash-dotted circle.
\par
Interior to the ionisation front the gas is highly ionised and
photo-heated to typical temperatures $T\approx 1.5 \times 10^4 \K$
(with maximum temperatures $T \approx 2 \times 10^4 \K$). The runs
that did not employ the resampling show slight deviations from the
expected spherical shape which depend on the angular resolution. As
discussed in Paper~I, the deviations are caused by the particular
arrangement of the SPH particles. Reducing this particle noise, which
is strongest when $N_{\rm c} \approx \tilde{N}_{\rm ngb}$, was the
motivation for introducing the resampling technique. Indeed, the
distribution of neutral fractions and temperatures from the simulation
that employed the resampling of the density field is spherically
symmetric to a high degree.
\par
In Fig.~\ref{Fig:Test2:Profiles} we compare the median profiles of the
neutral fraction (left-hand panel) and the temperature (right-hand panel) at time $t=100 \Myr$ obtained
from the three-dimensional simulations with \traphic\ (solid curves
with error bars) to the reference simulation obtained with our 1-d RT
code \testtraphic\ (dashed curves). The results of all simulations are
in excellent agreement with the reference result. The small deviations
that are present very close to the ionising source and in regions
where the profile gradients are steep are due to the finite spatial
resolution (indicated with horizontal error bars in the top left corners of the panels). The effect of
resampling in reducing noise can most clearly be seen when comparing
the simulations with angular resolution $N_{\rm c} = \tilde{N}_{\rm
ngb} = 32$ with each other (middle panels). Note that for
the simulation with the highest angular resolution that we have
considered here ($N_{\rm c} = 128$), the resampling slightly reduces
the agreement with the reference simulation because it introduces
additional scatter. This scatter is consistent with the
spatial resolution employed.
\par

\begin{figure}
\begin{center}
  \includegraphics[trim = 40mm 10mm 40mm 10mm, width=0.23\textwidth]{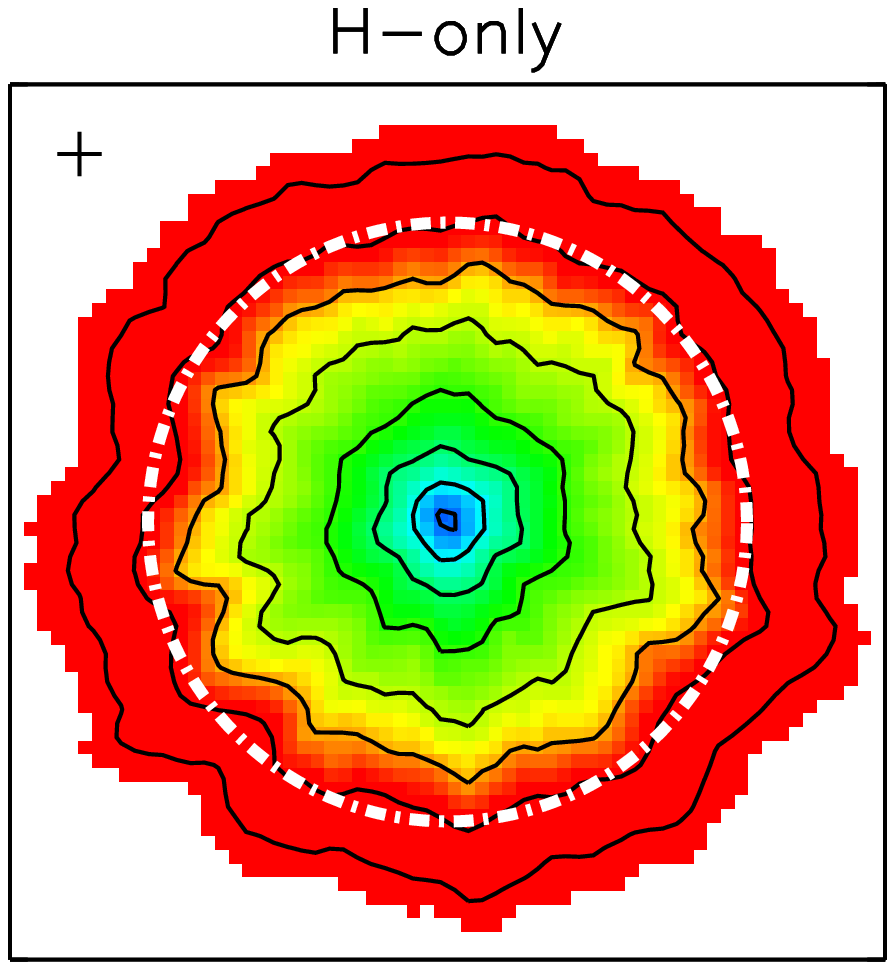} 
  \includegraphics[trim = 40mm 10mm 40mm 10mm,width=0.23\textwidth]{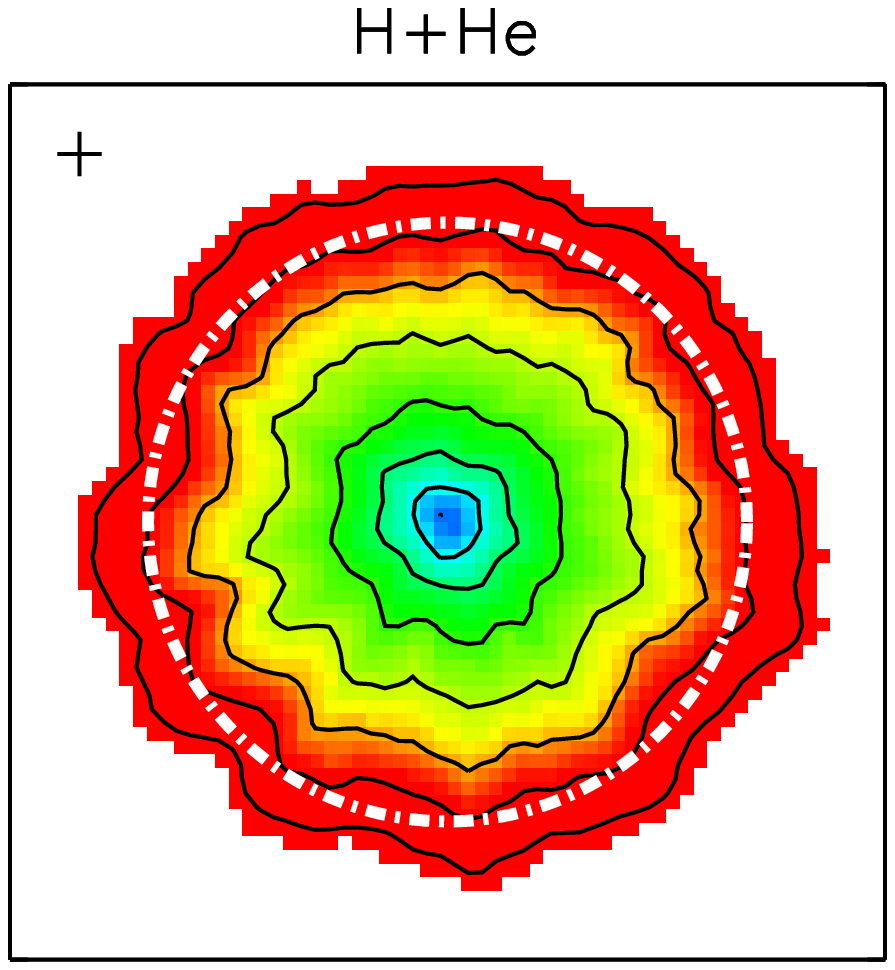} \\
  \includegraphics[trim = 40mm 10mm 40mm 10mm,width=0.23\textwidth]{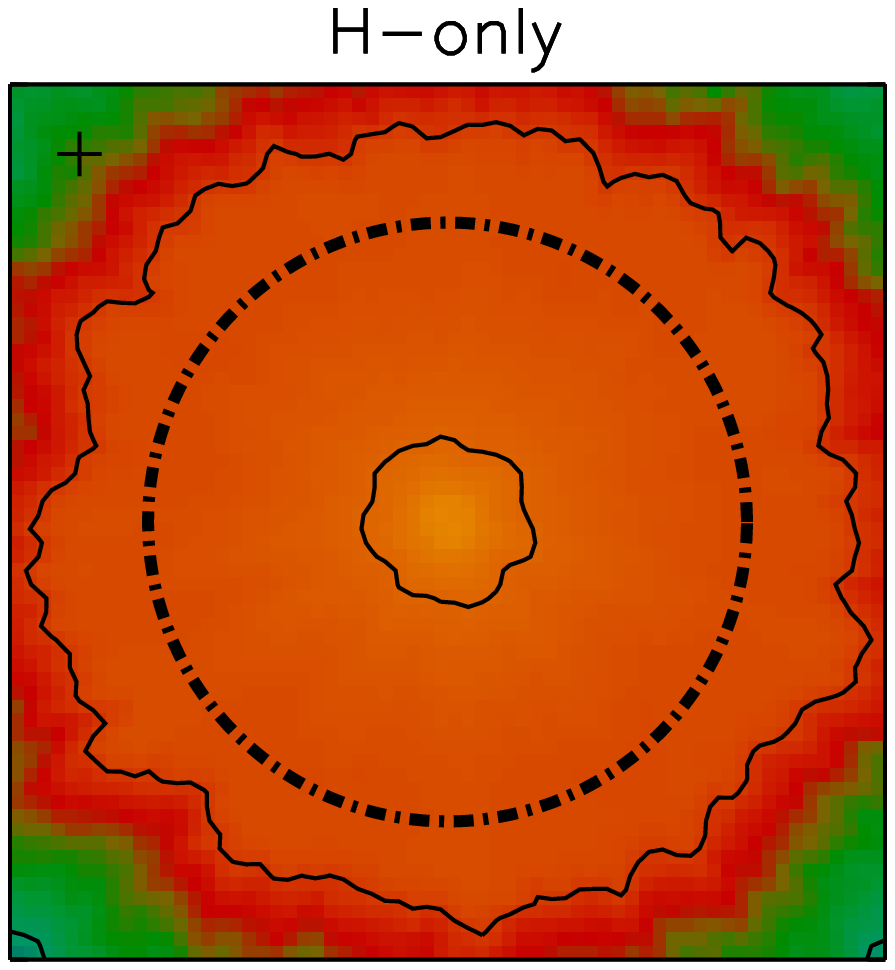}
  \includegraphics[trim = 40mm 10mm 40mm 10mm,width=0.23\textwidth]{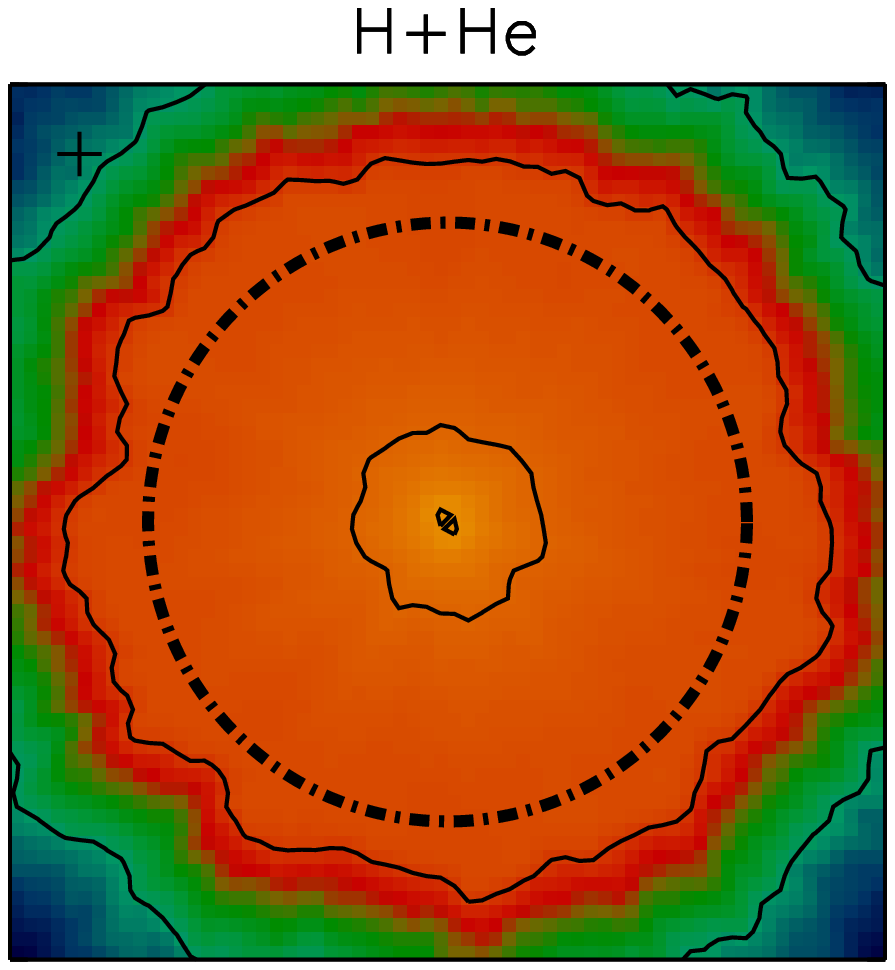} 
  \caption{Test 2 (using \traphic). Multi-frequency transport and
    effects of the inclusion of helium
    (Sec.~\ref{Sec:Tests:Test2:TraphicMultifreq}). Neutral hydrogen
    fraction $\eta_{\rm HI}$ ({\it top panels}) and temperature ({\it
    bottom panels}) at time $t = 100 \Myr$ in a slice through the
    centre of the simulation box. The {\it left} and {\it
    right} panels show, respectively, results from simulations of
    an ionising source in gas of pure hydrogen ($X=1$) and gas with
    primordial abundances ($X = 0.75$, $Y = 1-X$).  Except for this
    difference in the helium abundance, both simulations are
    identical; in particular, they both make use of five frequency
    bins to transport the emitted $10^5 \K$ blackbody photons.  The
    panels can be compared with the left-most panels of
    Fig.~\ref{Fig:Test2:Slices}, which show results from a simulation
    that is identical except that it used a single frequency bin and
    the grey optically thin approximation. As in
    Fig.~\ref{Fig:Test2:Slices}, the dot-dashed circle indicates the
    position of the ionisation front, calculated using the analytical
    approximation (Eq.~\ref{Eq:Ifront}). Contours show neutral
    fractions of $\eta_{\rm HI} = 0.9, 0.5$ , $\log_{10} \eta_{\rm HI}
    = -1, -1.5, -2, -2.5, -3, -3.5$ and temperatures $\log_{10} (T /
    \rm{K}) = (3, 4, 4.2, 4.4)$ (from the outside in). The colour
    coding is the same as in Fig.~\ref{Fig:Test2:Slices}. The crosses
    indicate the spatial resolution $\langle 2\tilde{h} \rangle$.}
  \label{Fig:Test2:Helium:Slices}
\end{center}
\end{figure}

\subsubsection{HII region expansion: multi-frequency, hydrogen and helium}

Next we demonstrate the ability of \traphic\ to accurately solve the
present multi-frequency problem in gas of primordial composition
(i.e., in the presence of helium) and using multiple frequency
bins. For brevity, we discuss only a
single simulation with particle number $N = 64^3$ and angular
resolution $N_{\rm c} = 8$. We have verified that
simulations with other choices for these parameters show the expected
behaviour. 
\par
We perform two simulations. The first simulation assumes a hydrogen
mass fraction $X=1$. The second simulation assumes a hydrogen
mass fraction $X = 0.75$ and a helium mass fraction $Y = 1-X$. We set
the initial ionised helium fractions to zero, $\eta_{\rm HeII} =
\eta_{\rm HeIII} = 0$. All other physical parameters are as in the
previous section. For both simulations we use the same number of
frequency bins, $N_{\nu} = 5$ (starting at $13.6 \eV$, $24.6 \eV$,
$35.5 \eV$, $54.4 \eV$ and $75.0 \eV$, with the last bin extending to
infinity).  The photoionisation cross-section and excess energy
associated with each bin are obtained from averaging over a blackbody
spectrum of temperature $10^5 \K$, assuming the optically thin limit 
(Eqs.~\ref{Eq:Crosssection} and \ref{Eq:AverageExcessEnergy}).
\par
The motivation behind our choice to use a small number of frequency
bins is that in realistic simulations that will be computationally
more expensive, limited resources will require the usage of as few
frequency bins as possible. Our results below show that a number as
low as five (and perhaps even as low as three, see
Sec.~\ref{Sec:Tests:Test3}) may be sufficient to capture the main
effects associated with multi-frequency radiation
transport.\footnote{While this statement is certainly true for the
present test problem, we caution that the answer to the question of
how many frequency bins are sufficient will be
problem-dependent. Hence, the minimum number of frequency bins that
can be employed while still capturing the main physical effects must
be determined by performing explicit convergence studies for the particular
problem at hand (see also, e.g., \citealp{McQuinn:2009}).}
\par
Fig.~\ref{Fig:Test2:Helium:Slices} shows the neutral hydrogen
fractions (top) and temperatures (bottom) in a slice through the
centre of the simulation box for the simulation without (left) and with
(right) helium. The panels can be compared with the left-most
panels of Fig.~\ref{Fig:Test2:Slices}, which show results from an
identical simulation except that it used a single frequency bin and
the grey, optically thin approximation. The effect of spectral
hardening is most visible in the panels showing the temperature, with
the multi-frequency simulations showing a substantial pre-heating
ahead of the ionisation front. Interestingly, the simulation that
includes helium shows slightly less pre-heating (and pre-ionisation)
than the one that assumes pure hydrogen.
\par
\begin{figure*}
\begin{center}
  \includegraphics[trim = 20mm 0mm 40mm 0mm,width=0.44\textwidth]{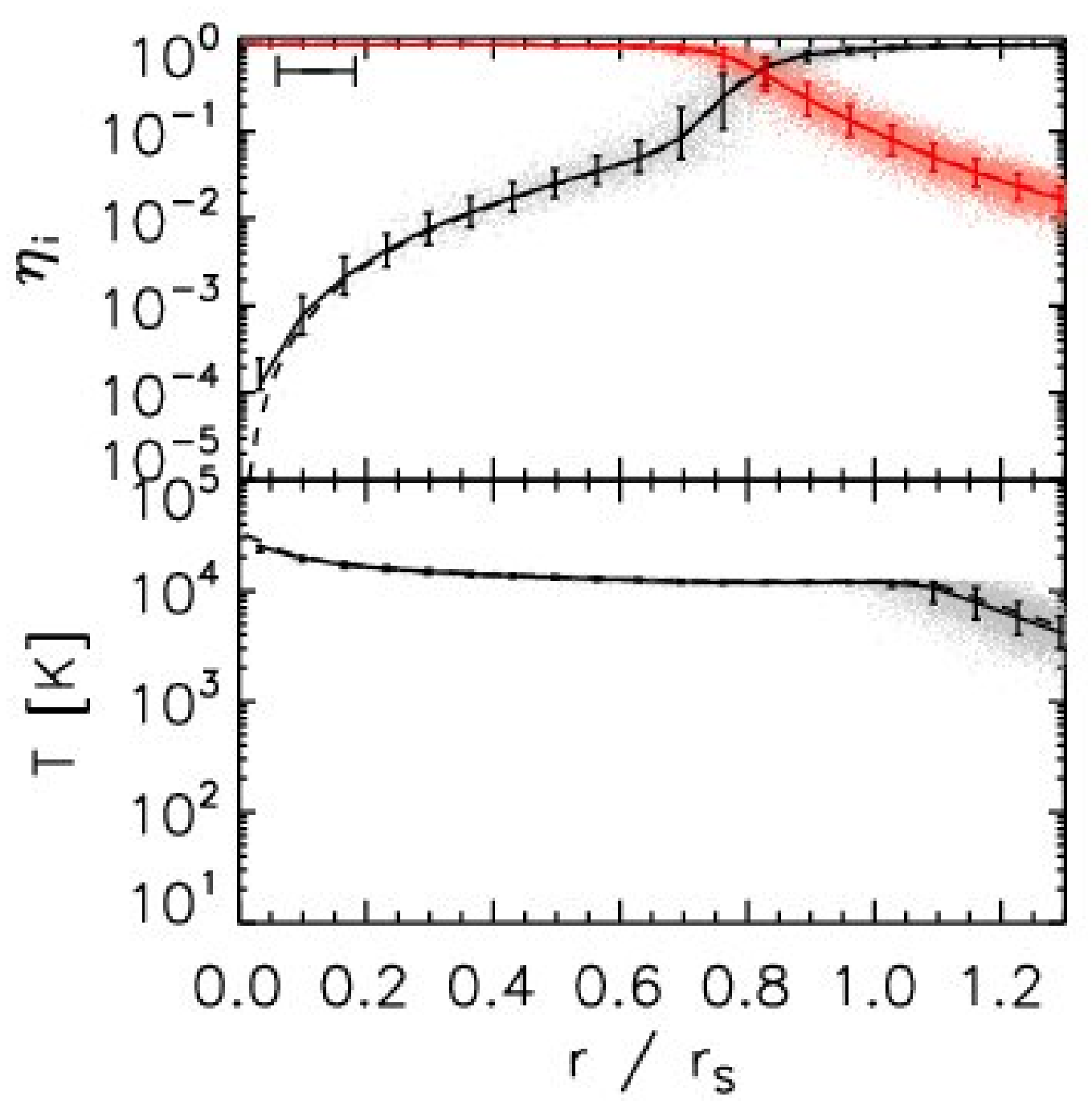}
  \includegraphics[trim = 10mm 0mm 50mm 0mm,width=0.44\textwidth]{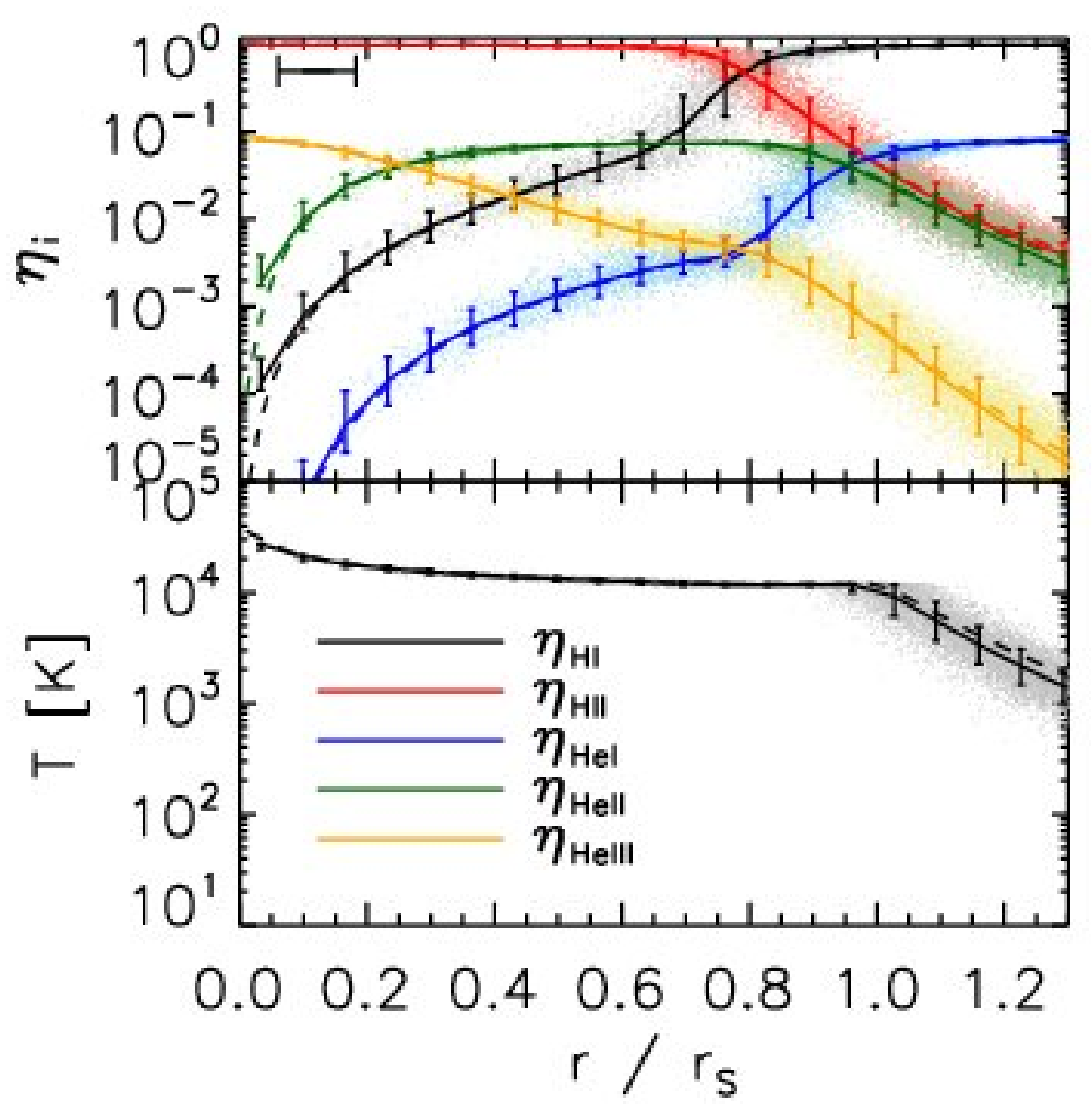}
  \caption{Test 2 (using \traphic). Multi-frequency transport and
    effects of the inclusion of helium
    (Sec~\ref{Sec:Tests:Test2:TraphicMultifreq}). Scatter plots and
    profiles of species fractions and temperature at time $ t = 100
    \Myr$. The {\it left} and {\it right} panels show, respectively,
    results from simulations of an ionising source in gas of pure
    hydrogen ($X=1$) and gas with primordial abundances ($X = 0.75$,
    $Y = 1-X$).  Except for this difference in the helium abundance,
    both simulations are identical; in particular, they both make use
    of five frequency bins to transport the emitted $10^5 \K$
    blackbody photons.  The panels could be compared with the
    left-most panels of Fig.~\ref{Fig:Test2:Profiles}, which show
    results from a simulation that is identical except that it used a
    single frequency bin and the grey, optically thin
    approximation. As in Fig.~\ref{Fig:Test2:Profiles}, each dot
    represents the species fraction (grey: $\eta_{\rm HI}$, red:
    $\eta_{\rm HII}$, blue: $\eta_{\rm HeI}$, green: $\eta_{\rm
    HeII}$, yellow: $\eta_{\rm HeIII})$ or temperature of a single SPH
    particle (only a randomly chosen subset of $10\%$ of all particles is shown). Solid curves show the median neutral fraction (ionised
    fraction, temperature) in spherical bins around the ionising
    source. The vertical error bars enclose $68.3\%$ of the particles
    in each bin. Dashed curves indicate the reference solution
    obtained with our 1-d RT code \testtraphic. The horizontal error
    bars in the upper left corners indicate the spatial resolution
    $\langle 2\tilde{h} \rangle$.}
  \label{Fig:Test2:Helium:Profiles}
\end{center}
\end{figure*}

The solid curves in Fig.~\ref{Fig:Test2:Helium:Profiles} show median
profiles of the species fractions and temperatures for both
simulations. For comparison, the (converged) reference results
obtained with \testtraphic\ are shown by dashed curves.  The results
of the simulations with \traphic\ are in excellent agreement with the
reference result, both with and without the inclusion of helium. The
small deviations that are present very close to the ionising source
and in regions where the profile gradients are steep are due to the
finite spatial resolution (indicated with horizontal error bars). The
fact that $\eta_{\rm HeII}$ shows reduced scatter is probably because
its value is not free but depends on $\eta_{\rm HeI}$ and $\eta_{\rm
HeIII}$ according to Eq.~\ref{Eq:NeutralFractions5}.
\label{Sec:Tests:Test2:TraphicMultifreq} 

\subsection{Test 3: Cosmological reionisation}
\label{Sec:Tests:Test3}

In this section we use our thermally coupled implementation of
\traphic\ to repeat Test~4 of the cosmological RT code
comparison project (\citealp{Iliev:2006a}) that we have discussed in
Paper~I for the case of fixed temperature ($T = 10^4 \K$). 
This test involves the simulation of the evolution of ionised regions
around multiple sources in a static cosmological density field at
redshift $z \approx 8.85$ and was designed to resemble
important aspects of state-of-the-art simulations of the epoch of
reionisation.  In contrast to our Test 4 simulations in Paper~I, 
we will here compute the evolution of the temperature along with that of
the ionisation state of the gas. We will perform both simulations assuming the 
grey approximations (optically thin and optically thick) and using 
multiple frequency bins and we will discuss the origin of the differences 
in the results that these simulations yield.
\par
The setup of this test is identical to that of Test 4 in Paper~I, to
which we refer the reader for a detailed description. Briefly, the
initial conditions are provided by a snapshot (at redshift $z \approx
8.85$) from a cosmological N-body and gas-dynamical uniform-mesh
simulation. The simulation box is $L_{\rm box} = 0.5\cMpch$ on a side,
uniformly divided into $N_{\rm cell} = 128^3$ cells. We Monte Carlo
sample this input density field to replace the mesh cells with $N_{\rm
SPH}=N_{\rm cell}=128^3$ SPH particles. The gas is assumed to be
initially fully neutral and to have an initial temperature of $T =
10^2\K$. The ionising sources are chosen to correspond to the 16 most
massive halos in the box. They are assumed to have blackbody spectra
$B_\nu(\nu, T_{\rm bb})$ with temperature $T_{\rm bb} = 10^5\K$. The
ionising photon production rate is taken to be constant and all
sources are switched on at the same time. The box boundaries are
photon-transmissive.
\par
We perform three RT simulations to solve the
time-indepen\-dent RT equation, all with an angular resolution of
$N_{\rm c} = 32$ (and setting $\tilde{N}_{\rm ngb}=32$). We have
demonstrated in Paper~I (Test~4) that for the current problem this
angular resolution is sufficiently high to obtain converged
results. To facilitate a direct comparison with the corresponding
simulation in Paper~I, we employ the same time step $\Delta t_{\rm r}
= 10^{-4} \Myr$ and transport photons only over a single
inter-particle distance per time step.  We note that the current
simulations do not employ the resampling technique to suppress noise
in the neutral fraction caused by the particular realisation of the
SPH density field. As discussed in Paper~I, for the present test this
noise is small.  For definiteness we mention that all simulations
include collisional ionisation and all relevant cooling processes
(including Compton cooling off the $z=8.85$ cosmic microwave
background), employing the rates listed in Table~\ref{Tab:References}.
\par
\begin{figure*}
\begin{center}
  \includegraphics[trim = 45mm 15mm 45mm 15mm, width=0.16\textwidth]{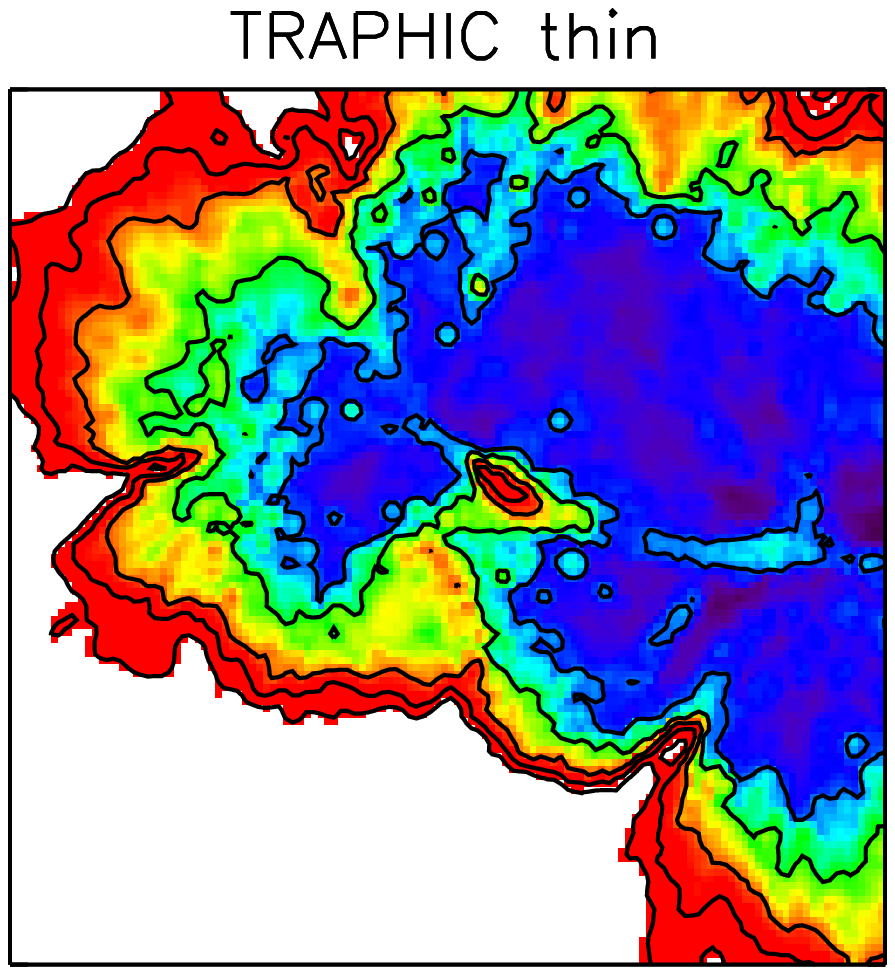} 
  \includegraphics[trim = 45mm 15mm 45mm 15mm, width=0.16\textwidth]{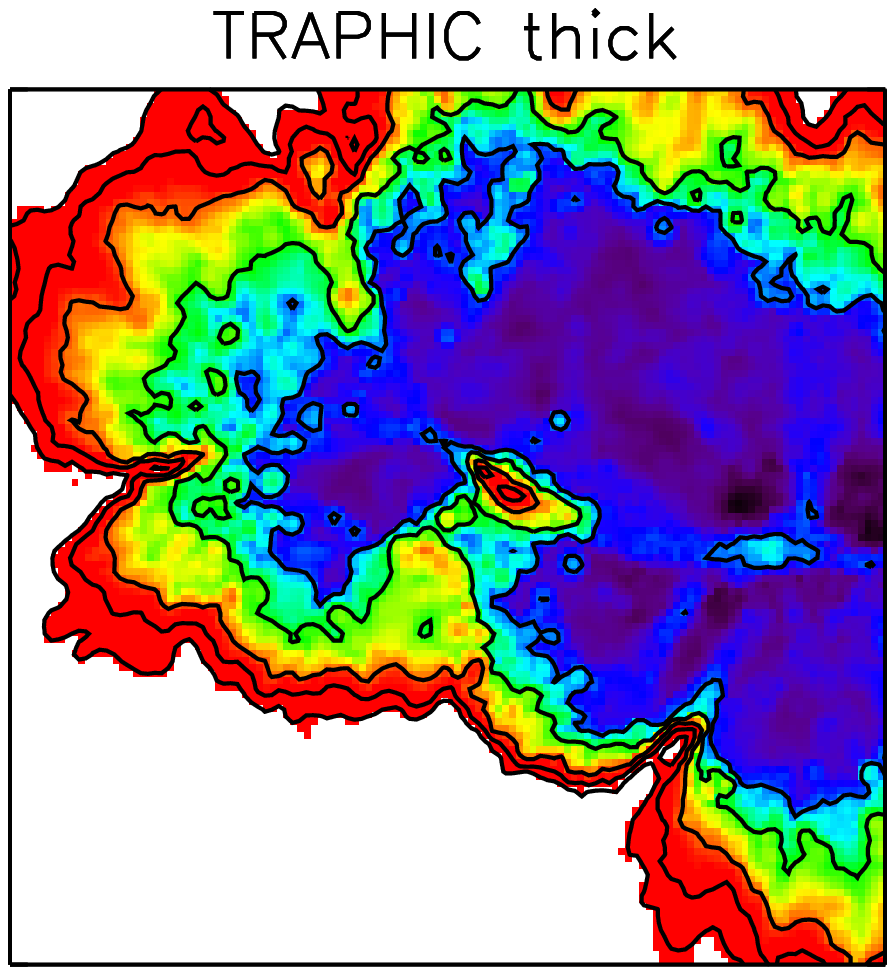} 
  \includegraphics[trim = 45mm 15mm 45mm 15mm, width=0.16\textwidth]{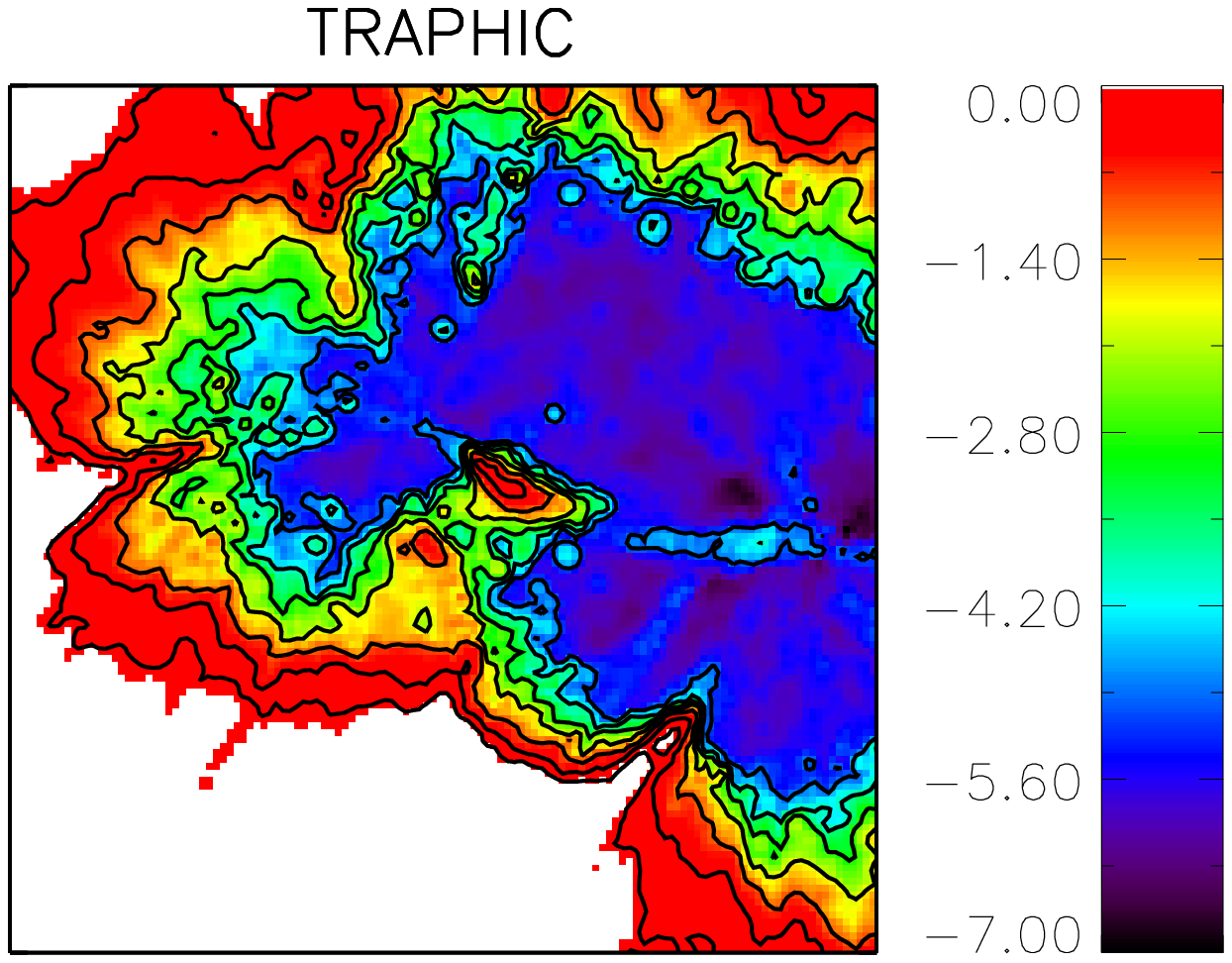} 
  \includegraphics[trim = 45mm 15mm 45mm 15mm, width=0.16\textwidth]{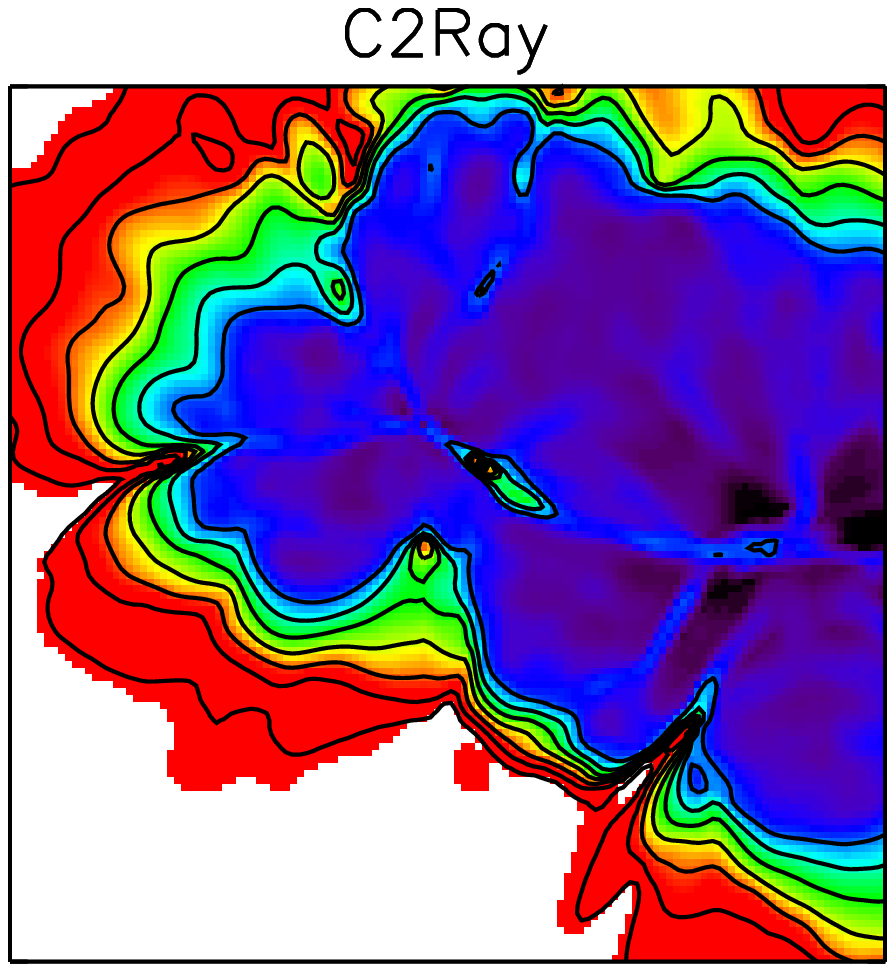} 
  \includegraphics[trim = 45mm 15mm 45mm 15mm, width=0.16\textwidth]{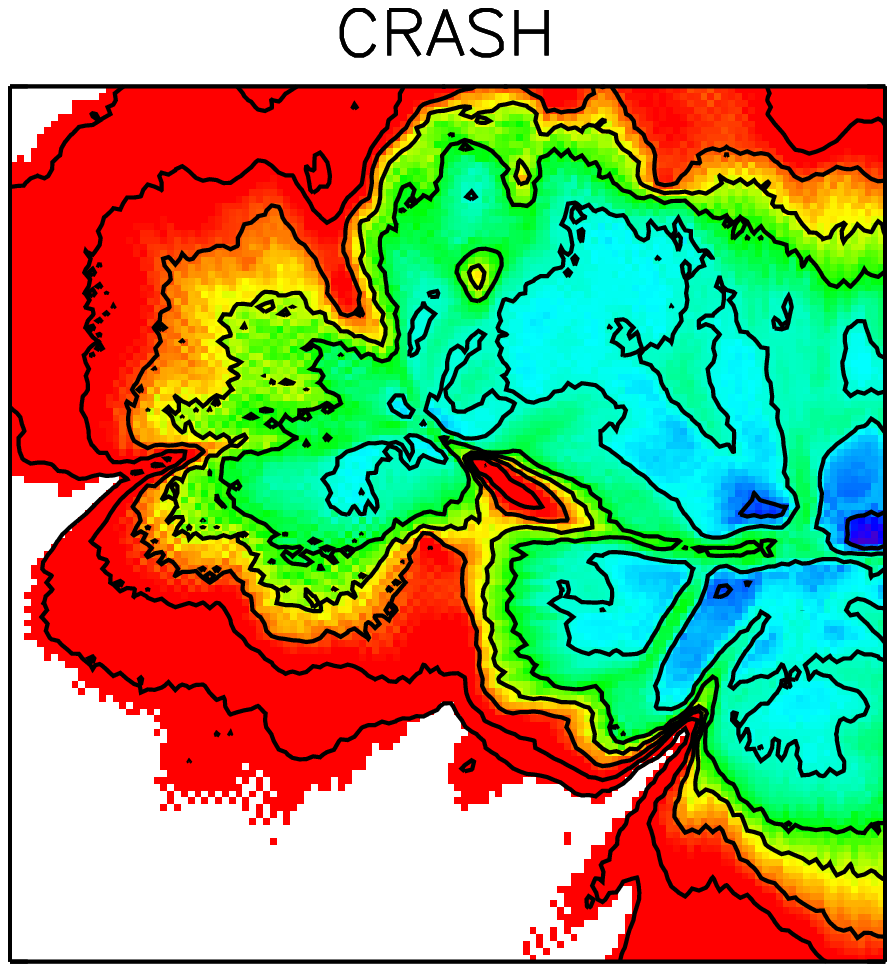} 
  \includegraphics[trim = 45mm 15mm 45mm 15mm, width=0.16\textwidth]{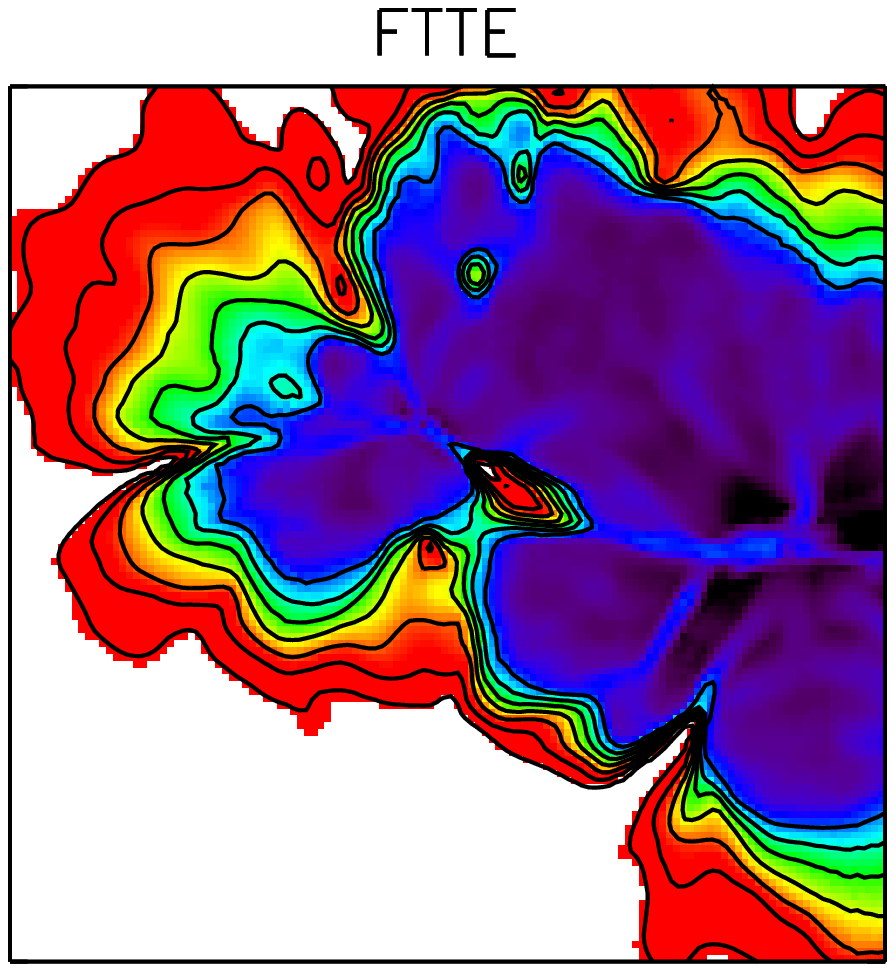}\\ 
  \includegraphics[trim = 0mm 100mm 0mm -5mm, width=0.4\textwidth]{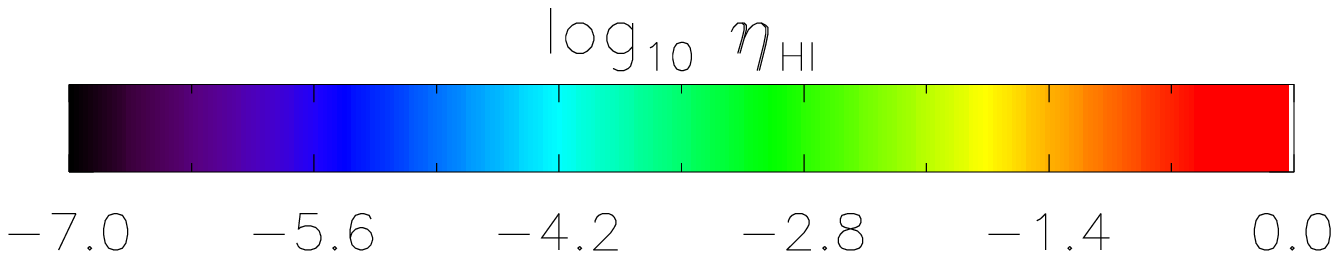} 

  \caption{Test 3. Neutral fraction in a slice through $z = L_{\rm
    box}/2$ at time $t = 0.2\Myr$. {\it From left to right:} \traphic\
    {\it thin} (assuming grey optically thin photoheating rates),
    \traphic\ {\it thick} (assuming grey optically thick photoheating
    rates), \traphic\ (using three frequency bins), \ctworay, \crash\
    and \ftte.  Contours show neutral hydrogen fractions $\eta_{\rm
    HI} = 0.9, 0.5$, $\log \eta_{\rm HI} = -1, -3$ and $-5$, from the
    outside in.  The results obtained with \traphic\ {\it thick} are
    in excellent agreement with those obtained with \ftte. They are
    also in excellent agreement with the results obtained with
    \ctworay\ in highly ionised regions, where the neutral fraction is
    unaffected by spectral hardening. The small differences in the
    neutral fractions obtained with \traphic\ {\it thick}, \traphic\
    {\it thin} and \traphic\ are mostly due to differences in the
    recombination rate, caused by differences in the gas temperatures
    (see~Fig.~\ref{Chapter:Heating:Test7:Temp}). \label{Chapter:Heating:Test7:HI} 
  }      
    
\end{center}
\end{figure*}

\begin{figure*}
\begin{center}

  \includegraphics[trim = 45mm 15mm 45mm 15mm, width=0.16\textwidth]{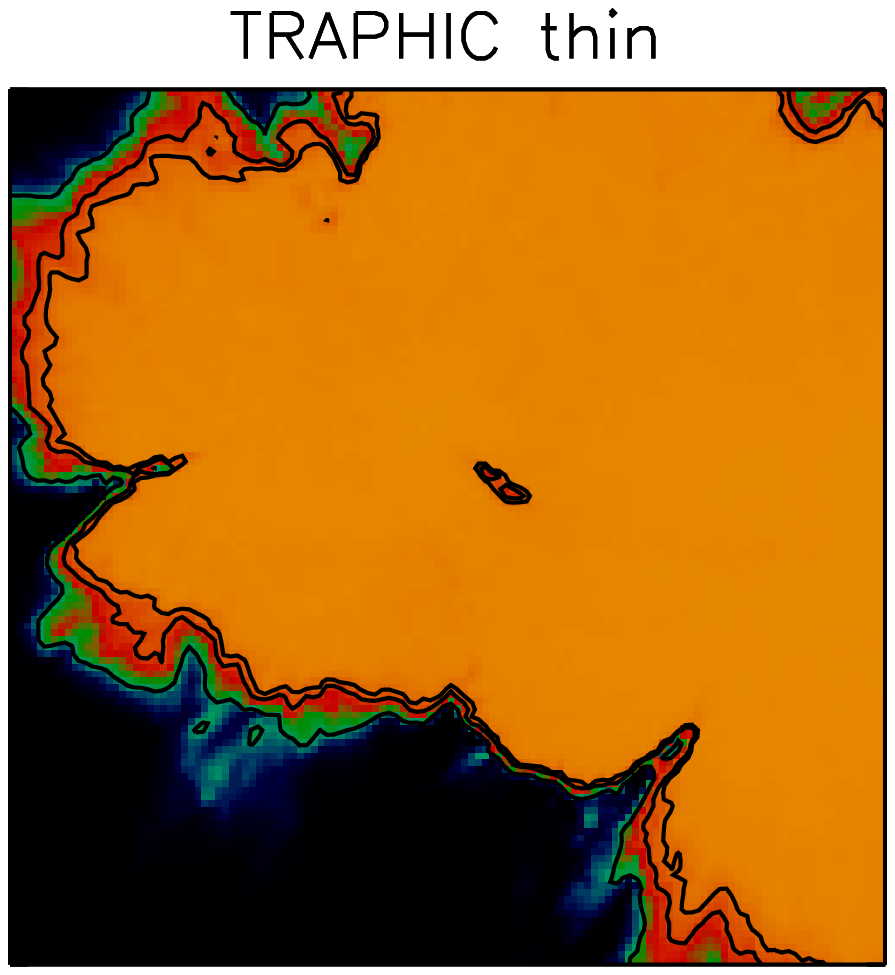} 
  \includegraphics[trim = 45mm 15mm 45mm 15mm, width=0.16\textwidth]{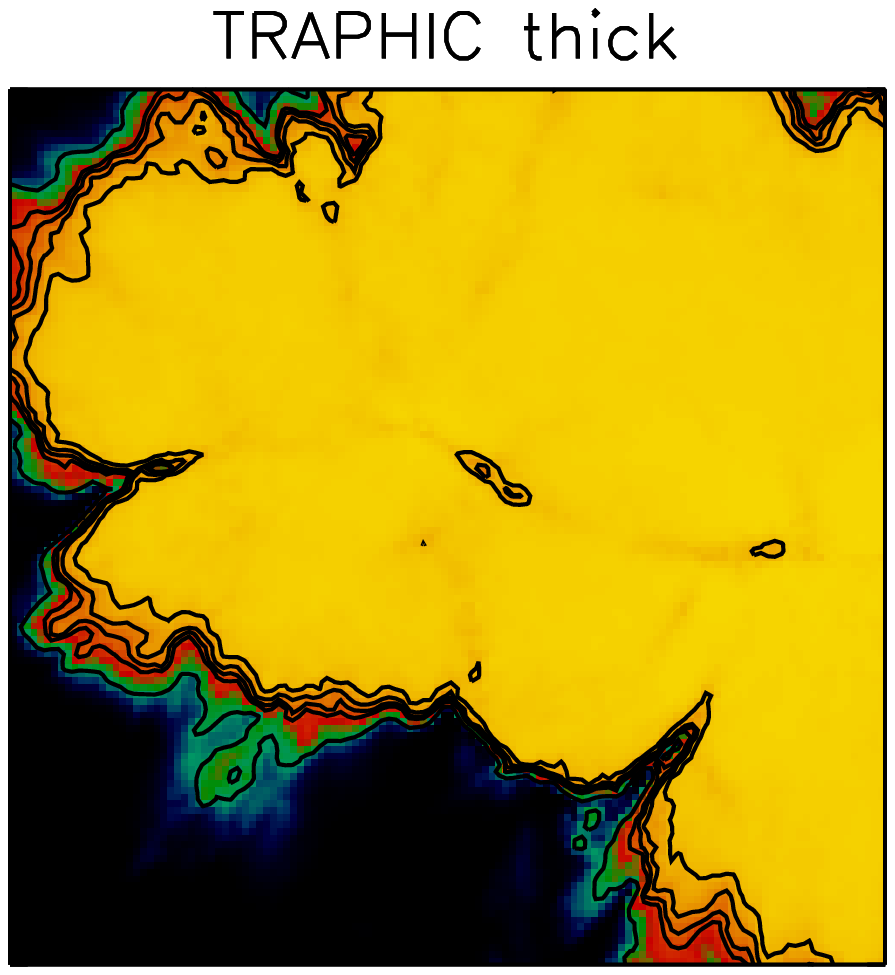} 
  \includegraphics[trim = 45mm 15mm 45mm 15mm, width=0.16\textwidth]{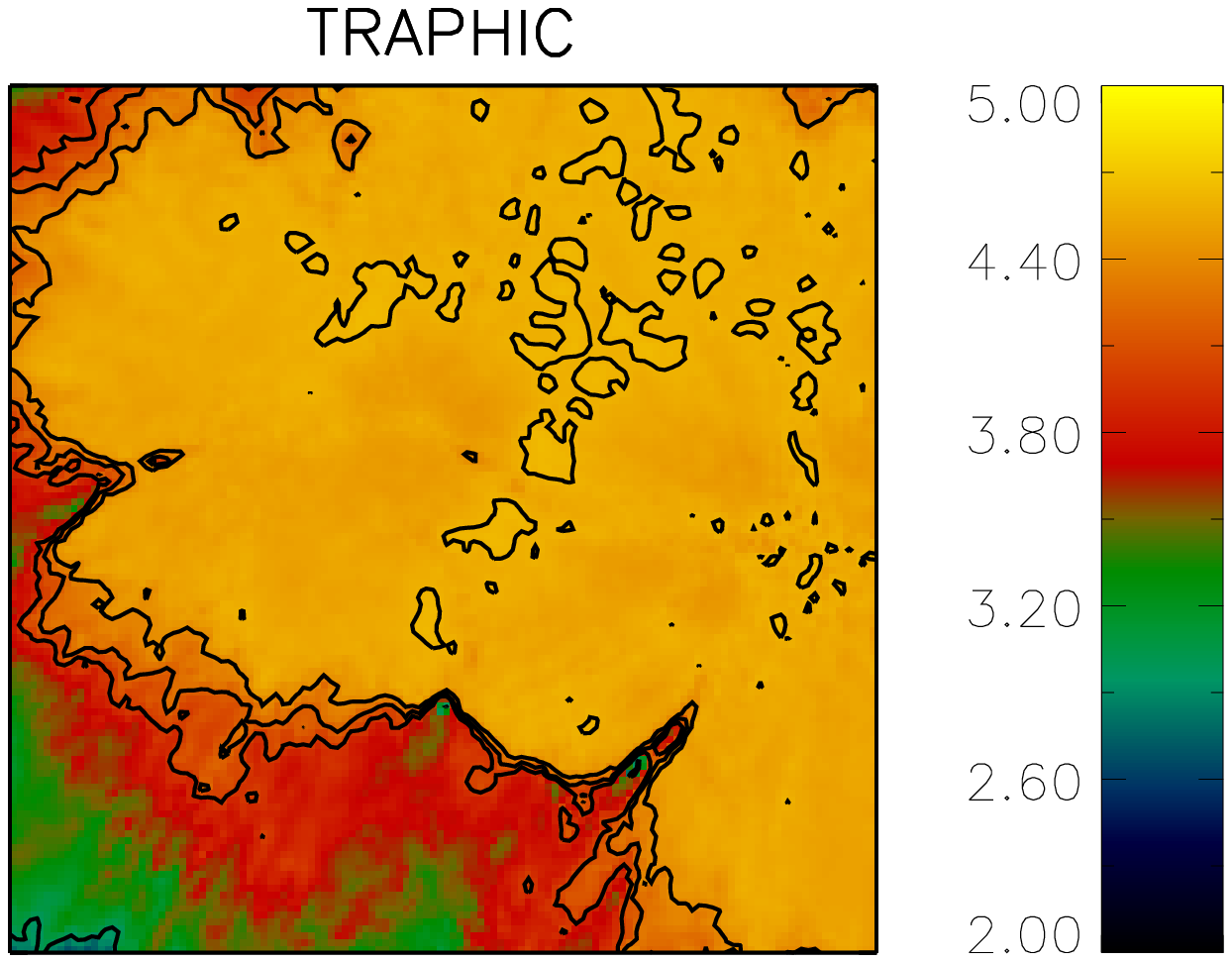} 
  \includegraphics[trim = 45mm 15mm 45mm 15mm, width=0.16\textwidth]{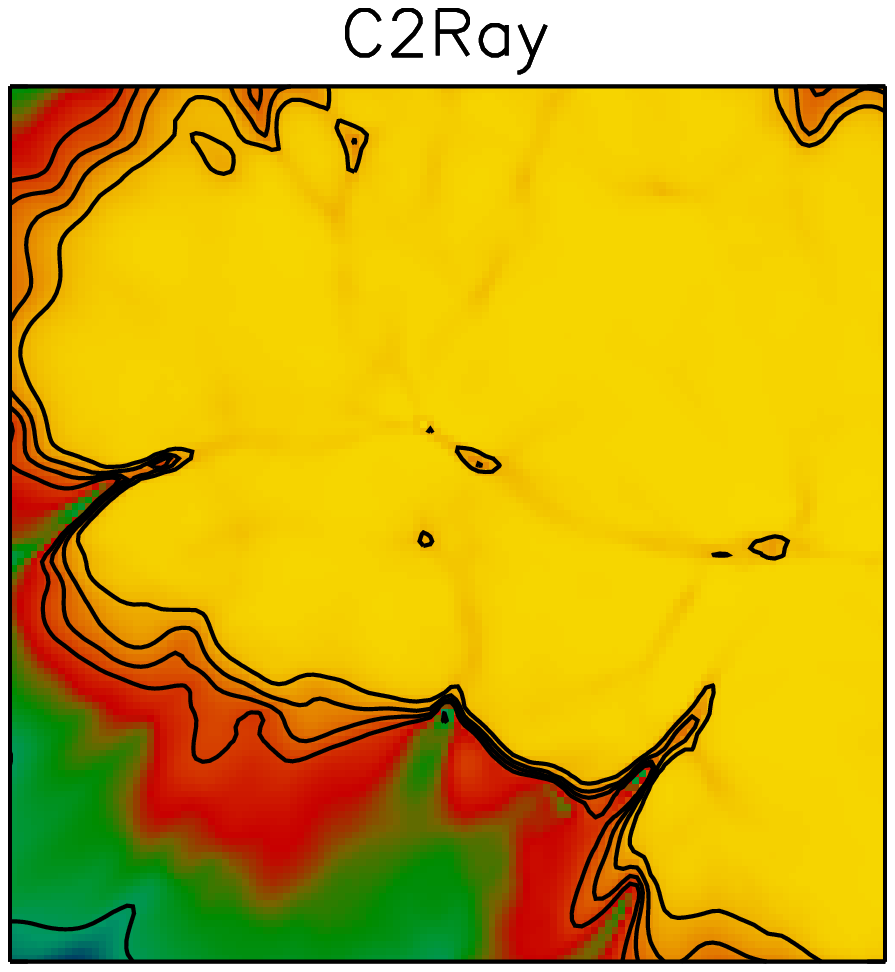} 
  \includegraphics[trim = 45mm 15mm 45mm 15mm, width=0.16\textwidth]{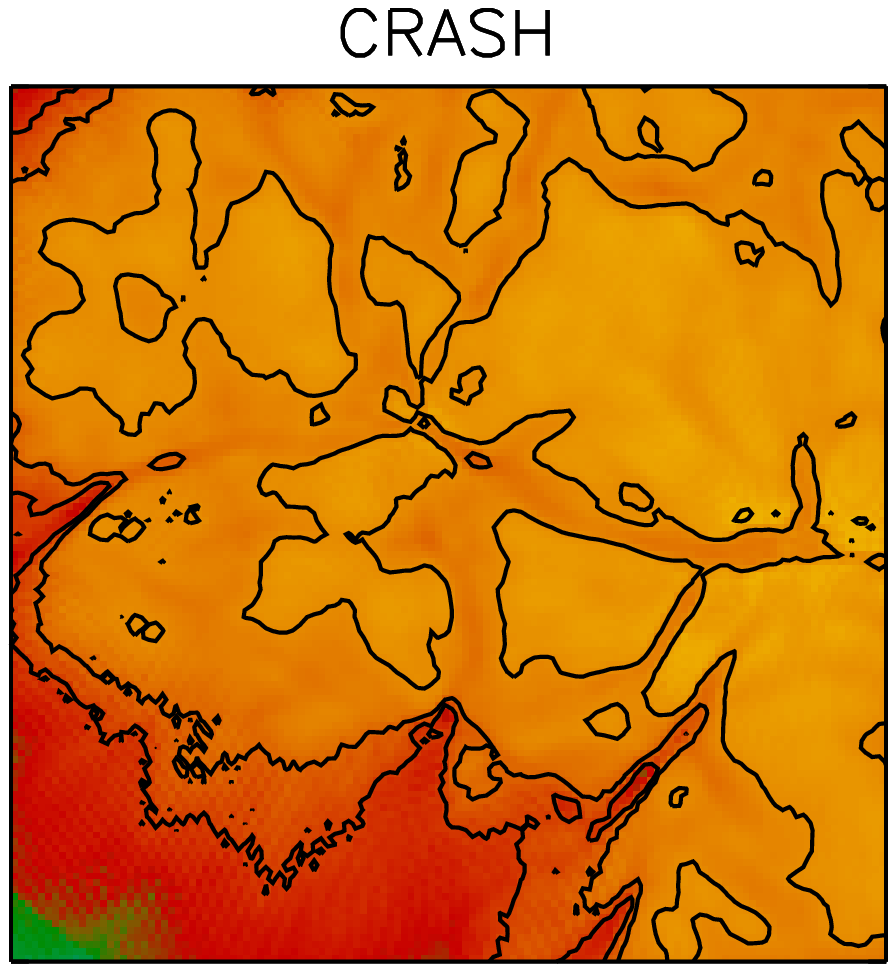} 
  \includegraphics[trim = 45mm 15mm 45mm 15mm, width=0.16\textwidth]{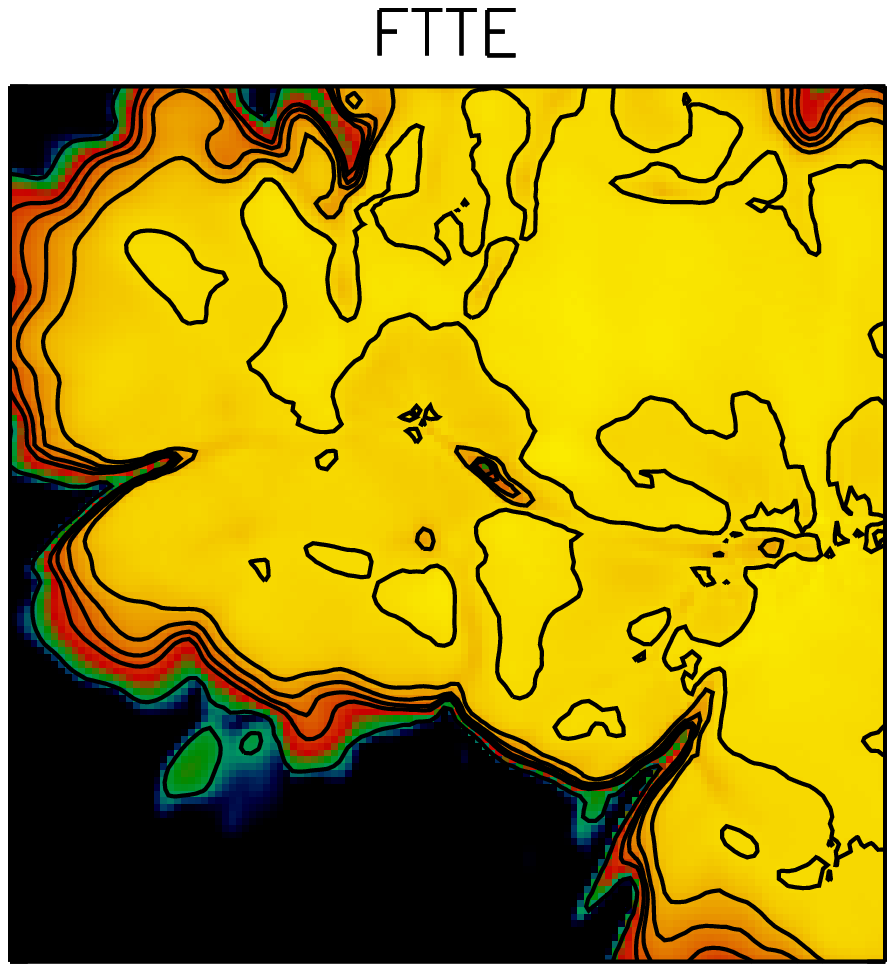}\\ 
  \includegraphics[trim = 0mm 100mm 0mm -5mm, width=0.4\textwidth]{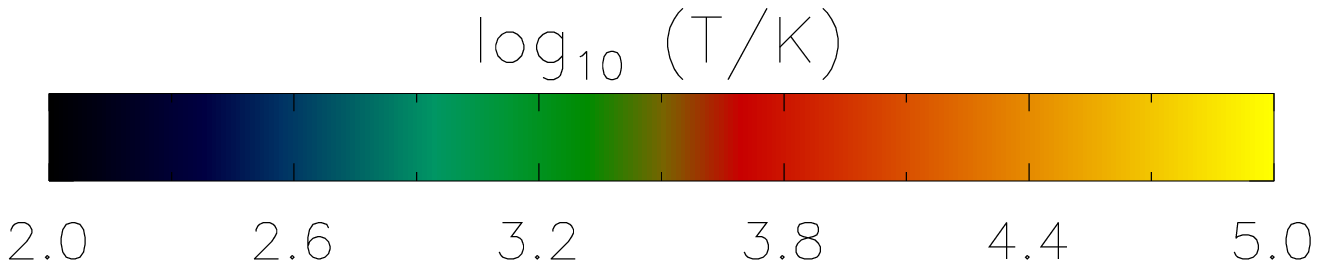} 

  \caption{Test 3. Temperature in a slice through $z = L_{\rm box}/2$
    at time $t = 0.2 \Myr$. {\it From left to right:} \traphic\ {\it
    thin} (assuming optically thin photoheating rates), \traphic\
    {\it thick} (assuming optically thick photoheating rates),
    \traphic\ (using three frequency bins), \ctworay, \crash\ and
    \ftte.  Contours show temperatures $\log_{10} (T /{\rm K})=
    3,4,4.2,4.4$ and $4.6$, from the outside in. Most of the
    morphological differences may be attributed to differences in the
    spectral hardening of the ionising radiation, with the
    multi-frequency codes \traphic, \ctworay\ and \crash\ yielding a
    substantial amount of pre-heating and the monochromatic (grey)
    codes \traphic\ {\it thin}, \traphic\ {\it thick} and \ftte\
    yielding sharp transitions between the hot ionised and the cold
    neutral phases. The differences in the maximum gas temperatures are
    mainly due to photoheating being computed in the optically thick
    limit (\traphic\ {\it thick}, \ctworay, \ftte), the optically thin
    limit (\traphic\ {\it thin}) or using multiple frequency bins
    (\traphic, \crash).
    \label{Chapter:Heating:Test7:Temp}}      
    
\end{center}
\end{figure*}

In the first two of the three simulations we transport radiation using
a single frequency bin, employing the grey photoionisation
cross-section $\langle \sigma_{\gamma \rm HI}\rangle = 1.63 \times 10^{-18}
\cms$ (Sec.~\ref{Sec:Ionisation}). The difference
between these two simulations is in the computation of the
photoheating rates used to evolve the gas temperatures. For one
simulation ({\it \traphic\ thin}) we compute photoheating in the
optically thin limit, assuming that each photoionisation adds
$\langle \epsilon_{\rm HI} \rangle = 6.32 \eV$ to the thermal energy
of the gas (Sec.~\ref{Sec:Heating}). In the other
simulation ({\it \traphic\ thick}) we compute photoheating in the
optically thick limit, assuming that each photoionisation on average
adds $\langle \epsilon^{\rm thick}_{\rm HI} \rangle = 16.01 \eV$ to
the thermal energy of the gas
(Sec.~\ref{Sec:Heating}).
\par
The third simulation (\traphic) differs from the first two in that we
transport photons using $N_{\nu}=3$ frequency bins (starting at $13.6
\eV$, $35 \eV$ and $50 \eV$, with the last bin extending to
infinity). The photoionisation cross-section and the excess energy
associated with each bin are obtained from averaging over a blackbody
spectrum of temperature $10^5 \K$, assuming the optically thin limit 
(Eqs.~\ref{Eq:Crosssection} and \ref{Eq:AverageExcessEnergy}). As in the previous section, our
choice in favour of a very small number of frequency bins has purely
practical reasons: computational efficiency. While we could have
performed this relatively small test simulation at higher spectral
resolution, we anticipate that applications of \traphic\ to large
simulations of reionisation will generally require us to choose a number of
frequency bins as small as possible. Using a small number of frequency
bins in the present test should thus give results that more closely 
resemble the results of future, larger simulations. 
\par
\begin{figure}
\begin{center}

  \includegraphics[trim = 25mm 5mm 35mm 15mm, width=0.49\textwidth]{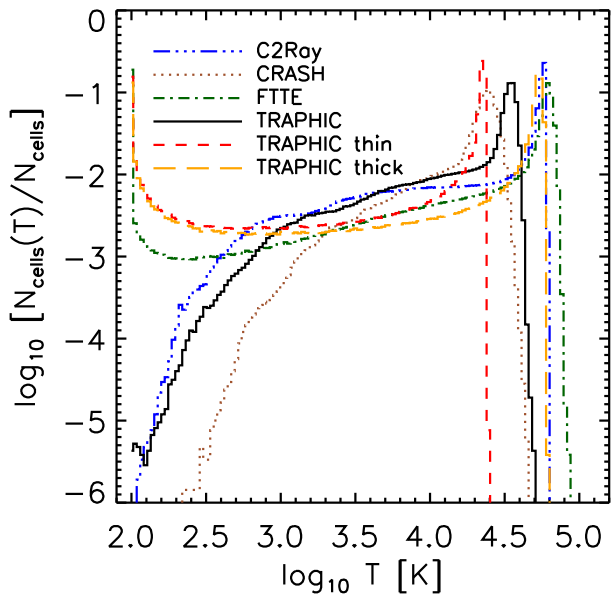} 
  \caption{Test 3. Histograms of the temperature at time $t = 0.2\Myr$
    from simulations with different RT codes.  At low
    temperatures the differences in the shapes of the histograms are
    mainly due to differences in the code-specific treatment of
    spectral hardening. The differences exhibited at high temperatures
    are mainly due to photoheating being computed in the optically
    thick limit (\traphic\ {\it thick}, \ctworay, \ftte), the
    optically thin limit (\traphic\ {\it thin}) or using multiple
    frequency bins (\crash,
    \traphic). \label{Chapter:Heating:Test7:PDF}
  }   
\end{center}
\end{figure}

Figs.~\ref{Chapter:Heating:Test7:HI}-\ref{Chapter:Heating:Test7:PDF}
show our results. Fig.~\ref{Chapter:Heating:Test7:HI} shows images of
the neutral fraction in slices through the centre of the simulation
box at time $t=0.2 \Myr$ (our conclusions also hold for other times).
The individual panels show results obtained with {\it \traphic\ thin},
{\it \traphic\ thick} and \traphic. For reference, we also show the
results obtained with other RT codes for the same test
problem as published in \cite{Iliev:2006a}. Neutral fraction contours
are shown to facilitate the comparison. While the simulation with
\crash\ treated the present problem by performing a multi-frequency
computation, the simulation with \ftte, as our simulation {\it
\traphic\ thick}, solved it in the grey approximation using optically
thick photoheating rates. Finally, \ctworay\ employed a hybrid method
that treats the transport of radiation with multiple frequency bins
but computes photoheating rates in the grey (optically thick)
approximation.
\par
The differences in the neutral fractions are generally small. The
simulations that employ photoheating rates in the optically thick
limit (\ftte, \ctworay, \traphic\ {\it thick}) yield smaller minimum
neutral fractions than the simulations that compute photoheating
rates in the optically thin limit (\traphic\ {\it thin}) or using
multiple frequency bins (\crash, \traphic).  This is the result of
lower recombination rates caused by the higher temperatures these
simulations yield\footnote{As noted earlier, the main reason why
\crash\ finds significantly larger neutral fractions may be an
insufficient sampling of the photon field, see, e.g., Fig.~2 in
\citealp{Maselli:2003}.}  (see
Fig.~\ref{Chapter:Heating:Test7:Temp}). The regions with low
ionisation ($\eta_{\rm HI}> 0.5$) found with \traphic\ are slightly
smaller than those found with \ctworay\ and \crash, which indicates
that three frequency bins are not sufficient for obtaining highly
accurate multi-frequency solution. Still, the simulations with
\traphic\ seem to capture the main effects (see the discussion on
pre-heating below).
\par
Fig.~\ref{Chapter:Heating:Test7:Temp} shows images of the gas
temperature in slices through the centre of the simulation box that
correspond to the images of the neutral fraction shown in
Fig.~\ref{Chapter:Heating:Test7:HI}. There are significant differences
in both the morphologies of the photo-heated regions and the typical
temperatures attained by the photoionised gas between the different
simulations.  Outside the ionisation fronts, these differences can
mostly be attributed to differences in the spectral hardening of the
ionising radiation. \ctworay, \crash\ and \traphic\ all yield a
substantial pre-heating of the gas ahead of the ionisation
fronts. This pre-heating is not seen in the simulations with {\it
\traphic\ thin}, {\it \traphic\ thick} and \ftte\ since they all
assume the grey approximation. In Sec.~\ref{Sec:Tests:Test2:Reference:T}
we have already discussed, for the same set of codes, the differences
between a multi-frequency treatment and its grey approximations in
idealised simulations of the evolution of a single, spherically
symmetric, ionised region. The results here are in close qualitative
agreement with that discussion.

\par
The results obtained with the different codes also exhibit significant
variations in the gas temperature in regions well inside the
ionisation fronts. While \crash, \traphic\ and \traphic\ {\it thin}
yield typical temperatures\footnote{\citealp{Maselli:2009} have
repeated this test with a more recent version of \crash\ with improved
sampling of the Monte Carlo photon field. They find slightly larger
($20-30\%$ at $t = 0.05 \Myr$) peak temperatures (their Fig.~5), which
improves the agreement with the peak temperatures found with
\traphic. \label{Footnote:Crash}} of $T\approx 2-3 \times 10^4 \K$,
the typical temperatures obtained with \ctworay, \ftte\ and \traphic\
{\it thick} are $T\approx 6\times 10^4 \K$, i.e., substantially
higher. Note that there is also disagreement between codes which
incorporate the detailed treatment of multi-frequency radiation
(\ctworay, \crash, \traphic) and between codes in which the radiation is
treated in the grey approximation (\ftte, \traphic\ {\it thin},
\traphic\ {\it thick}). Spectral hardening may thus only provide an
explanation for part of the differences between the simulations.
\par
We recall that in Sec.~\ref{Sec:Tests:Test2:Reference:T}, where we
simulated the evolution of a single, spherically symmetric,
photoionised region, we found qualitatively similar differences
between the results obtained with \ctworay, \crash\ and \ftte. In the
(nearly optically thin) region close to the ionising source, the
simulations that employed \ctworay\ and \ftte\ yielded gas
temperatures that were substantially larger than those in 
the simulation that employed \crash. By comparing with results
obtained with our 1-d RT code \testtraphic, we were able to explain
most of these temperature differences in terms of differences in the
assumptions underlying the computation of photoheating rates. The
results presented in Fig.~\ref{Chapter:Heating:Test7:Temp} are another
manifestation of this explanation.
\par
In Fig.~\ref{Chapter:Heating:Test7:PDF} we compare histograms of the
temperature at time $t= 0.2\Myr$. For the simulations with \ctworay,
\crash\ and \ftte\ we have computed these histograms directly from the
$N_{\rm cell} = 128^3$ values of temperatures published in
\cite{Iliev:2006a}. For the simulations with \traphic, \traphic\ {\it
thin} and \traphic\ {\it thick} we assigned the temperatures to a
corresponding uniform mesh with $N_{\rm cell} = 128^3$ cells using SPH
interpolation before we computed the histograms.
\par
The histograms provide a quantitative confirmation of our qualitative
discussion above. The simulations with \traphic\ and \traphic\ {\it
thin} yield, in close agreement\footnote{See
footnote~\ref{Footnote:Crash}} with the simulations performed with
\crash, typical temperatures of $T\approx 2-3\times 10^4\K$. On the
other hand, the simulations performed with \ctworay, \ftte\ and
\traphic\ {\it thick} closely agree on typical temperatures of
$T\approx 6\times 10^4\K$. The differences between the histograms at
low values of the temperature are mostly caused by the differences in
spectral hardening. Due to the pre-heating of gas ahead of the
ionisation fronts obtained with the multi-frequency codes \traphic,
\ctworay\ and \crash, the number of cells that are still at their
initial temperature $T = 100 \K$ is much smaller than found by \ftte,
\traphic\ {\it thin} and \traphic\ {\it thick}, which only employ a
single frequency bin.
\par
In summary, in this section we have repeated the simulation of the
expansion of multiple ionised regions in a cosmological density field
that we discussed in Test 4 in Paper~I, but this time we explicitly
computed the evolution of the temperature of the photoionised gas. We
performed three simulations: one simulation computed photoheating in
the grey, optically thin limit (\traphic\ {\it thin}), one computed
photoheating in the grey, optically thick limit (\traphic\ {\it
thick}) and one employed multiple (i.e., three) frequency bins. 
All three simulations showed only small differences in the neutral
fractions when compared with each other, which could be plausibly
explained with the differences in the temperatures. The 
temperatures differed due to the computation of
photoheating rates in different limits (grey optically thin, grey
optically thick and multi-frequency). In particular, employing 
the grey approximation in the optically thick limit yields 
temperatures that are typically too large by factors $2-3$. 
We also compared the results of our thermally coupled simulations with results obtained with other RT
codes for the same test problem (\citealp{Iliev:2006a}). We found very
good agreement between these and our results when comparing
simulations that employed similar assumptions for computing
photoionisation and photoheating rates. 
\par
The results of the
multi-frequency simulation (neutral fractions and temperatures
interpolated to a $128^3$ uniform mesh using SPH interpolation) are 
available for download at the website of the Cosmological RT Code
Comparison Project (\citealp{Iliev:2006a}; \citealp{Iliev:2009}).
\par

\section{Summary}
\label{Sec:Summary}
In this work we described an extension of the implementation of
\traphic, the radiative transfer (RT) method for use with Smoothed
Particle Hydrodynamics (SPH) simulations that we have introduced in
Pawlik \& Schaye (2008, Paper~I), in a modified version of the SPH
code {\sc gadget} (\citealp{Springel:2005}). The new implementation of
\traphic\ can be used to solve multi-frequency RT problems in
primordial gas consisting of hydrogen and helium. It also allows for
the computation of the non-equilibrium evolution of the gas
temperature due to photoheating and radiative cooling.
\par
As part of the new implementation we introduced a 
numerical method that allows us to accurately compute the 
coupled evolution of the ionisation balance and temperature 
of gas parcels exposed to ionising radiation and that works 
independently of the size of the chosen RT time step. This 
decoupling of the RT time step from the time scales that govern 
the evolution of the species fractions and temperatures, 
i.e. the decoupling from the ionisation, recombination and 
radiative cooling time scales, is an important pre-requisite 
for performing efficient RT simulations. The alternative, a 
RT time step limited by the values for the ionisation, recombination or cooling time scales, could quickly become 
computationally infeasible since these time scales may become very small. 
\par
We discussed the performance of the new, thermally coupled
implementation of \traphic\ in three-dimensional multi-frequency RT
test simulations of spherically symmetric expanding HII regions and 
non-spherical expanding HII regions in a cosmological reionisation setting. We 
treated the multi-frequency radiation both using a single frequency bin in the grey approximation
and using multiple frequency bins. We distinguished two types of grey
approximations by computing photoheating both in the optically thin
and optically thick limits. We compared the results of our test simulations to
results obtained with other RT codes for identical test
problems. We found excellent agreement in the morphologies
and gas temperatures of the photoionised and photoheated regions
when comparing simulations that employed similar assumptions for
computing photoionisation and photoheating rates. 
\par
We used the new implementation to demonstrate and pinpoint the
differences in the results obtained from grey simulations and
simulations that use multiple frequency bins. Close to and ahead of
ionisation fronts these differences are mostly due to the spectral
hardening of the radiation field caused by the dependence of the
absorption cross-section on photon energy. Spectral hardening
significantly increases the widths of ionisation fronts and implies a
substantial preheating of the gas ahead of
them. Additional significant differences between the simulations are
caused by the choice of the limit in which grey photoheating rates are
computed. Simulations that use grey photoheating rates computed in the
optically thick limit yield typical gas temperatures that are too
large by factors $2-3$ when compared with the exact multi-frequency solution. Simulations 
that use grey photoheating rates
computed in the optically thin limit yield typical gas temperatures
that asymptote to the multi-frequency result with decreasing distances
from the ionising sources.
\par
The additions presented in this work are crucial for applications of
\traphic\ to simulations of the (re-)ionisation of both hydrogen and
helium that also wish to account for the preheating of gas ahead of
ionisation fronts due to spectral hardening. We plan to perform such
simulations in the future. Note that we have limited our considerations
to RT simulations on pre-computed static density
fields. But photoheating increases also the gas pressure and
hence affects the hydrodynamical evolution of the gas. An important goal
for the future will therefore be to present an implementation of
\traphic\ that allows one to perform radiation-hydrodynamical simulations.  

\section*{Acknowledgments} 
We thank Benedetta Ciardi, Antonella Maselli, Garrelt Mellema, Alexei Razoumov and
Dominique Aubert for helpful discussions. We thank the referee for a 
careful reading of the manuscript and the many excellent suggestions that significantly improved the presentation 
of this work. This research was sponsored by
the National Computing Facilities Foundation (NCF) for the use of
supercomputer facilities, with financial support from the Netherlands
Organization for Scientific Research (NWO), and by an NWO VIDI grant. 
This research was furthermore supported by NSF grants AST-0708795
and AST-1009928, as well as NASA through Astrophysics Theory and
Fundamental Physics Program grants NNX08AL43G and NNX09AJ33G.

\appendix

\section{A new treatment of absorptions by virtual particles}
\label{Sec:Appendix}
In this appendix we show that the treatment of virtual particles
(ViPs) in the implementation of \traphic\ that we have used to perform
the (hydrogen-only) simulations published in Pawlik \& Schaye (2008,
hereafter Paper~I) and that we will refer to as the old
implementation, results in a temporary underestimate of the neutral
hydrogen fraction just behind evolving ionisation fronts in
simulations that use a high angular resolution. We will show that this
underestimate is absent in simulations that employ our new
implementation (presented in the current work). Moreover, in
simulations that employ this new implementation, the numerical scatter
in the neutral hydrogen fraction is significantly reduced. For clarity
of the presentation and because we will compare results obtained with
the new implementation with results obtained with the old
implementation presented in Paper~I that lacked the treatment of
helium, we will assume that photons are transported in gas that
consists only of hydrogen (i.e., $X = 1$). Our discussion generalizes
straightforwardly to the transport of photons in gas consisting of
both hydrogen and helium.
\par
The number of hydrogen-ionising photons a ViP absorbs depends on its
neutral hydrogen density. As explained in Paper~I, the computation of
this number is performed in exactly the same manner as for SPH
particles. The only difference between the treatment of photons
absorbed by SPH particles and ViPs is that the latter distribute the
absorbed photons amongst their SPH neighbours. For this distribution
of absorbed photons one must specify the fraction of the total that is
given to each of the SPH neighbours. In the old implementation of
\traphic\, this fraction was taken to be proportional to the value of
the SPH kernel $W$ of the distributing ViP at the position of the SPH
neighbour. In the new version this fraction is taken to be
proportional to the contribution of the SPH neighbour to the SPH estimate of the ViP's neutral hydrogen density. 
\par
\begin{figure*}

  \begin{center}
 
 \includegraphics[trim = 20mm 0mm 20mm 0mm, width=0.3\textwidth]{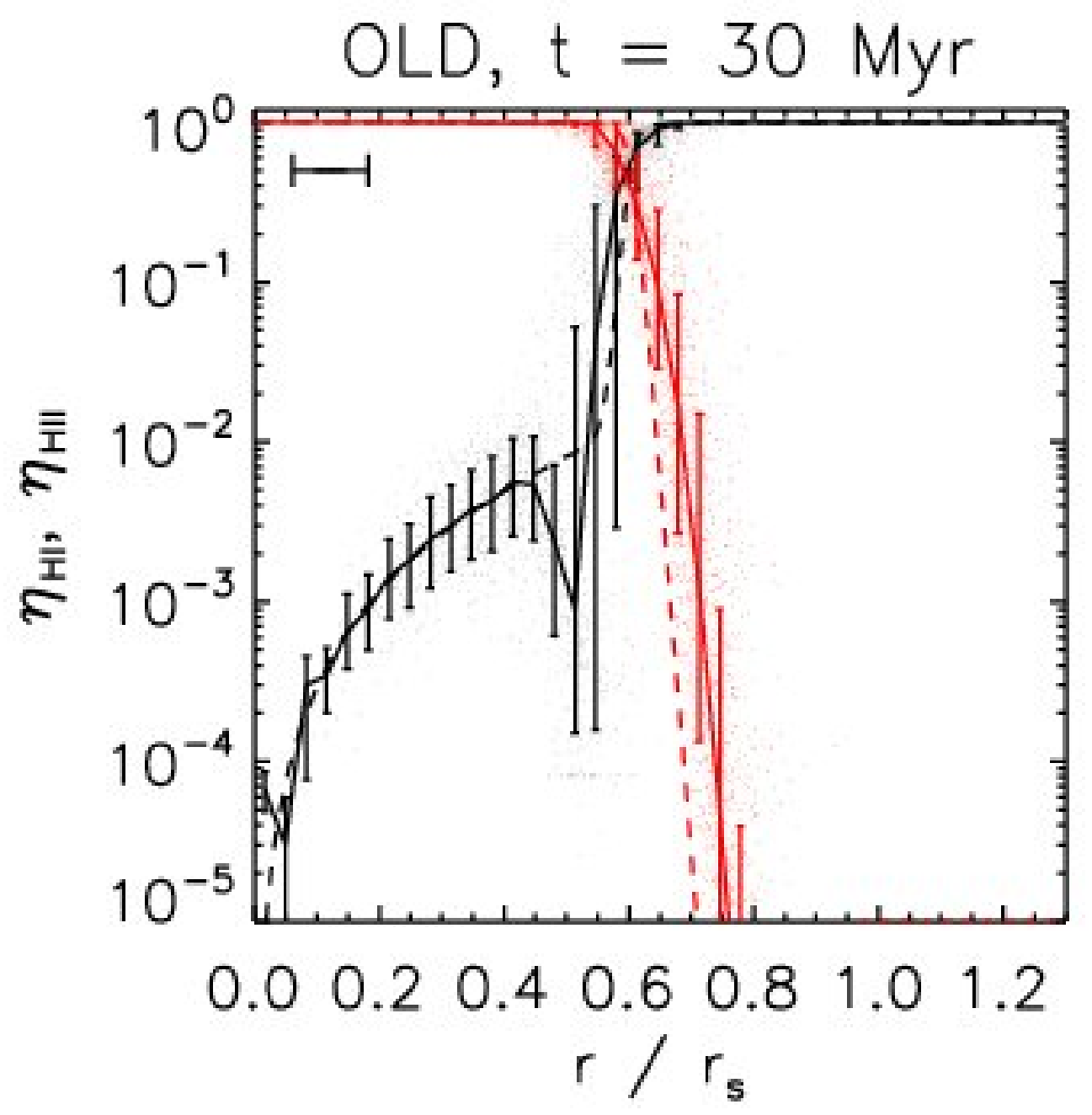}      
  \includegraphics[trim = 20mm 0mm 20mm 0mm, width=0.3\textwidth]{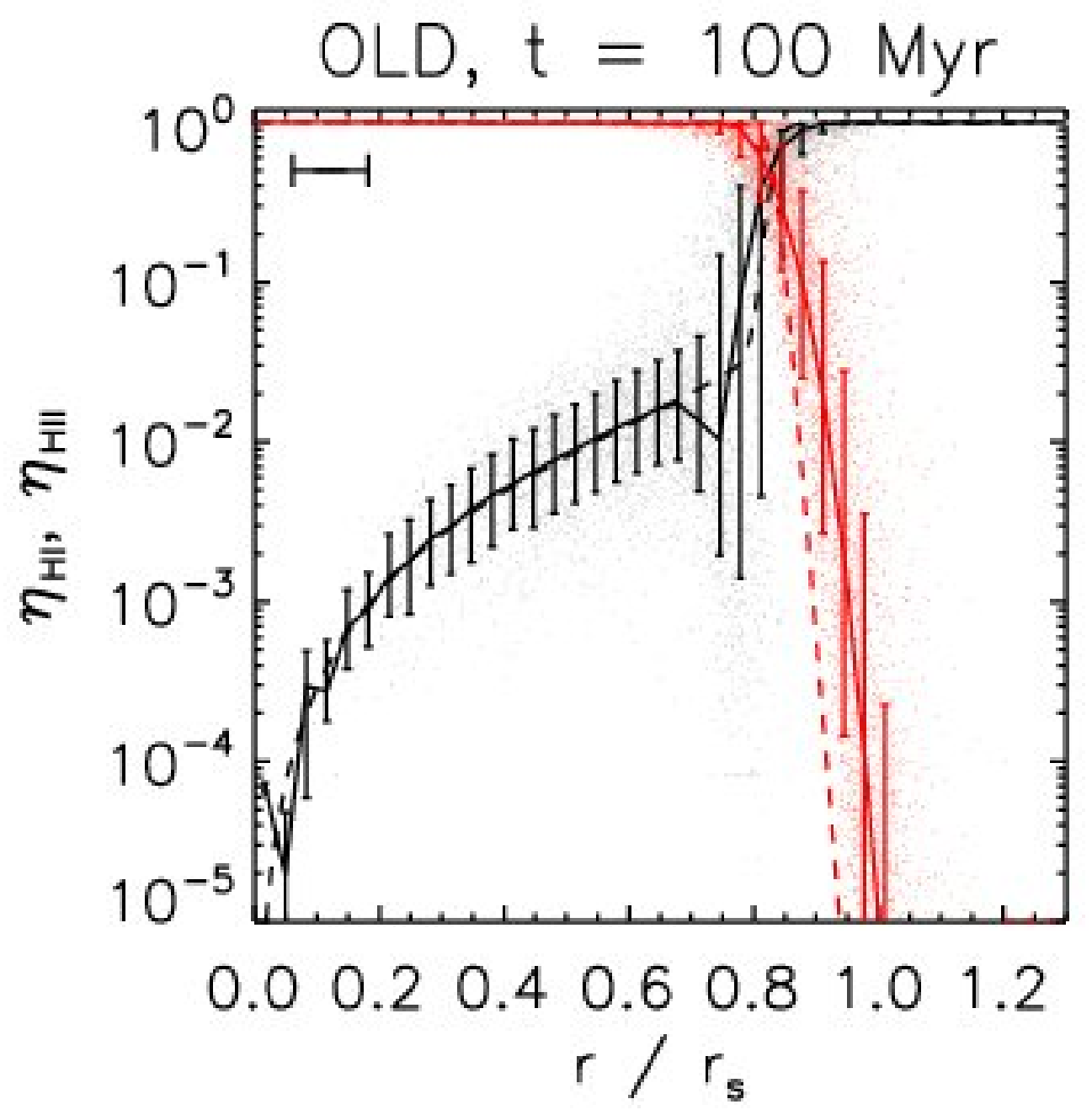}      
  \includegraphics[trim = 20mm 0mm 20mm 0mm, width=0.3\textwidth]{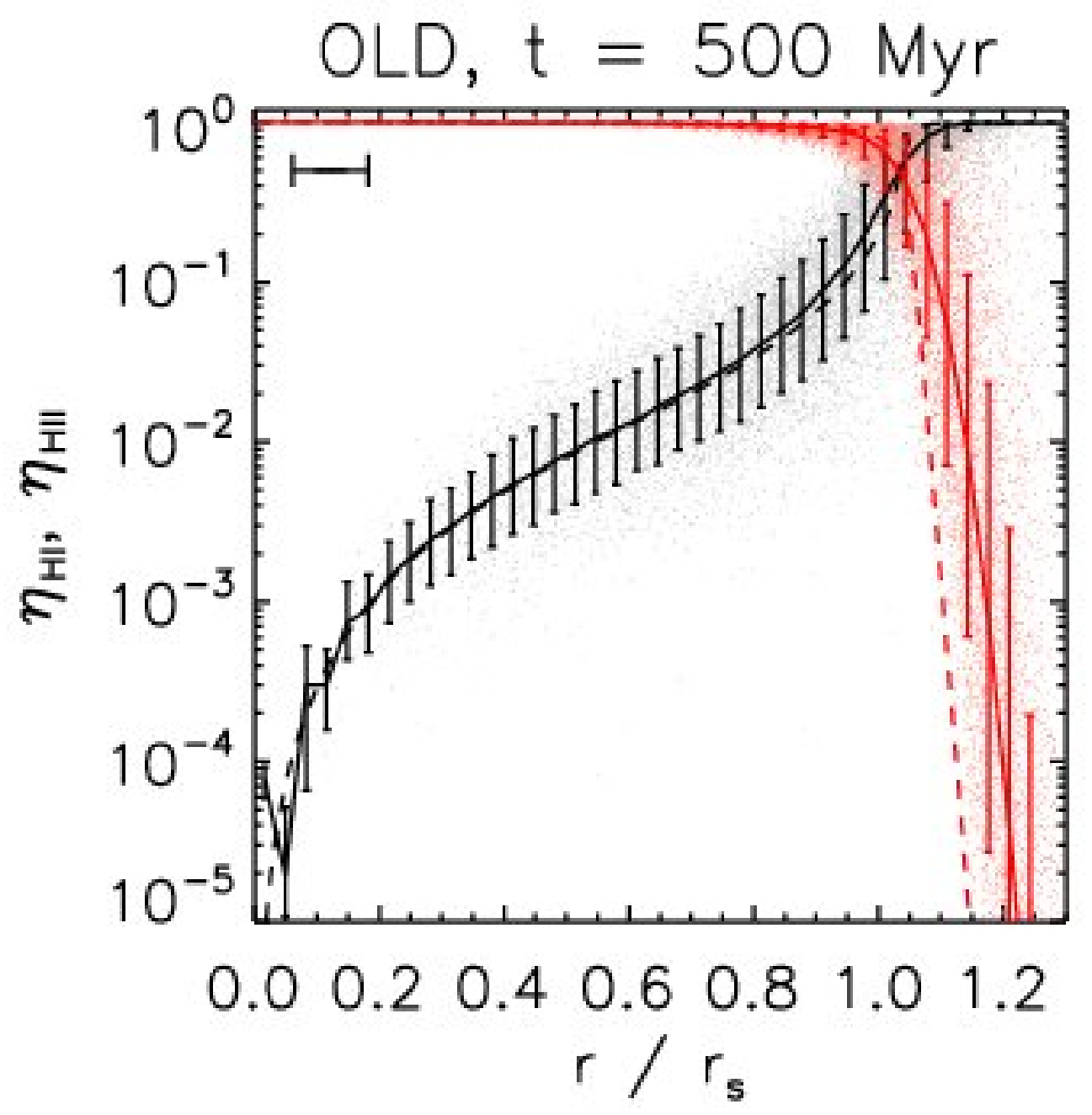}     \\ 
  \includegraphics[trim = 20mm 0mm 20mm 0mm, width=0.3\textwidth]{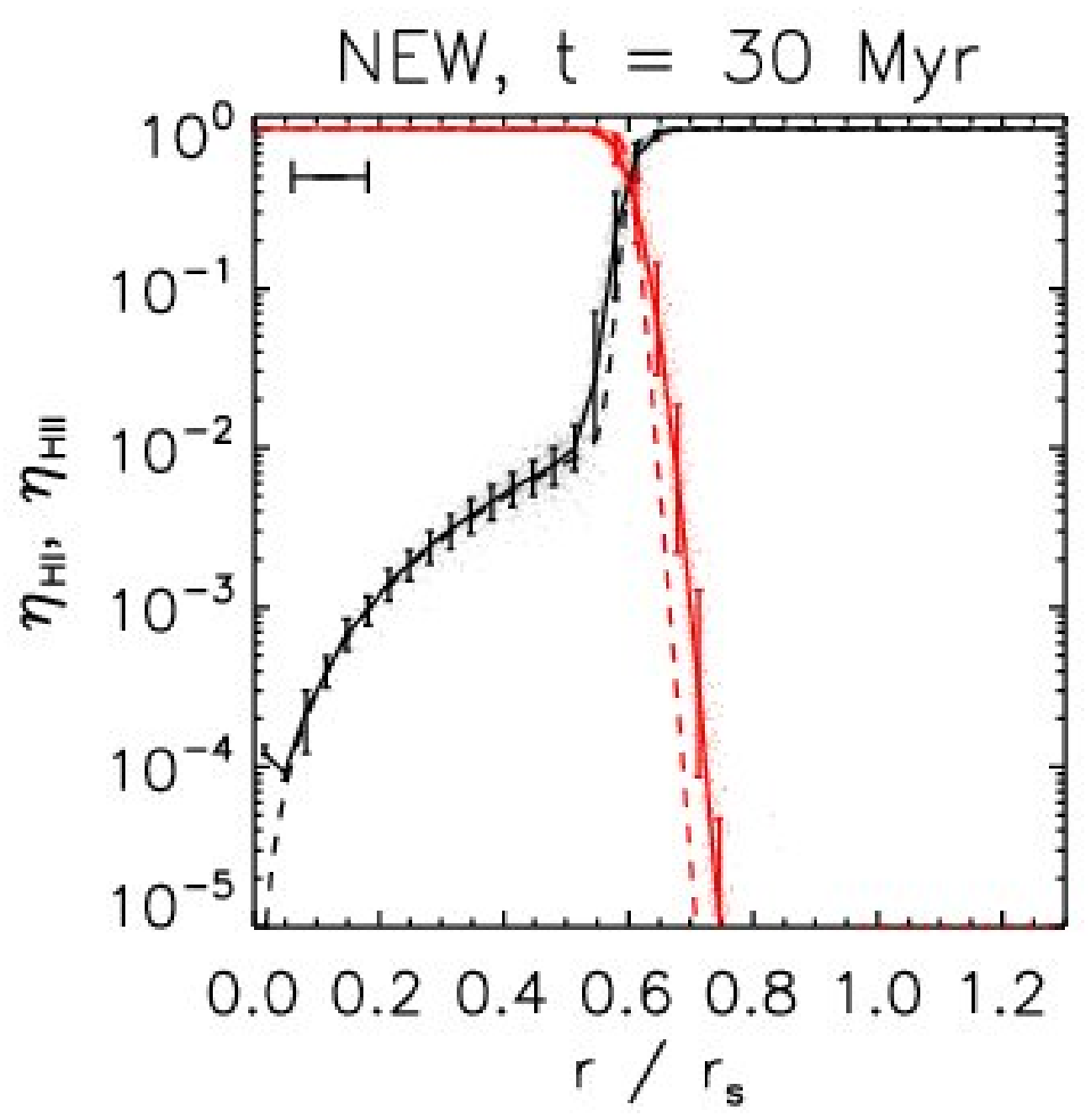}      
  \includegraphics[trim = 20mm 0mm 20mm 0mm, width=0.3\textwidth]{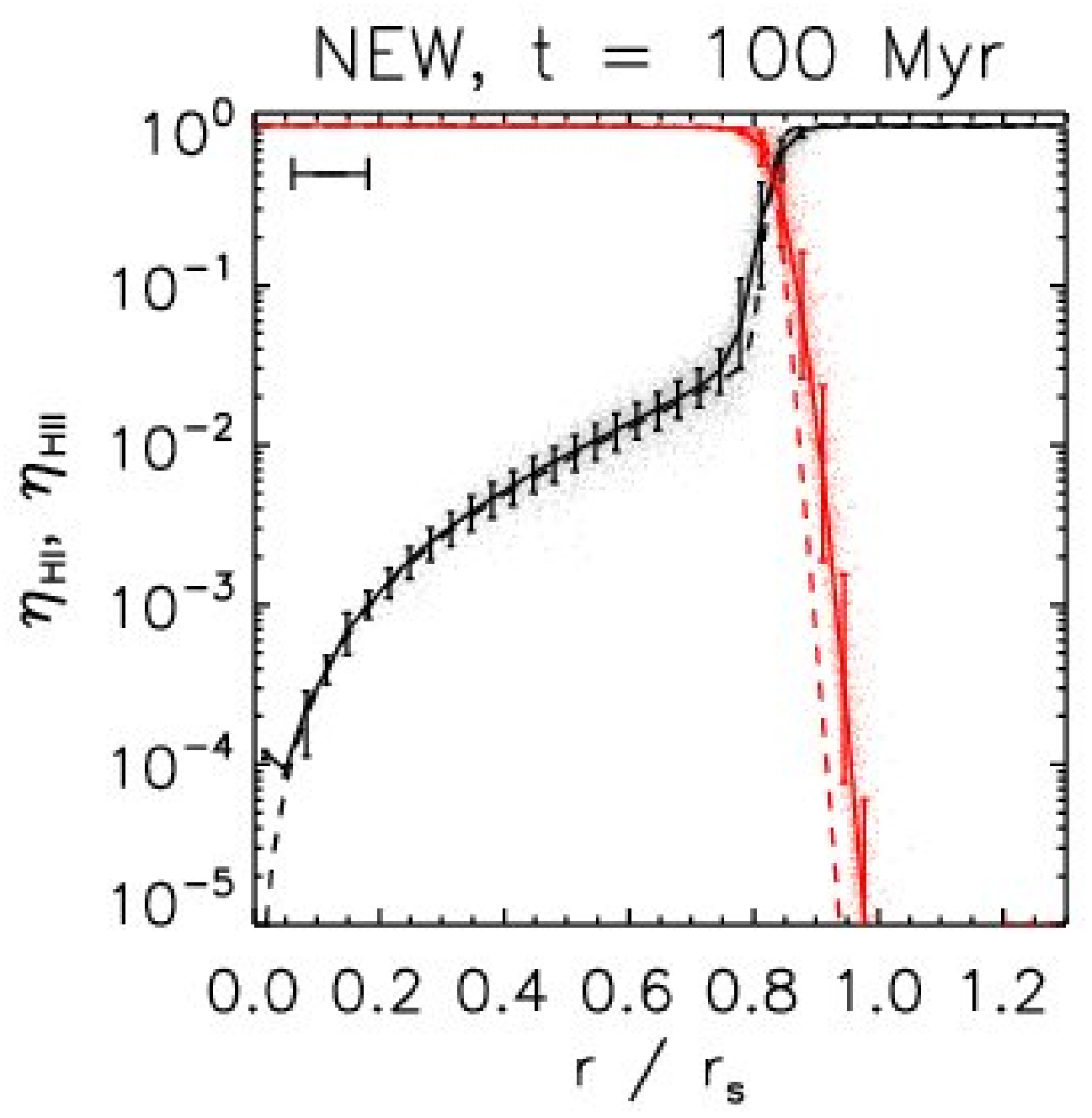}      
  \includegraphics[trim = 20mm 0mm 20mm 0mm, width=0.3\textwidth]{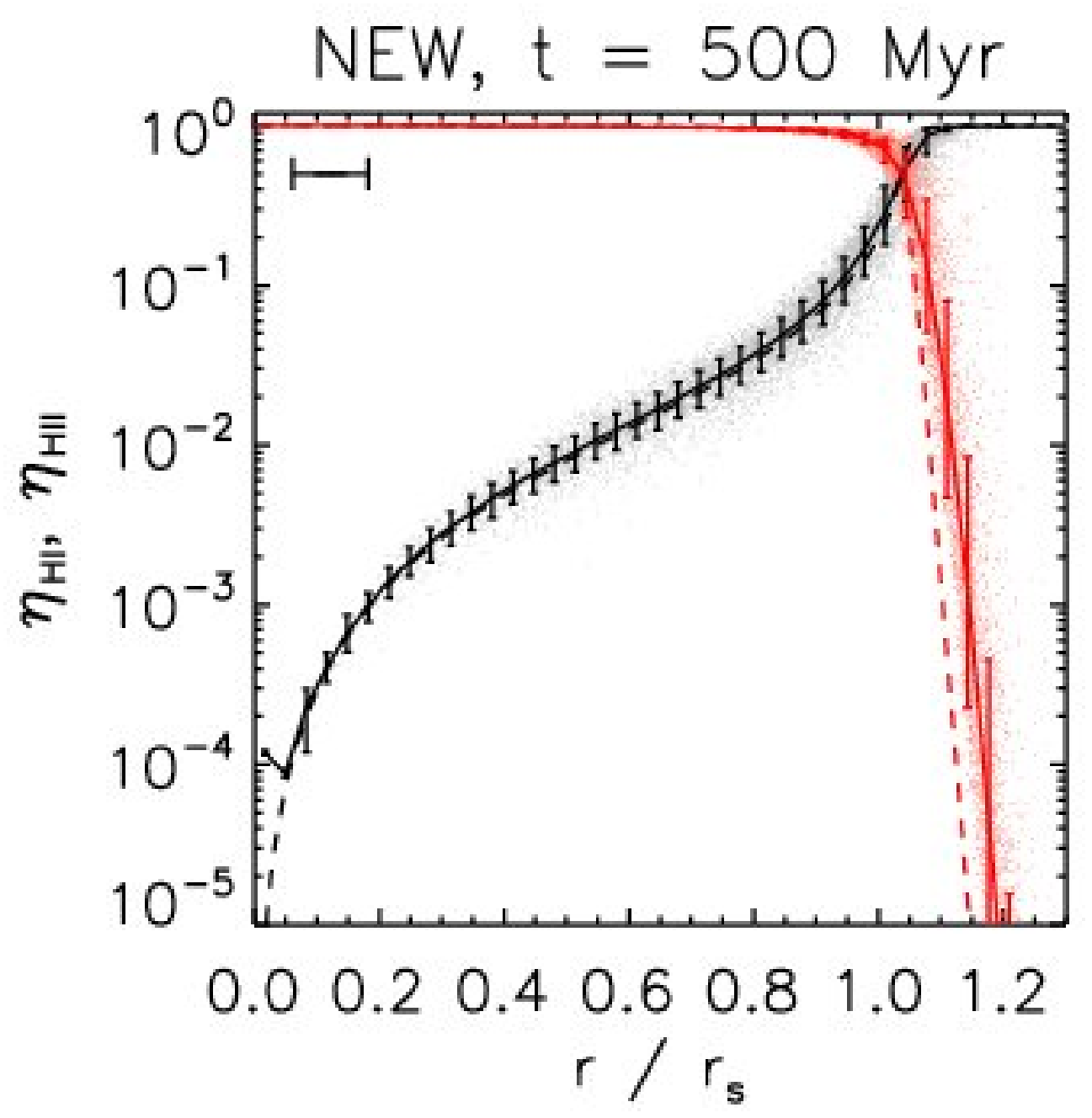}      \\
  \includegraphics[trim = 20mm 0mm 20mm 0mm, width=0.3\textwidth]{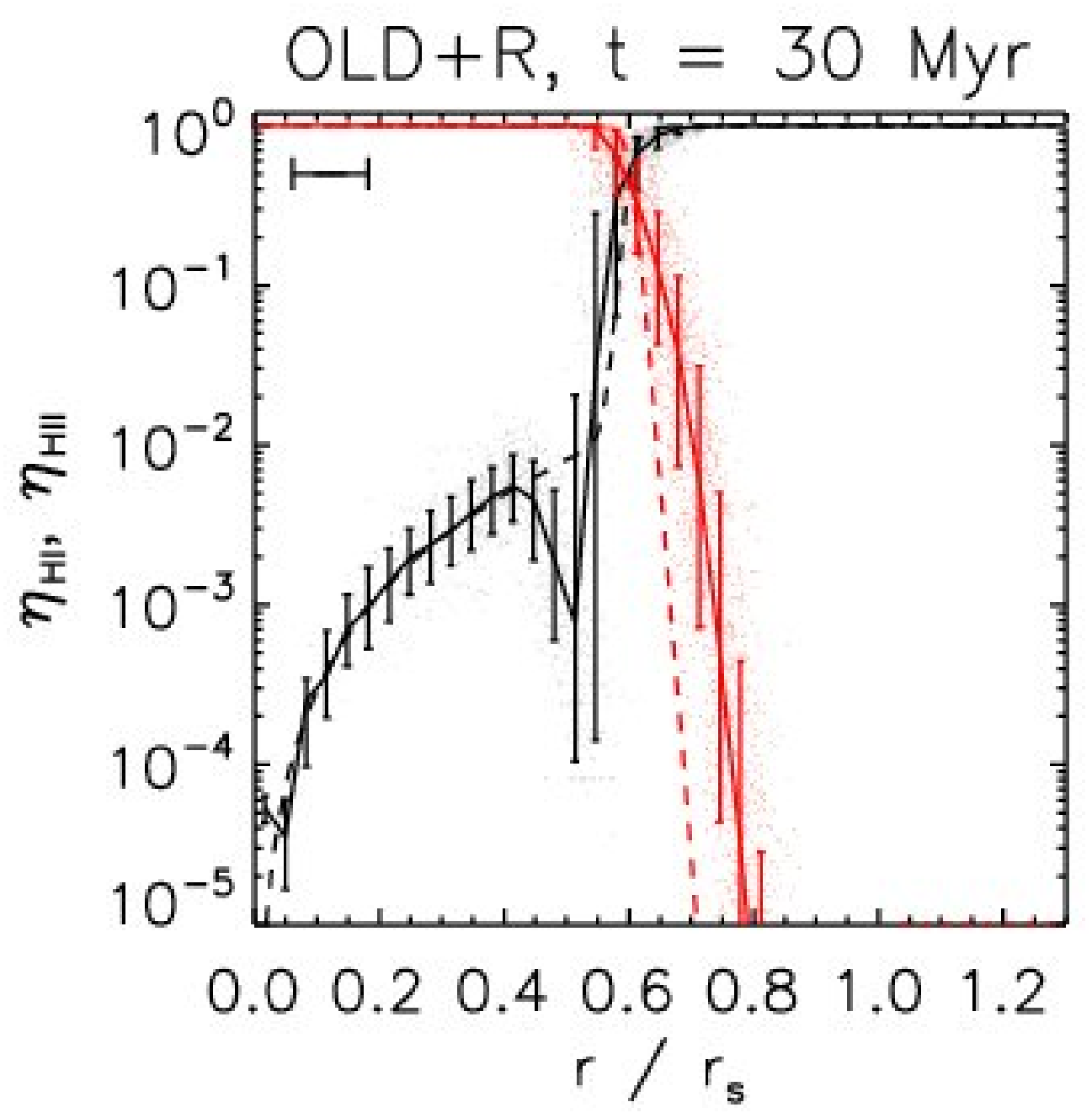}      
  \includegraphics[trim = 20mm 0mm 20mm 0mm, width=0.3\textwidth]{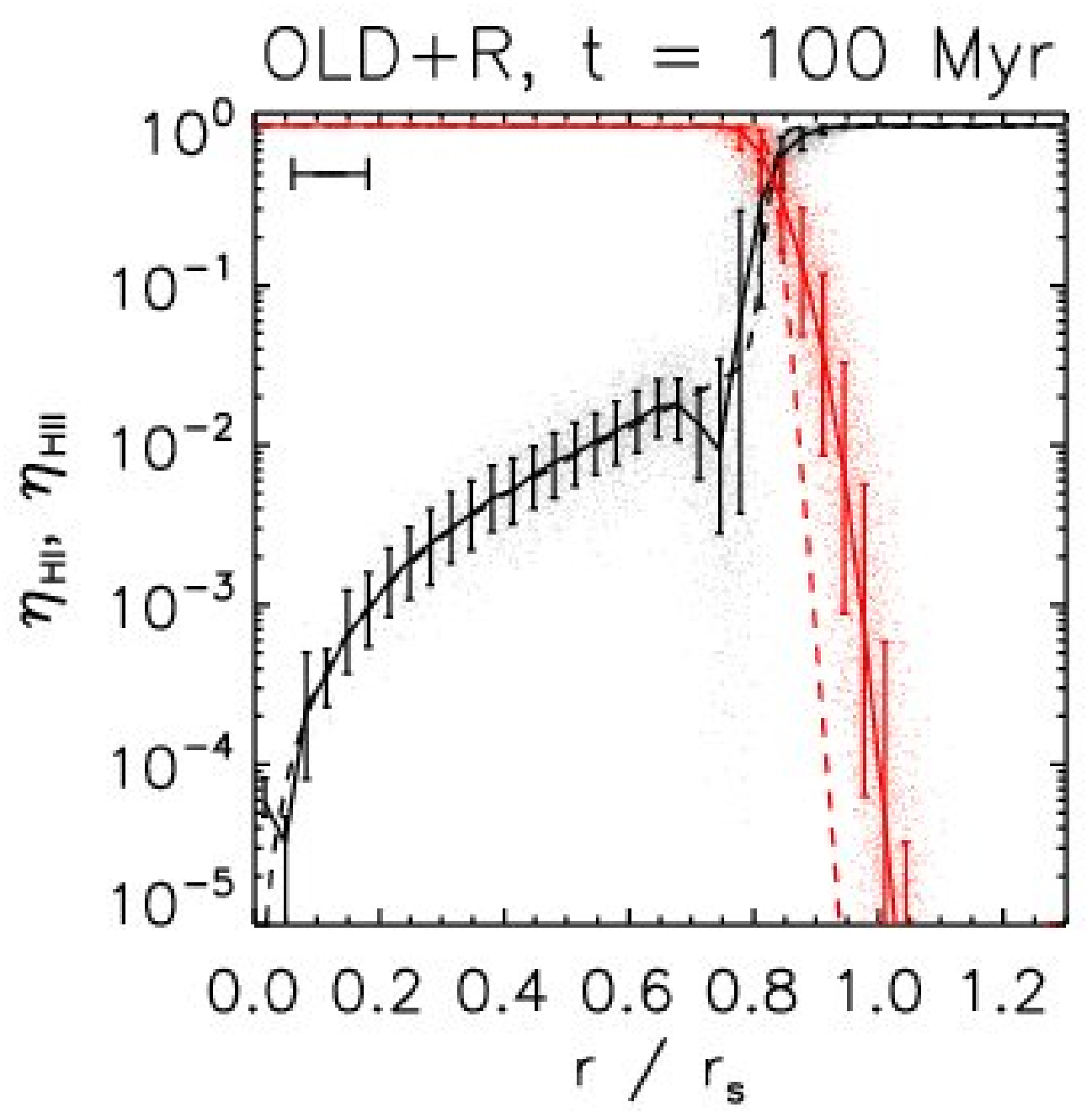}      
  \includegraphics[trim = 20mm 0mm 20mm 0mm, width=0.3\textwidth]{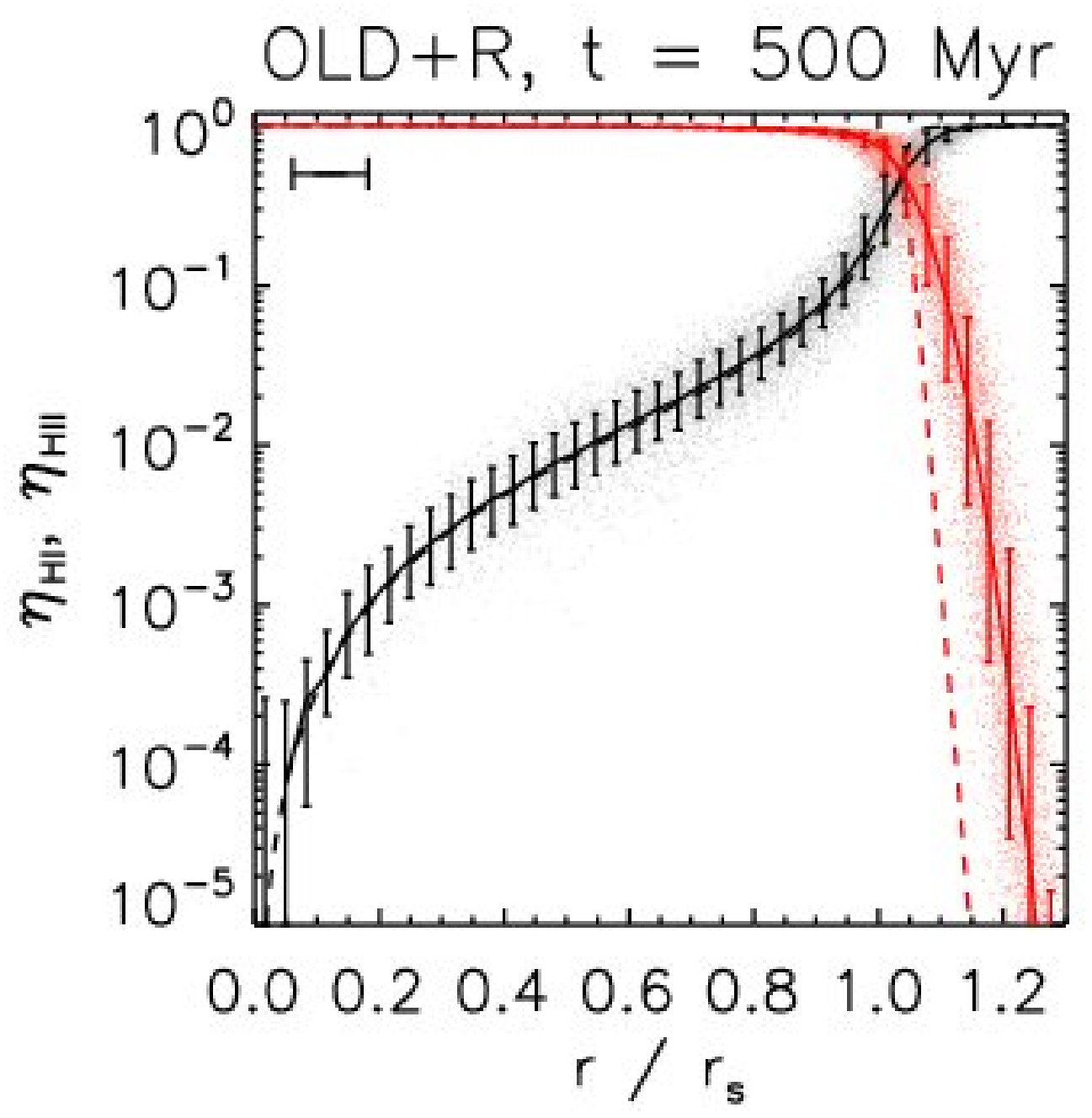}     \\ 
  \includegraphics[trim = 20mm 0mm 20mm 0mm, width=0.3\textwidth]{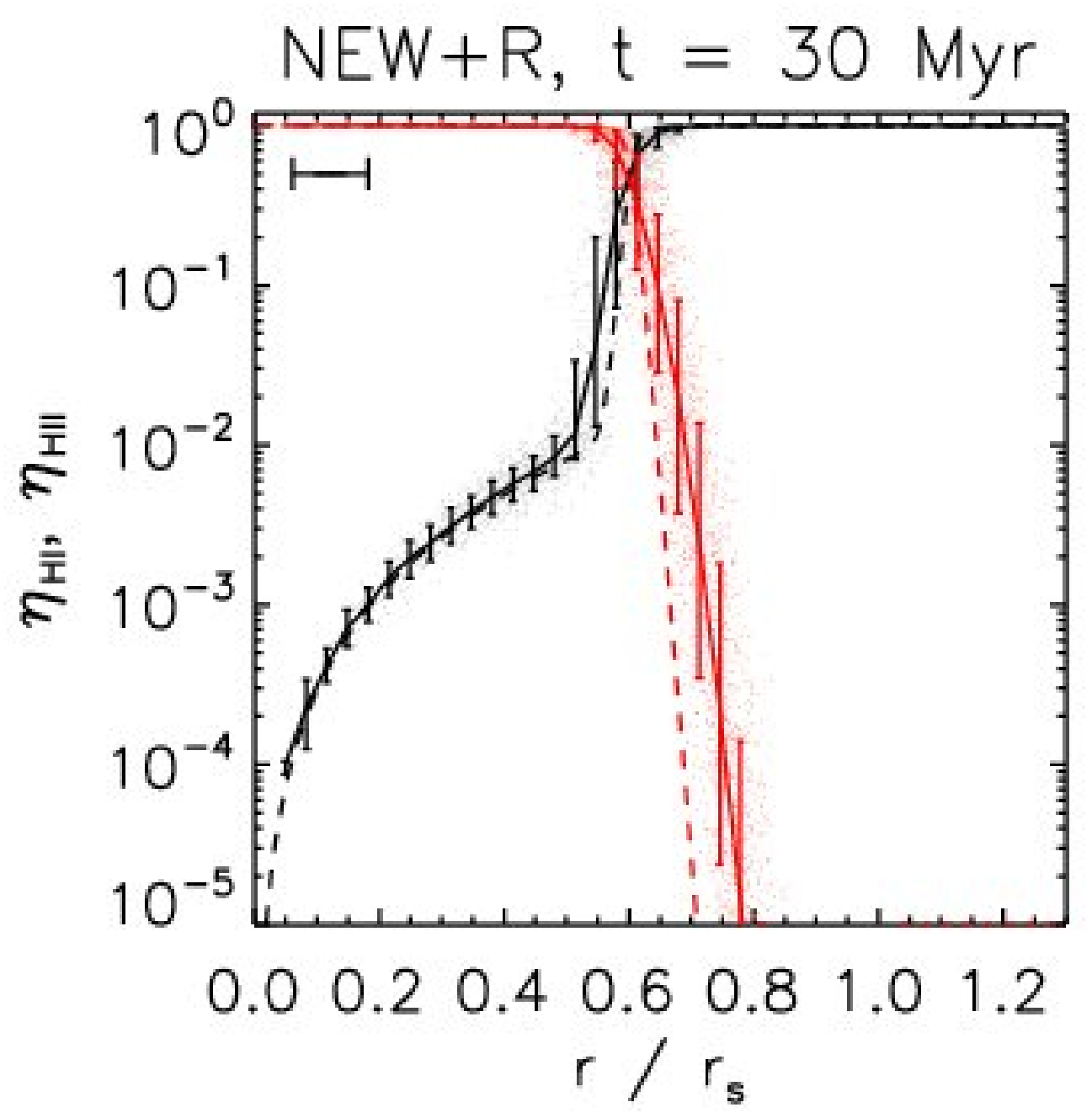}      
  \includegraphics[trim = 20mm 0mm 20mm 0mm, width=0.3\textwidth]{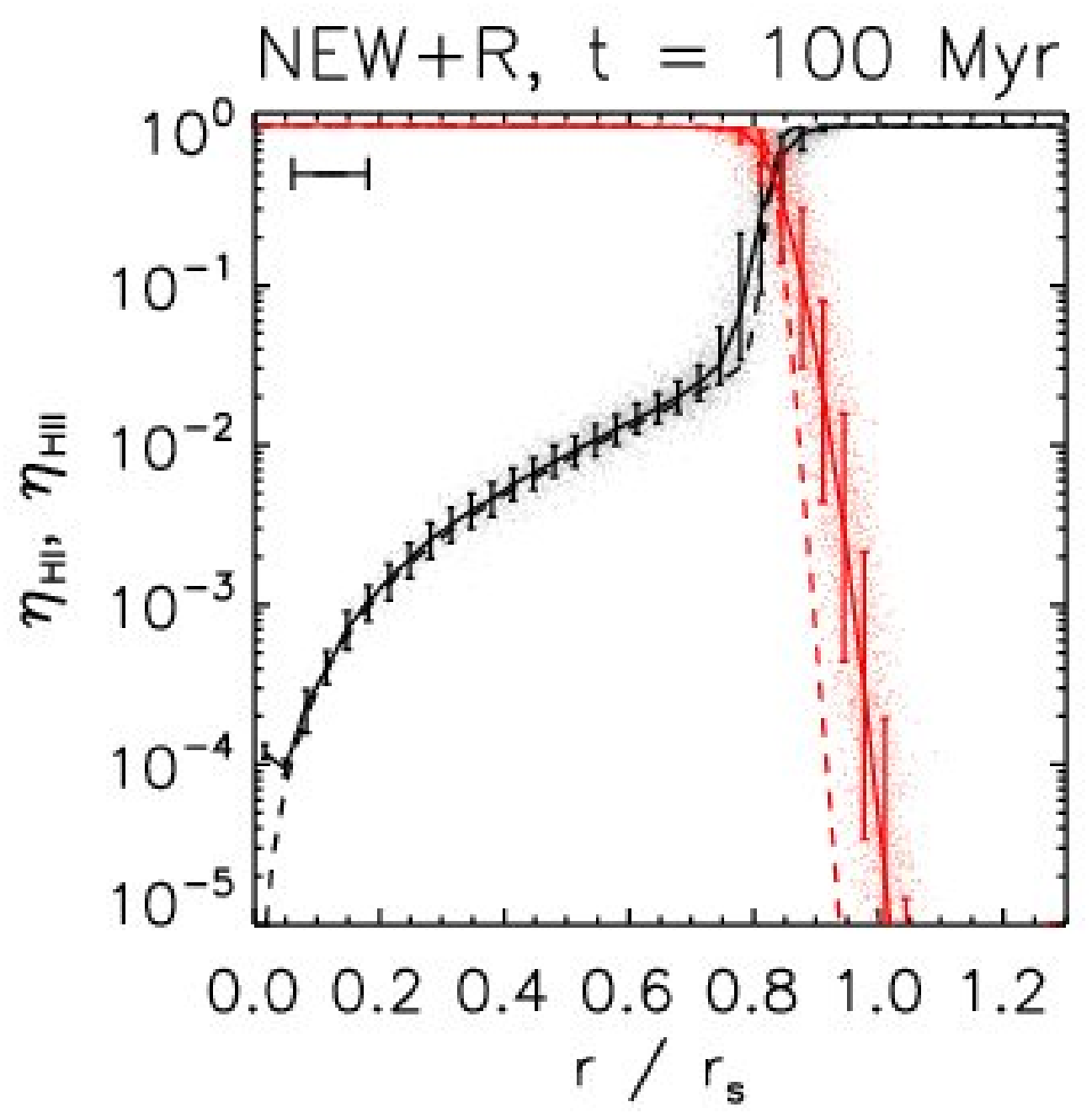}      
  \includegraphics[trim = 20mm 0mm 20mm 0mm, width=0.3\textwidth]{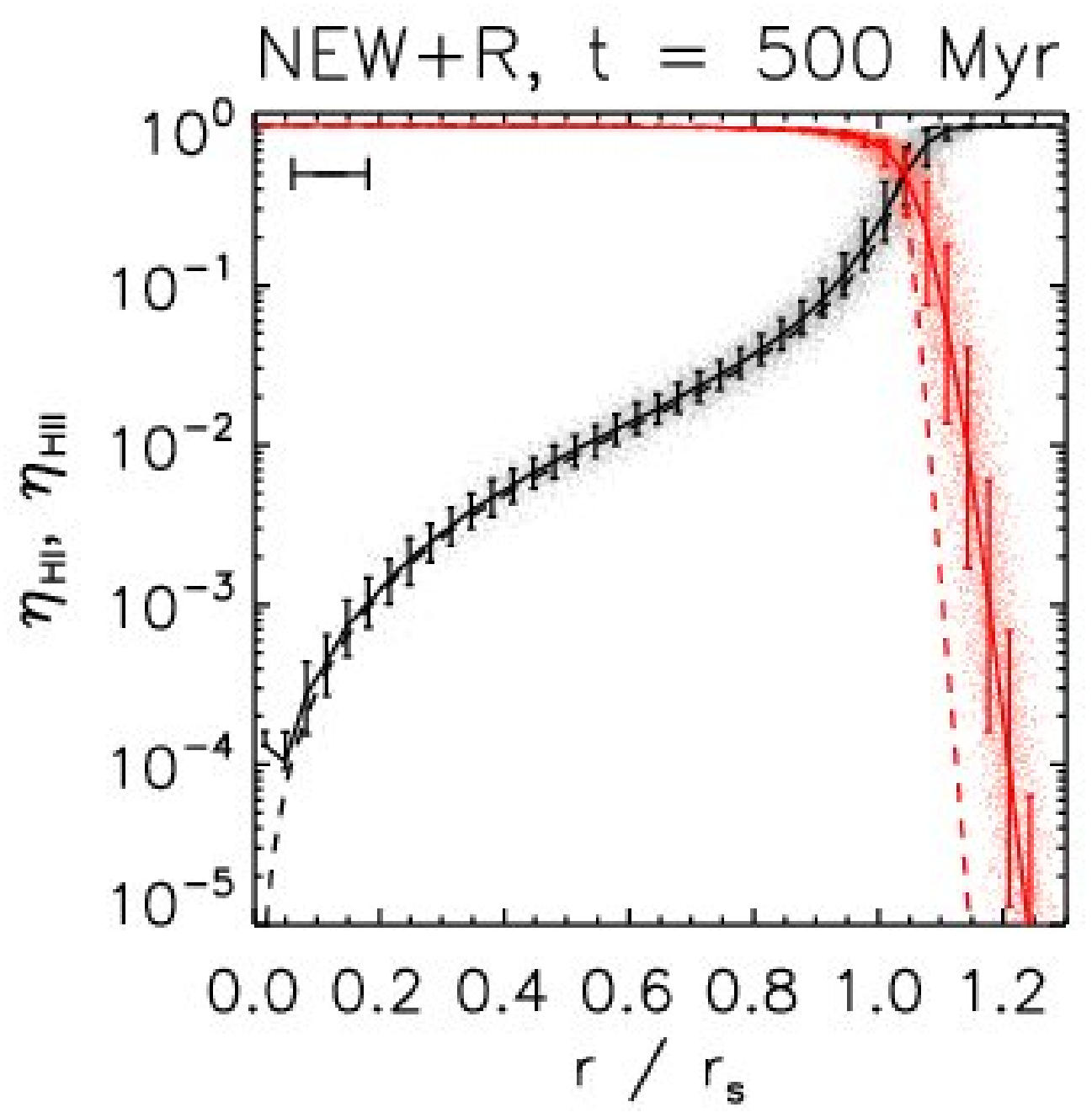}      \\

  \caption{Test 1. Neutral and ionised hydrogen fractions obtained in
    simulations with the old (\citealp{Pawlik:2008}; first and third
    row) and new (second and fourth row) implementations of
    \traphic. Shown are profiles of neutral and ionised fractions at times $t=30$ (left panel), $100$ (middle panel)
    and $500\Myr$ (right panel), for simulations with (second and
    fourth row) and without (first and third row) resampling of the
    density field. The spatial resolution is fixed to $N_{\rm SPH} =
    64^3$, $\tilde{N}_{\rm ngb} = 32$ and is indicated by the
    horizontal error bar in the upper left corner of each panel. The
    angular resolution is $N_{\rm c} = 128$.  The grey (light red)
    points show the neutral (ionised) hydrogen fraction for 
    a randomly chosen subset of $10\%$ of all particles. 
    The solid black (red) curve shows the median neutral
    (ionised) hydrogen fraction in spherical bins and the error bars
    enclose $68.3 \%$ of the particles in each bin.  The dashed black
    (red) curves show the exact solutions, obtained with our reference
    code \testtraphic. The underestimate of the non-equilibrium
    neutral fraction exhibited in simulations with the old
    implementation of \traphic\ is absent in the simulations that
    employ our new implementation, thanks to a new self-consistent
    manner of distributing photons absorbed by ViPs. The new
    implementation also reduces the scatter in the ionisation
    balance. \label{Fig:Test1:Artefact}} 

  \label{Fig:Artefact} 
\end{center}
\end{figure*}

The old treatment of ViPs results in an underestimate of the simulated
non-equilibrium neutral hydrogen
fractions. Fig.~\ref{Fig:Test1:Artefact} serves to demonstrate
this. Its panels show the neutral and ionised hydrogen fractions
around a single ionising source in a homogeneous hydrogen-only medium
at times $t=30, 100$ and $500 \Myr$ (from left to right) obtained with
the old (first and third rows) and new (second and fourth rows)
implementation. The setup and parameters for the simulations presented
here are identical to the setup and parameters used for the $N_{\rm
SPH}=64^3$, $N_{\rm c}=128$ simulation presented in Test~1 in
Sec.~5.3.1 of Paper~I. In addition to the neutral (grey dots) and
ionised (light red dots) fractions of each particle,
Fig.~\ref{Fig:Test1:Artefact} shows the median neutral (black solid
curves) and ionised (red solid curves) fractions in spherical
bins, which are compared to the exact solution obtained with our 1-d
RT code \testtraphic\
(Sec.~\ref{Sec:Tests:Test2}; dashed curves of the corresponding
colour). The error bars indicate the $68.3\%$ confidence intervals in
each bin.  For each implementation we have performed simulations both
with and without resampling the density field, as indicated by the
presence or absence of the letter `R' in the panel titles.
\par
In the simulations employing the old implementation of \traphic\ the
neutral hydrogen fractions at times $t = 30$ and $100 \Myr$ are
underestimated at radii slightly smaller than the radius of the
ionisation front.  In the simulations that employ the new
implementation this underestimate is no longer present, thanks to the
new manner in which the photons absorbed by ViPs are distributed.  At
$t = 500 \Myr$, i.e.\ when the ionised region has (nearly) reached its
equilibrium size, the underestimate is also absent in the simulations
that employ the old implementation. However, at this time these old
simulations still exhibit an increased scatter around the median when
compared to the corresponding snapshots from the simulations that
employ the new implementation.
\par
The underestimate of the neutral hydrogen fraction just behind
evolving ionisation fronts in simulations employing the old
implementation is caused by the fact that in this implementation the
distribution of the photons absorbed by ViPs does not respect the
spatial distribution of the neutral gas in their
surroundings. It mainly affects the neutral fraction of
particles close to evolving ionisation fronts, because the number of
photons absorbed and subsequently distributed by ViPs near the
ionisation front is significantly larger than the number of photons
that are absorbed by the SPH particles behind the ionisation front and
because the ViPs distribute the absorbed photons irrespective of the
neutral hydrogen mass with which the corresponding SPH particles
contributed to the computation of its neutral hydrogen density. 
\par
We did not notice the described temporary underestimate of the neutral
fraction just behind non-equilibrium ionisation fronts in the
simulations that we have presented in our original publication
(Paper~I), since there we only discussed profiles of the neutral
fraction at $t=500\Myr$. The reason why we limited ourselves
to discussion of equilibrium results in that publication, was that we were still lacking accurate non-equilibrium
reference solutions at that time (our 1-d reference RT code \testtraphic\ was still
under development). The discovery of the underestimate of the neutral
fraction was triggered by scatter plots of the neutral and
ionised hydrogen fractions like those presented in
Fig.~\ref{Fig:Test1:Artefact} that we have performed more recently.
\par

%\bsp
\label{lastpage}
\end{document}